\documentclass[twoside,11pt]{article}

\usepackage{jmlr2e}
\usepackage{graphicx}
\usepackage{amsmath,amssymb,amsbsy}

\newcommand{\frobn}[1]{\left\|#1\right\|_{F}}
\newcommand{\ltwon}[1]{\left\|#1\right\|_{2}}
\newcommand{\lonen}[1]{\left\|#1\right\|_{1}}

\newcommand{\infn}[1]{\left\|#1\right\|_{\infty}}
\newcommand{\opn}[1]{\left\|#1\right\|_{op}}

\newcommand{\bx}{\mbox{\bf x}}

\newcommand{\bH}{\mbox{\bf H}}

\newcommand{\bpi}{\mbox{\boldmath $\pi$}}

\newcommand{\bX}{\mbox{\bf X}}

\newcommand{\bY}{\mbox{\bf Y}}
\newcommand{\bZ}{\mbox{\bf Z}}

\newcommand{\bxi}{\mbox{\boldmath $\xi$}}

\newcommand{\bmu}{\mbox{\boldmath $\mu$}}

\newcommand{\bomega}{\mbox{\boldmath $\omega$}}

\newcommand{\bGamma}{\mbox{\boldmath $\Gamma$}}
\newcommand{\bepsilon}{\mbox{\boldmath $\epsilon$}}
\newcommand{\bigO}[1]{\mbox{$\mathcal{O}(#1)$}}
\newcommand{\littleo}[1]{\mbox{$o(#1)$}}
\newcommand{\Ex}[1]{\mbox{$\mathbb{E}(#1)$}}

\newcommand{\bgamma}{\mbox{\boldmath $\gamma$}}

\newcommand{\bweight}{\mbox{\boldmath $\omega$}}

\newcommand{\phivarmin}[3][]{\mbox{$\phi^{#1}_{#2}(#3)$}}

\newcommand{\sto}[1][]{\mbox{$V^{#1}_{k,21}$}}
\newcommand{\soo}[1][]{\mbox{$V^{#1}_{k,11}$}}
\newcommand{\hsto}[1][]{\mbox{$\hat{V}^{#1}_{k,21}$}}
\newcommand{\hsoo}[1][]{\mbox{$\hat{V}^{#1}_{k,11}$}}

\newcommand{\var}{\mathrm{var}}

\newtheorem{thm}{Theorem}[section]
\newtheorem{prp}{Proposition}[section]

\ShortHeadings{Two-Stage Penalized Least Squares}{Chen, Ren, Zhang and Zhang}
\firstpageno{1}

\begin{document}

\title{A Two-Stage Penalized Least Squares Method for Constructing Large Systems of Structural Equations}

\author{\name Chen Chen\thanks{The first two authors contribute equally.} \email chen1167@stat.purdue.edu
       \AND
       \name Min Ren\footnotemark[1] \email ren80@stat.purdue.edu
       \AND
       \name Min Zhang \email minzhang@stat.purdue.edu
       \AND
       \name Dabao Zhang \email zhangdb@stat.purdue.edu \\
       \addr Department of Statistics\\
       Purdue University\\
       West Lafayette, IN 47907, USA
       }

\editor{}

\maketitle

\begin{abstract}
We propose a two-stage penalized least squares method to build large systems of structural equations based on the instrumental variables view of the classical two-stage least squares method. We show that, with large numbers of endogenous and exogenous variables, the system can be constructed via consistent estimation of a set of conditional expectations at the first stage, and consistent selection of regulatory effects at the second stage. While the consistent estimation at the first stage can be obtained via the ridge regression, the adaptive lasso is employed at the second stage to achieve the consistent selection. This method is computationally fast and allows for parallel implementation. We demonstrate its effectiveness via simulation studies and real data analysis.\\
\end{abstract}


\section{Introduction}
\label{Sec-Intro}

We consider a linear system with $p$ endogenous and $q$ exogenous variables. With a sample of $n$ observations from this system, we denote the observed values of endogenous and exogenous variables by $\mathbf{Y}_{n\times p} = (\mathbf{Y}_1, \cdots, \mathbf{Y}_p)$ and $\mathbf{X}_{n\times q} = (\mathbf{X}_1, \cdots, \mathbf{X}_q)$, respectively. The interactions among endogenous variables and the direct causal effects by exogenous variables can be described by a system of structural equations,
\begin{eqnarray} \label{Eqn-FullInfo}
\mathbf{Y}=\mathbf{Y}\boldsymbol{\Gamma} + \mathbf{X}\boldsymbol{\Psi} + \boldsymbol{\epsilon},
\end{eqnarray}
where the $p\times p$ matrix $\boldsymbol{\Gamma}$ has zero diagonal elements and contains regulatory effects, the $q\times p$ matrix $\boldsymbol{\Psi}$ contains causal effects, and $\mathbf{\boldsymbol{\epsilon}}$ is an $n\times p$ matrix of error terms. We assume that $\mathbf{X}$ and $\mathbf{\boldsymbol{\epsilon}}$ are independent of each other, and each component of $\mathbf{\boldsymbol{\epsilon}}$ is independently distributed as normal with zero mean while rows of $\mathbf{\boldsymbol{\epsilon}}$ are identically distributed.

With gene expression levels and genotypic values as endogenous and exogenous variables, respectively, the model (\ref{Eqn-FullInfo}) has been used to represent a gene regulatory network with each equation modeling the regulatory genetic effects as well as the causal genomic effects from cis-eQTL (i.e., expression quantitative trait loci located within the regions of their target genes) on a given gene, see \citet{Xiong2004}, and \citet{Liu2008}, among others. Genetical genomics experiments, which collect genome-wide gene expressions and genotypic values, have been widely undertaken to construct gene regulatory networks \citep{Jansen2001, Schadt2003}. However, fitting a system of structural equations in (\ref{Eqn-FullInfo}) to genetical genomics data for the purpose of revealing a whole-genome gene regulatory network is still hindered by lack of an effective statistical method which addresses issues brought by large numbers of endogenous and exogenous variables.

Several efforts have been made to construct the system (\ref{Eqn-FullInfo}) with genetical genomics data. \citet{Xiong2004} proposed to use a genetic algorithm to search for genetic networks which minimize the Akaike Information Criterion (AIC; \citealp{Akaike1974}), and \citet{Liu2008} instead proposed to minimize the Bayesian Information Criterion (BIC; \citealp{Schwartz1978}) and its modification \citep{Broman2002} for the optimal genetic networks. Both AIC and BIC are applicable to inferring networks for only a small number of endogenous variables. For a large system with many endogenous and exogenous variables, \citet{Cai2013} proposed to maximize a penalized likelihood to construct a sparse system. However, it is computationally formidable to fit a large system based on the likelihood function of the complete model. \citet{Logsdon2010} instead proposed to apply the adaptive lasso \citep{Zou2006} to fitting each structural equation separately, and then recover the network relying on additional assumption on unique exogenous variables. However, \citet{Cai2013} demonstrated its inferior performance via simulation studies, which is consistent with our conclusion.

Instead of the full information model specified in (\ref{Eqn-FullInfo}), we seek to establish the large system via constructing a large number of limited information models, each for one endogenous variable \citep{Schmidt1976}. For example, when the $k$-th endogenous variable is concerned, we focus on the $k$-th structural equation in (\ref{Eqn-FullInfo}) which models the regulatory effects of other endogenous variables and direct causal effects of exogenous variables, and ignore the system structures contained in other structural equations, leading to the following limited-information model,
\begin{eqnarray} \label{Eqn-LimitedInfo}
\left\{\begin{array}{l} \mathbf{Y}_k = \mathbf{Y}_{-k}\boldsymbol{\gamma}_k + \mathbf{X}\boldsymbol{\psi}_k+\boldsymbol{\epsilon}_k,\\
\mathbf{Y}_{-k} = \mathbf{X}\mathbf{\boldsymbol{\pi}}_{-k} + \mathbf{\boldsymbol{\xi}}_{-k}. \end{array}\right.
\end{eqnarray}
Here $\mathbf{Y}_{-k}$ refers to  $\mathbf{Y}$ excluding the $k$-th column, $\boldsymbol{\gamma}_k$ refers to the $k$-th column of $\boldsymbol{\Gamma}$ excluding the diagonal zero, and $\boldsymbol{\psi}_k$ and $\boldsymbol{\epsilon}_k$ refer to the $k$-th columns of $\boldsymbol{\Psi}$ and $\boldsymbol{\epsilon}$ respectively. The second part of the model (\ref{Eqn-LimitedInfo}) is from the following reduced model by excluding the $k$-th reduced-form equation, with $\boldsymbol{\pi} = \boldsymbol{\Psi}(\mathbf{I}-\boldsymbol{\Gamma})^{-1}$ and $\boldsymbol{\xi} = \boldsymbol{\epsilon}(\mathbf{I}-\boldsymbol{\Gamma})^{-1}$,
\begin{eqnarray} \label{Eqn-ReducedForm}
\mathbf{Y} = \mathbf{X}\boldsymbol{\pi} + \boldsymbol{\xi}.
\end{eqnarray}

In a classical low-dimensional setting, applying the ordinary least squares method to the first equation in (\ref{Eqn-LimitedInfo}) leads to underestimated $\boldsymbol{\gamma}_k$ and $\boldsymbol{\psi}_k$ due to correlated $\mathbf{Y}_{-k}$ and $\boldsymbol{\epsilon}_k$. Instead, the reduced-form equations in (\ref{Eqn-LimitedInfo}) are fitted to obtain least squares estimator $\hat{\mathbf{\boldsymbol{\pi}}}_{-k}$ of $\mathbf{\boldsymbol{\pi}}_{-k}$, and least squares estimators of $\boldsymbol{\gamma}_k$ and $\boldsymbol{\psi}_k$ are further obtained by regressing $\mathbf{Y}_k$ against $\hat{\mathbf{Y}}_{-k}=\mathbf{X}\hat{\mathbf{\boldsymbol{\pi}}}_{-k}$ and $\mathbf{X}$. This procedure is widely known as the two-stage least squares (2SLS) method which can produce consistent estimates of the parameters when the system is identifiable. The 2SLS estimator was originally proposed by \citet{Theil1953a, Theil1953b, Theil1961} and, independently, \citet{Basmann1957}, and can be restated as the instrumental variables estimator \citep{Reiersol1941, Reiersol1945}.

As in a typical genetical genomics experiment, we are interested in constructing a large system with the number of endogenous variables $p$ possibly larger than the sample size $n$. Such a high-dimensional and small sample size data set makes it infeasible to directly apply the 2SLS method. Indeed, $p\ge n$ may result in perfect fits of reduced-form equations at the first stage, which implies that we regress against the observed values of endogenous variables at the second stage and therefore obtain ordinary least squares estimates of the parameters. It is well known that such ordinary least squares estimates are inconsistent. Furthermore, constructing a large system demands, at the second stage, selecting regulatory endogenous variables among massive candidates, i.e., variable selection in fitting high-dimensional linear models.

In the setting of selecting instrumental variables (IVs) among a large number of candidates, $L_1$ regularized least squares estimators have been recently proposed to replace the ordinary least squares estimator at the first stage of 2SLS \citep{Belloni2012, Lin2015, Zhu2015}. \citet{Belloni2012} applied lasso-based methods to select IVs and obtain consistent estimations at the first stage when the first stage is approximately sparse. For sparse instrumental variables models, \citet{Zhu2015} proposed to replace with lasso-based methods at both stages of 2SLS and \citet{Lin2015} considered the representative $L_1$ regularization methods and a class of concave regularization methods for both stages. All of these methods assume that each endogenous variable is only associated to a relatively small set of exogenous variables, i.e., each row of $\boldsymbol{\pi}$ in (\ref{Eqn-ReducedForm}) only has a small set of nonzero components.

Here we consider to construct a general system of structural equations, which allows us to model nonrecursive or even cyclic relationships between endogenous variables. With the instrumental variables view of the two-stage approach, we observed that successful identification and consistent estimation of model parameters rely on consistent estimation of a set of conditional expectations which are optimal instruments. Therefore, establishing the system (\ref{Eqn-FullInfo}) in a high-dimensional setting is contingent on obtaining consistent estimation of these conditional expectations at the first stage, and effectively selecting and estimating of regulatory effects out of a large number of candidates at the second stage. Accordingly, we propose a two-stage penalized least squares (2SPLS) method to fit regularized linear models at each stage, with $L_2$ regularized linear models at the first stage and $L_1$ regularized linear models at the second stage.

The proposed method addresses three challenging issues in constructing a large system of structural equations, i.e., memory capacity, computational time, and statistical power. First, the limited information models are considered to develop the algorithm. In this way, we avoid working with full information models which may consist of many subnetworks and involve a massive number of endogenous variables. Second, allowing us to fit one linear model for each endogenous variable at each stage makes the algorithm computationally fast. It also makes it feasible to parallelize the large number of model fittings at each stage. Finally, the oracle properties of the resultant estimates show that the proposed method can achieve optimal power in identifying and estimating regulatory effects. Furthermore, the efficient computation makes it feasible to use the bootstrap method to evaluate the significance of regulatory effects.

The rest of this paper is organized as follows. First, we state an identifiable model in Section~\ref{Sec-SEM}. Section~\ref{Sec-2SLS} revisits the instrumental variables view on the classical 2SLS method, which motivates our development of the 2SPLS method in Section~\ref{Sec-2SPLS}. We show in Section~\ref{Sec-Theory} the theoretical properties of the estimates from 2SPLS, with the proof included in the Appendix. Simulation studies are carried out in Section~\ref{Sec-Simu} to evaluate the performance of 2SPLS. An application to a real data set to infer a yeast gene regulatory network is presented in Section~\ref{Sec-RData}. We conclude this paper with a discussion in Section~\ref{Sec-Disc}.

\section{The Identifiable Model} \label{Sec-SEM}

We follow the practice of constructing system (\ref{Eqn-FullInfo}) in analyzing genetical genomics data \citep{Logsdon2010, Cai2013}, and assume that each endogenous variable is affected by a unique set of exogenous variables, that is, the structural equation in (\ref{Eqn-LimitedInfo}) has known zero elements of $\boldsymbol{\psi}_k$. Explicitly, we use $\mathcal{S}_k$ to denote the set of row indices of known nonzero elements in $\boldsymbol{\psi}_k$. Then we have known sets $\mathcal{S}_k, k=1, 2, \cdots, p$, which dissect the set $\{1, 2, \cdots, q\}$. We explicitly state this assumption in the below.

\begin{description}
\item[]{\bf Assumption A.} $\mathcal{S}_k\ne\emptyset$ for $k=1,\cdots, p$, but $\mathcal{S}_j\cap\mathcal{S}_k=\emptyset$ as long as $j\ne k$.
\end{description}

The above assumption satisfies the rank condition \citep{Schmidt1976}, which is a sufficient condition for model identification. Since each $\boldsymbol{\psi}_k$ has a set of known zero components, from this point forward we ignore them and rewrite the structural equation in the model (\ref{Eqn-LimitedInfo}) as,
\begin{eqnarray} \label{Eqn-YkStructural}
\mathbf{Y}_k = \mathbf{Y}_{-k}\boldsymbol{\gamma}_k + \mathbf{X}_{\mathcal{S}_k}\boldsymbol{\psi}_{\mathcal{S}_k} + \boldsymbol{\epsilon}_k, \ \ \ \ \ \boldsymbol{\epsilon}_k\sim N(\mathbf{0}, \sigma_k^2 \mathbf{I}_n),
\end{eqnarray}
where $\mathbf{X}_{\mathcal{S}_k}$ refers to $\mathbf{X}$ including only columns indicated by $\mathcal{S}_k$, and $\boldsymbol{\psi}_{\mathcal{S}_k}$ refers to $\boldsymbol{\psi}_k$ including only elements indicated by $\mathcal{S}_k$.

\section{The Instrumental Variables View of the Two-Stage Least Squares Method} \label{Sec-2SLS}

Because $\mathbf{Y}_{-k}$ and $\boldsymbol{\epsilon}_k$ are correlated, fitting merely the model (\ref{Eqn-YkStructural}) results in biased estimates of $\boldsymbol{\gamma}_k$ and $\boldsymbol{\psi}_{\mathcal{S}_k}$. However, the following two sets of variables are independent,
\begin{eqnarray*}
\left\{\begin{array}{l}
\mathbf{Z}_{-k} =  E[\mathbf{Y}_{-k}|\mathbf{X}] = \mathbf{X}\boldsymbol{\pi}_{-k},\\
\boldsymbol{\varepsilon}_k = \boldsymbol{\epsilon}_k+\boldsymbol{\xi}_{-k}\boldsymbol{\gamma}_k. \end{array}\right.
\end{eqnarray*}
Consequently, consistent estimates of $\boldsymbol{\gamma}_k$ and $\boldsymbol{\psi}_{\mathcal{S}_k}$ can be obtained by applying least squares method to the following model,
\begin{eqnarray} \label{Eqn-IdealModel}
\mathbf{Y}_k = \mathbf{Z}_{-k}\boldsymbol{\gamma}_k + \mathbf{X}_{\mathcal{S}_k}\boldsymbol{\psi}_{\mathcal{S}_k} + \boldsymbol{\varepsilon}_k.
\end{eqnarray}

Observing $\mathbf{Y}_{-k}$ instead of $\mathbf{Z}_{-k} =  E[\mathbf{Y}_{-k}|\mathbf{X}]$ naturally leads to application of the instrumental variables method \citep{Reiersol1941, Reiersol1945}, that is, replacing $\mathbf{Z}_{-k} = \mathbf{X}\boldsymbol{\pi}_{-k}$ with its estimate  $\hat{\mathbf{Z}}_{-k} = \mathbf{X}\hat{\boldsymbol{\pi}}_{-k}$ in fitting the linear model (\ref{Eqn-IdealModel}). When a $\sqrt{n}$-consistent least squares estimator of $\boldsymbol{\pi}_{j}$ is obtained by fitting each equation in (\ref{Eqn-ReducedForm}) for $j=1, \cdots, p$, the resultant estimators of $\boldsymbol{\gamma}_k$ and $\boldsymbol{\psi}_{\mathcal{S}_k}$ are exactly the 2SLS estimators by \citet{Theil1953a, Theil1953b, Theil1961} and \cite{Basmann1957}.

Suppose that the matrix $\mathbf{X}$ satisfies the assumption in the below. It is easy to prove that, in a low-dimensional setting, we can obtain consistent estimators for the model (\ref{Eqn-IdealModel}) with any consistent estimate of $\boldsymbol{\pi}_{-k}$.

\begin{description}
\item[] {\bf Assumption B.} $n^{-1}\mathbf{X}^T\mathbf{X}\to\mathbf{C}$, where $\mathbf{C}$ is a positive definite matrix.
\end{description}

\begin{prp} \label{Thm-2SLS} Suppose Assumptions A and B are satisfied for the system (\ref{Eqn-FullInfo}) with fixed $p\ll n$ and $q\ll n$. When there exists a consistent estimator $\hat{\boldsymbol{\pi}}_{-k}$ of $\boldsymbol{\pi}_{-k}$, the ordinary least squares estimators of $(\boldsymbol{\gamma}_k, \boldsymbol{\psi}_{\mathcal{S}_k})$ obtained by regressing $\mathbf{Y}_k$ against $(\mathbf{X}\hat{\boldsymbol{\pi}}_{-k}, \mathbf{X}_{\mathcal{S}_k})$ are also consistent.
\end{prp}

The above instrumental variables view implies that the conditional expectation $\mathbf{Z}_{-k} = E[\mathbf{Y}_{-k}|\mathbf{X}]$ serves as the optimal instrument for $\mathbf{Y}_{-k}$. Although, in a low-dimensional setting, any consistent estimator $\hat{\boldsymbol{\pi}}_{-k}$ leads to the instrument $\hat{\mathbf{Z}}_{-k} = \mathbf{X}\hat{\boldsymbol{\pi}}_{-k}$, an efficient estimate of $\boldsymbol{\pi}_{-k}$ should be used to produce efficient estimates of $\boldsymbol{\gamma}_k$ and $\boldsymbol{\psi}_{\mathcal{S}_k}$. In the following section, we build up on this view and construct the high-dimensional system (\ref{Eqn-FullInfo}) by first fitting high-dimensional linear models to consistently estimate the conditional expectations of endogenous variables given exogenous variables.

\section{The Two-Stage Penalized Least Squares Method} \label{Sec-2SPLS}

To construct the limited-information model (\ref{Eqn-LimitedInfo}), we can obtain consistent estimates of the conditional expectations of endogenous variables given exogenous variables by fitting high-dimensional linear models, and then conduct a high-dimensional variable selection following our view on the model (\ref{Eqn-IdealModel}). Accordingly, we propose a two-stage penalized least squares (2SPLS) procedure to construct each model in (\ref{Eqn-LimitedInfo}) so as to establish the large system (\ref{Eqn-FullInfo}).

\subsection{The Method}

At the first stage, we use the ridge regression to fit each reduced-form equation in (\ref{Eqn-ReducedForm}) to obtain consistent estimates of the conditional expectations of endogenous variables given exogenous variables, that is, for each $j=1, 2, \cdots, p$, we obtain the ridge regression estimator of $\boldsymbol{\pi}_j$ by minimizing the following penalized sum of squares,
\begin{eqnarray} \label{Est-pi}
 \|\mathbf{Y}_j-\mathbf{X}\boldsymbol{\pi}_j\|_2^2 + \tau_j \|\boldsymbol{\pi}_j\|_2^2,
\end{eqnarray}
where $\|\cdot\|_2$ is the $L_2$ norm, and $\tau_j>0$ is a tuning parameter that controls the strength of the penalty. The solution to the minimization problem is  $\hat{\boldsymbol{\pi}}_j=(\mathbf{X}^T\mathbf{X}+\tau_j\mathbf{I})^{-1}\mathbf{X}^T\mathbf{Y}_j$, which leads to a consistent estimate of $\mathbf{Z}_j$,
\begin{eqnarray*} 
\hat{\mathbf{Z}}_j = \mathbf{P}_{\tau_j} \mathbf{Y}_j,
\end{eqnarray*}
where $\mathbf{P}_{\tau_j} = \mathbf{X}(\mathbf{X}^T\mathbf{X}+\tau_j\mathbf{I})^{-1}\mathbf{X}^T$. With a proper choice of $\tau_j$, the ridge regression has a good estimation performance as shown in the next section.

At the second stage, we replace $\mathbf{Z}_{-k}$ with $\hat{\mathbf{Z}}_{-k}$ in model (\ref{Eqn-IdealModel}) to derive estimates of $\boldsymbol{\gamma}_k$ and $\boldsymbol{\psi}_{\mathcal{S}_k}$, specifically, we minimize the following penalized error squares to obtain estimates of $\boldsymbol{\gamma}_k$ and $\boldsymbol{\psi}_{\mathcal{S}_k}$,
\begin{eqnarray} \label{Est-gammapsi}
\frac{1}{2} \|\mathbf{Y}_k-\hat{\mathbf{Z}}_{-k}\boldsymbol{\gamma}_k-\mathbf{X}_{\mathcal{S}_k}\boldsymbol{\psi}_{\mathcal{S}_k}\|_2^2 + \lambda_k \boldsymbol{\omega}_{k}^T|\boldsymbol{\gamma}_k|,
\end{eqnarray}
where $|\boldsymbol{\gamma}_k|$ denotes componentwise absolute value of $\boldsymbol{\gamma}_k$, $\boldsymbol{\omega}_k$ is a known weight vector, and $\lambda_k>0$ is a tuning parameter.

Minimizing for $\boldsymbol{\psi}_{\mathcal{S}_k}$ in (\ref{Est-gammapsi}) leads to
\begin{eqnarray*} 
\hat{\boldsymbol{\psi}}_{\mathcal{S}_k}=(\mathbf{X}_{\mathcal{S}_k}^T \mathbf{X}_{\mathcal{S}_k})^{-1} \mathbf{X}_{\mathcal{S}_k}^T (\mathbf{Y}_k-\hat{\mathbf{Z}}_{-k}\boldsymbol{\gamma}_{k}),
\end{eqnarray*}
where $\mathbf{X}_{\mathcal{S}_k}$ is usually of low dimension, and the above least squares estimator of $\boldsymbol{\psi}_{\mathcal{S}_k}$ is easy to obtain.

Plugging $\hat{\boldsymbol{\psi}}_{\mathcal{S}_k}$ into (\ref{Est-gammapsi}), we can solve the following minimization problem to obtain an estimate of $\boldsymbol{\gamma}_k$,
\begin{eqnarray} \label{Est-gamma}
\hat{\boldsymbol{\gamma}}_{k} = \arg\min_{\boldsymbol{\gamma}_{k}} \left\{\frac{1}{2}(\mathbf{Y}_k-\hat{\mathbf{Z}}_{-k}\boldsymbol{\gamma}_{k})^T\mathbf{H}_k(\mathbf{Y}_k-\hat{\mathbf{Z}}_{-k}\boldsymbol{\gamma}_{k}) +\lambda_k \mathbf{\boldsymbol{\omega}}_{k}^T|\boldsymbol{\gamma}_k|\right\},
\end{eqnarray}
where $\mathbf{H}_k=\mathbf{I}-\mathbf{X}_{\mathcal{S}_k} (\mathbf{X}_{\mathcal{S}_k}^T \mathbf{X}_{\mathcal{S}_k})^{-1} \mathbf{X}_{\mathcal{S}_k}^T$, this is equivalent to a variable selection problem in regressing $\mathbf{H}_k\mathbf{Y}_k$ against high-dimensional $\mathbf{H}_k\hat{\mathbf{Z}}_{-k}$. We will resort to adaptive lasso to select nonzero components of $\boldsymbol{\gamma}_k$ and estimate them. Specifically, picking up a $\delta>0$ and obtaining $\tilde{\boldsymbol{\gamma}}_k$ as a $\sqrt{n}$-consistent estimate of $\boldsymbol{\gamma}_k$, we calculate the weight vector $\boldsymbol{\omega}_k$ with components inversely proportional to components of $|\tilde{\boldsymbol{\gamma}}_k|^\delta$. The above minimization problem (\ref{Est-gamma}) is a convex optimization problem which is computationally efficient.

\subsection{Tuning Parameter Selection}

In this method, we need to select tuning parameters at each stage. At the first stage, we propose to choose each $\tau_j$ in (\ref{Est-pi}) by the method of generalized cross-validation (GCV; \citealp{Golub1979}), that is,
\[
\tau_j = \arg\min_{\tau>0} G_j(\tau) = \arg\min_{\tau>0}\frac{(\mathbf{Y}_j-\mathbf{P}_{\tau}\mathbf{Y}_j)^T(\mathbf{Y}_j-\mathbf{P}_{\tau}\mathbf{Y}_j)} {(n-\mathrm{tr}\{\mathbf{P}_{\tau}\})^2}.
\]
It is a rotation-invariant version of ordinary cross-validation, and leads to an approximately optimal estimate of the conditional expectation $\mathbf{Z}_j$. At the second stage, the tuning parameter $\lambda_k$ in (\ref{Est-gamma}) is obtained via $K$-fold cross validation.

\section{Theoretical Properties} \label{Sec-Theory}

\subsection{The Number of Endogenous Variables is Fixed}\label{Sec-Fixed}

As an extension of the classical 2SLS method to high dimensions, the proposed 2SPLS method also has some good theoretical properties. In this section, we will show that the 2SPLS estimates enjoy the oracle properties. As the second-stage estimation relies on the ridge estimates $\hat{\mathbf{Z}}_{-k}$ obtained from the first stage, we start with the theoretical properties of $\hat{\mathbf{Z}}_{-k}$.

As mentioned previously, each $\tau_j$ in (\ref{Est-pi}) is obtained by GCV. Interestingly, as stated by \citet{Golub1979}, such a $\tau_j$ is closely related to the one minimizing
\[
T_j(\tau) = (\mathbf{Z}_j-\mathbf{P}_{\tau}\mathbf{Y}_j)^T(\mathbf{Z}_j-\mathbf{P}_{\tau}\mathbf{Y}_j).
\]
We have the following result similar to Theorem 2 of \citet{Golub1979}.

\begin{thm} \label{Thm-GCV} Suppose that all components of $\boldsymbol{\pi}_j$ are i.i.d. with mean zero and variance $\sigma_{\boldsymbol{\pi}}^2$, then
\[
\arg\min_{\tau>0} E\left[E[G_j(\tau)|\boldsymbol{\pi}_j]\right] = \arg\min_{\tau>0} E\left[E[T_j(\tau)|\boldsymbol{\pi}_j]\right] = \sigma_{\boldsymbol{\xi}_j}^2\big/\sigma_{\boldsymbol{\pi}}^2,
\]
where $\sigma_{\boldsymbol{\xi}_j}^2$ is the variance component of $\boldsymbol{\xi}_j$ in model (\ref{Eqn-LimitedInfo}).
\end{thm}

This theorem implies that the GCV estimate $\hat{\mathbf{Z}}_j = \mathbf{P}_{\tau_j}\mathbf{Y}_j$ is approximately the optimal estimate of the conditional expectation $\mathbf{Z}_j$; furthermore, as the optimal tuning parameter approximates a constant determined by the variance components ratio, we make the following assumption on $\tau_j$.

\begin{description}
\item[] {\bf Assumption C.} $\tau_j/\sqrt{n}\to 0$ as $n\to \infty$, for $j=1, \cdots, p$.
\end{description}

We then have the following properties on $\hat{\mathbf{Z}}_{-k}$.

\begin{thm} \label{Thm-RidgeEst}
For $k=1,\dots,p$, let $\mathbf{M}_k = \mathbf{\boldsymbol{\pi}}_{-k}^T (\mathbf{C}-\mathbf{C}_{\bullet\mathcal{S}_k} \mathbf{C}_{\mathcal{S}_k,\mathcal{S}_k}^{-1} \mathbf{C}_{\mathcal{S}_k\bullet})\mathbf{\boldsymbol{\pi}}_{-k}$ where each $\mathbf{C}_{\mathcal{S}_r\mathcal{S}_c}$ is a submatrix of $\mathbf{C}$ identified with row indices in $\mathcal{S}_r$ and column indices in $\mathcal{S}_c$ (the dot implies all rows or columns).
Then, under Assumptions A, B, and C,
\begin{enumerate}
\item[a.] $n^{-1}\hat{\mathbf{Z}}_{-k}^T\mathbf{H}_k\hat{\mathbf{Z}}_{-k}\to_p \mathbf{M}_k$, as $n\to\infty$;
\item[b.] $n^{-1/2} (\mathbf{Y}_k-\hat{\mathbf{Z}}_{-k}\boldsymbol{\gamma}_k)^T \mathbf{H}_k\hat{\mathbf{Z}}_{-k} \to_d N(\mathbf{0},\sigma_k^2\mathbf{M}_k)$, as $n\to\infty$.
\end{enumerate}
\end{thm}

Since $n^{-1}\mathbf{Z}_{-k}^T\mathbf{H}_k\mathbf{Z}_{-k}\to\mathbf{M}_k$, Theorem~\ref{Thm-RidgeEst}.a states that $\hat{\mathbf{Z}}_{-k}^T\mathbf{H}_k\hat{\mathbf{Z}}_{-k}$ is a good approximation to $\mathbf{Z}_{-k}^T\mathbf{H}_k\mathbf{Z}_{-k}$. On the other hand,
$\mathbf{H}_k(\mathbf{Y}_k-\hat{\mathbf{Z}}_{-k}\boldsymbol{\gamma}_k)$ is the error term in regressing $\mathbf{H}_k\mathbf{Y}_k$ against $\mathbf{H}_k\hat{\mathbf{Z}}_{-k}$, and Theorem~\ref{Thm-RidgeEst}.b implies that $n^{-1}(\mathbf{Y}_k-\hat{\mathbf{Z}}_{-k}\boldsymbol{\gamma}_k)^T \mathbf{H}_k\hat{\mathbf{Z}}_{-k} \to_d 0$. Thus $\hat{\mathbf{Z}}_{-k}$ results in regression errors with good properties, i.e., the error effects on the 2SPLS estimators will vanish when the sample size gets sufficiently large.

In summary, the above theorem indicates that $\hat{\mathbf{Z}}_{-k}$ behaves the same way as $\mathbf{Z}_{-k}$ asymptotically, which makes it reasonable to replace $\mathbf{Z}_{-k}$ with $\hat{\mathbf{Z}}_{-k}$ at the second stage. Denote the $j$-th elements of $\boldsymbol{\gamma}_{k}$ and $\hat{\boldsymbol{\gamma}}_{k}$ as $\gamma_{kj}$ and $\hat{\gamma}_{kj}$, respectively. Then, the properties of $\hat{\mathbf{Z}}_{-k}$ in Theorem~\ref{Thm-RidgeEst}, together with the oracle properties of the adaptive lasso, will lead to the following oracle properties of our proposed estimates.

\begin{thm}{(Oracle Properties)} \label{Thm-Oracle}
Let $\mathcal{A}_k=\left\{j: \gamma_{kj}\neq 0, j\ne k \right\}$ and $\hat{\mathcal{A}}_k=\left\{j:\hat{\gamma}_{kj}\neq 0, j\ne k\right\}$. Further index both rows and columns of $\mathbf{M}_k$ with $1, \cdots, k-1, k+1, \cdots, p$, and let $\mathbf{M}_{k,\mathcal{A}_k}$ be the submatrix of $\mathbf{M}_k$ identified with both row and column indices in $\mathcal{A}_k$.
Suppose that $\lambda_k/\sqrt{n} \to 0$ and $\lambda_k n^{(\delta-1)/2} \to \infty$. Then, under Assumptions A, B, and C, the estimates from the proposed 2SPLS method satisfy the following properties,
\begin{enumerate}
\item[a.] Consistency in variable selection: $\lim_{n\rightarrow\infty} P(\hat{\mathcal{A}}_k=\mathcal{A}_k)=1$;
\item[b.] Asymptotic normality: $\sqrt{n}(\hat{\boldsymbol{\gamma}}_{k,\mathcal{A}_k}-\boldsymbol{\gamma}_{k,\mathcal{A}_k}) \to_d N(\mathbf{0},\sigma_k^2\mathbf{M}_{k,\mathcal{A}_k}^{-1})$, as $n\rightarrow\infty$.
\end{enumerate}
\end{thm}

It is worth mentioning that Theorem~\ref{Thm-RidgeEst} plays an essential role in establishing the oracle properties of 2SPLS. In fact, as long as the properties in Theorem~\ref{Thm-RidgeEst} hold true for the first-stage estimates of $\mathbf{Z}_{-k}$, the oracle properties can be expected from the adaptive lasso \citep{Zou2006} at the second stage. On the other hand, we can also generalize the second-stage regularization to a wide class of regularization methods \citep{Fan2001, Huang2011, Zhang2010}, the theoretical properties, of which, can still be inherited due to the results in Theorem~\ref{Thm-RidgeEst}.

\subsection{The Number of Endogenous Variables is Divergent} \label{Sec-Divergent}

In this section, we investigate the theoretical properties of 2SPLS with a divergent $p$. That is, per Assumption A, both $p$ and $q$ may  grow with sample size $n$ at the the same order. The theoretical properties will be described by a prespecified sequence $f_n =\littleo{n}$ but $f_n \rightarrow \infty$.

We first update Assumptions B and C for the divergent $p$ and $q$.

\begin{description}
\item[] {\bf Assumption B$^\prime$.} Both $p$ and $q$ grow at the same order of $o(n)$, i.e., $p \asymp q = o(n)$. Furthermore, the singular values of $\mathbf{I}-\boldsymbol{\Gamma}$ are positively bounded from below, and there exist positive constants $c_1$ and $c_2$ such that, for any vector $\delta$ with $\ltwon{\delta}=1$,  $c_1\ge n^{-1/2}\ltwon{\bX\delta} \ge c_2$.
\end{description}

\begin{description}
\item[] {\bf Assumption C$^\prime$.} $r_{nk} \triangleq \tau_k^2 \ltwon{\pi_k}^2/n = \littleo{n}$.
\end{description}

We have the following properties on the ridge regression estimator of $\bpi_k$ from the first stage.

\begin{thm} \label{ridgetheoryk}
Under Assumptions A, B$^{\prime}$, and C$^{\prime}$, for each ridge regression estimator $\hat{\bpi}_k$, there exist constants $C_1$ and $C_2$ such that, with probability at least $1-e^{-f_n}$,
\begin{enumerate}
\item[(a)] $\ltwon{\hat{\bpi}_{k} - \bpi_{k}  }^2 \le C_1 \left(r_{nk} \lor q \lor f_n\right)/n$;
\item[(b)] $n^{-1}\ltwon{\bX(\hat{\bpi}_{k}  - \bpi_{k} )}^2 \le C_2\left(r_{nk} \lor q \lor f_n\right)/n$.
\end{enumerate}
\end{thm}

Denote $r_{\max} = \max_{1\le k\le p} r_{nk}$. Then the system-wise losses in both $\ltwon{\hat{\bpi}_{k}-\bpi_{k}}^2$ and $n^{-1}\ltwon{\bX(\hat{\bpi}_{k}-\bpi_{k})}^2$ have upper bounds in the same order as $(r_{\max} \lor q \lor f_n)/n$, with probability at least $1- e^{-(f_n -\log(p))}$. With $p = o(n)$, we henceforth select $f_n$ to dominate $log(p)$, i.e. $f_n-\log(p) \rightarrow \infty$, to guarantee the well-controlled losses over the whole system.

Denote $\mathcal{A}_k=\left\{j: \gamma_{kj}\neq 0, j\ne k\right\}$. Indexing all rows and columns with only $j=1, \cdots, k-1, k+1, \cdots, p$, we define the restricted eigenvalue for a $(p-1)\times(p-1)$ matrix $\mathbf{M}$ as
\begin{eqnarray*}
\phivarmin{k}{\mathbf{M}} = \text{min}\left\{ n^{-1/2}\ltwon{\mathbf{M}\gamma}\ltwon{\gamma_{\mathcal{A}_{k}}}^{-1}:
\lonen{\gamma_{\mathcal{A}_{k}^c}} \le 3\lonen{\gamma_{\mathcal{A}_{k}}} \right\}.
\end{eqnarray*}
We further define $\|\cdot\|_{\infty}$ and $\|\cdot\|_{-\infty}$ to be the maximum and minimum absolute values of the components of a vector, respectively. For a matrix, $\|\cdot\|_{\infty}$ is defined to be the maximum absolute row sum of the matrix.

We further make the following assumption on the tuning parameter $\lambda_{k}$ of the adaptive lasso at the second stage.

\begin{description}
\item[] {\bf Assumption D.} The adaptive tuning parameter $\lambda_k$ is at the same order as $\|\omega_{k}\|_{-\infty}^{-1} \lonen{\bGamma}$ $\lonen{\bpi} \sqrt{n (r_{\text{max}}\lor q \lor f_n) \log p}$.
\end{description}

We then have the consistency property of estimator $\hat{\bgamma}_k$.

\begin{thm}\label{theoremAdaConsisit} (Estimation Consistency)
Suppose that, for each node $k$, both inequalities $\|\omega_{k,\mathcal{A}_{k}}\|_{\infty} \|\omega_{k,\mathcal{A}_{k}^{c}}\|_{-\infty}^{-1} \le1$ and $\sqrt{(r_{\text{max}}\lor q \lor f_n)/n} +c_1\lonen{\bpi} \le \sqrt{c_1^2\lonen{\bpi}^2+\phi_0^2\big/64C_2|\mathcal{A}_{k}|}$ hold, and there exists a positive constant $\phi_0$ such that $\phivarmin{k}{\bH_k\bX\bpi_{-k}} \ge \phi_0$. Denote $h_n = (\lonen{\bGamma}^2 \land 1) \left[  (\frac{n}{q} \lonen{\bpi}^2) \land (r_{\text{max}} \lor  q \lor f_n)  \right]\log p$. Under Assumptions A, B$^{\prime}$, C$^{\prime}$, and D, there exist constants $C_3>0$ and $C_4>0$ such that, with probability at least $1- e^{-C_3 h_n+\log(4pq)}-e^{-f_n+\log(p)}$, each 2SPLS estimator $\hat{\bgamma}_k$ satisfies that
\begin{enumerate}
\item $\lonen{\hat{\bgamma}_k - \bgamma_{k } } \le 8 C_4  \frac{\|\omega_{k,\mathcal{A}_{k}}\|_{\infty}  \lonen{\bpi} \lonen{\bGamma}}{\phi_0^2 \|\omega_{k}\|_{-\infty}}  |\mathcal{A}_{k}| \sqrt{\frac{(r_{\text{max}}\lor q\lor f_n )\log p}{n}}$;
\item $n^{-1}\ltwon{ \bH_k\hat{\bZ}_{-k} (\hat{\bgamma}_k -  \bgamma_{k } )}^2 \le  \frac{C_4^2 \|\omega_{k,\mathcal{A}_{k}}\|_{\infty}^2 \lonen{\bpi}^2\lonen{\bGamma}^2}{\phi_0^2 \|\omega_{k}\|_{-\infty}^2}   |\mathcal{A}_{k}|  \frac{(r_{\text{max}}\lor q\lor f_n )\log p}{n}$.
\end{enumerate}
\end{thm}

Note that the system-wide upper bounds, defined by replacing $|\mathcal{A}_k|$ with $\max_k |\mathcal{A}_k|$, can also be achieved with probability at least $1- e^{-C_3 h_n+\log(4q)+2\log(p)}-e^{-f_n+2\log(p)}$.

Let $W_k = diag\{\bweight_{k}\}$ and $V_{k} = (v_{ij})_{(p-1)\times(p-1)} \triangleq \frac{1}{n} \bpi_{-k}^T\bX^T\bH_k\bX\bpi_{-k}$. Further denote $W_{k,\mathcal{A}_k} = diag\{\bweight_{k,\mathcal{A}_k}\}$, $W_{k,\mathcal{A}_k^c} = diag\{\bweight_{k,\mathcal{A}_k^c}\}$, $\sto = (v_{ij})_{i\in \mathcal{A}_{k}^c,j\in \mathcal{A}_{k}}$, $\soo = (v_{ij})_{i\in \mathcal{A}_{k},j\in \mathcal{A}_{k}}$, and $\theta_{k} = \infn{\soo[-1]  W_{k,\mathcal{A}_{k}}}$. We then have the following selection property.

\begin{thm} \label{selectionconsistency}(Selection Consistency) Suppose that, for each node $k$,  $\soo$ is invertible, and $\sqrt{(r_{\text{max}}\lor q \lor f_n)/n} +c_1\lonen{\bpi} \nonumber\le\sqrt{c_1^2\lonen{\bpi}^2+\min(\phi_0^2/64,\zeta(4-\zeta)^{-1}\|\bomega_{k}\|_{-\infty}/\theta_{k})/(C_2|\mathcal{A}_{k}|)}$.  Further assume that there exists a positive constant $\zeta\in (0,1)$ such that $\underset{j\in \mathcal{A}_{k}}{\text{min}} |\gamma_{kj}| > \frac{2\lambda_{k}\theta_{k}}{n(2-\zeta)}$ and $\infn{W_{k,\mathcal{A}_{k}^c}^{-1} \sto \soo[-1] W_{k,\mathcal{A}_{k}}} < 1-\zeta$. Under Assumptions A, B$^{\prime}$, C$^{\prime}$, and D, there exists a 2SPLS estimator $\hat{\gamma}_k$ satisfying that, with probability at least $1- e^{-C_5 h_n+\log(4pq)}-e^{-f_n+\log(p)}$ for some constant $C_5>0$, $\hat{\mathcal{A}}_k = \mathcal{A}_k$ with $\hat{\mathcal{A}}_k=\{j: \hat{\gamma}_{kj}\ne 0, j\ne k\}$.
\end{thm}

\section{Simulation Studies} \label{Sec-Simu}

We conducted simulation studies to compare 2SPLS with the adaptive lasso based algorithm (AL) by \citet{Logsdon2010}, and the sparsity-aware maximum likelihood algorithm (SML) by \citet{Cai2013}. To investigate whether it is necessary to select instrumental variables at the first stage as proposed in \citet{Belloni2012}, \citet{Lin2015}, and \citet{Zhu2015}, we also consider a method which replaces the ridge regression at the first stage of 2SPLS with the adaptive lasso, that is, the two-stage adaptive lasso (2SAL) method. Both acyclic networks and cyclic networks were simulated, each involving $300$ endogenous variables. Each endogenous variable was simulated to have, on average, one regulatory effect for sparse networks, or three regulatory effects for dense networks. The regulatory effects were independently simulated from a uniform distribution over  $(-1,-0.5)\cup(0.5,1)$. To allow the use of AL and SML, every endogenous variable in the same network was simulated to have the same number (either one or three) of nonzero exogenous effects (EEs) by the exogenous variables, with all effects equal to one. Each exogenous variable was simulated to take values 0, 1 and 2 with probabilities 0.25, 0.5 and 0.25, respectively, emulating genotypes of an F2 cross in a genetical genomics experiment. All error terms were independently simulated from $N(0,0.1^2)$, and the sample size $n$ varied from $100$ to $1,000$. For each network setup, we simulated 100 data sets and applied all four algorithms to calculate the power and false discovery rate (FDR).

\begin{figure}[!ht]
\begin{minipage}[h]{0.5\linewidth}
\centering a. Power of Sparse Networks
\end{minipage}
\begin{minipage}[h]{0.5\linewidth}
\centering b. FDR of Sparse Networks
\end{minipage}

\begin{minipage}[h]{0.5\linewidth}
\centering
\makebox{\includegraphics[width=3in, height=2.25in]{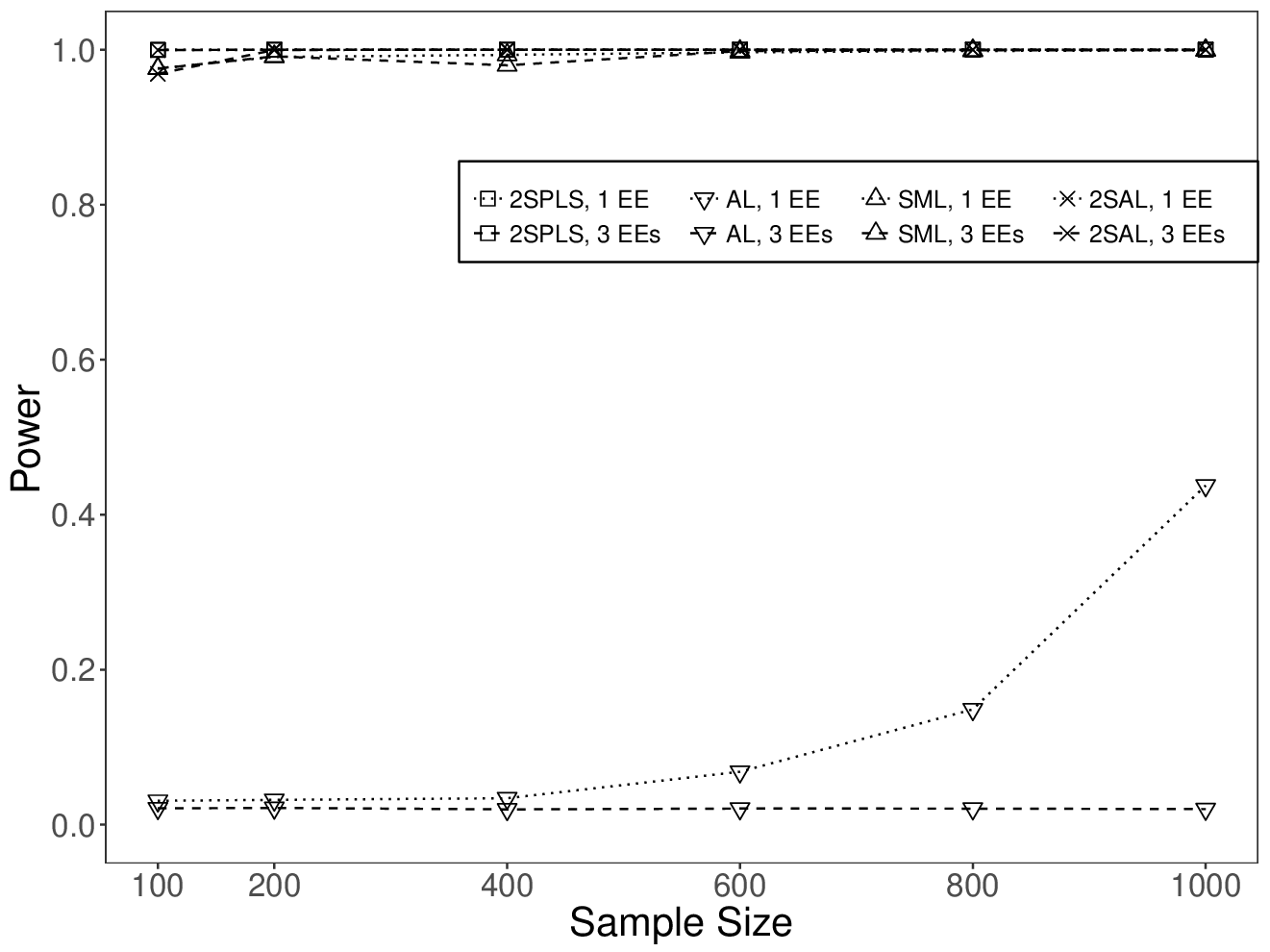}}
\end{minipage}
\begin{minipage}[h]{0.5\linewidth}
\centering
\makebox{\includegraphics[width=3in, height=2.25in]{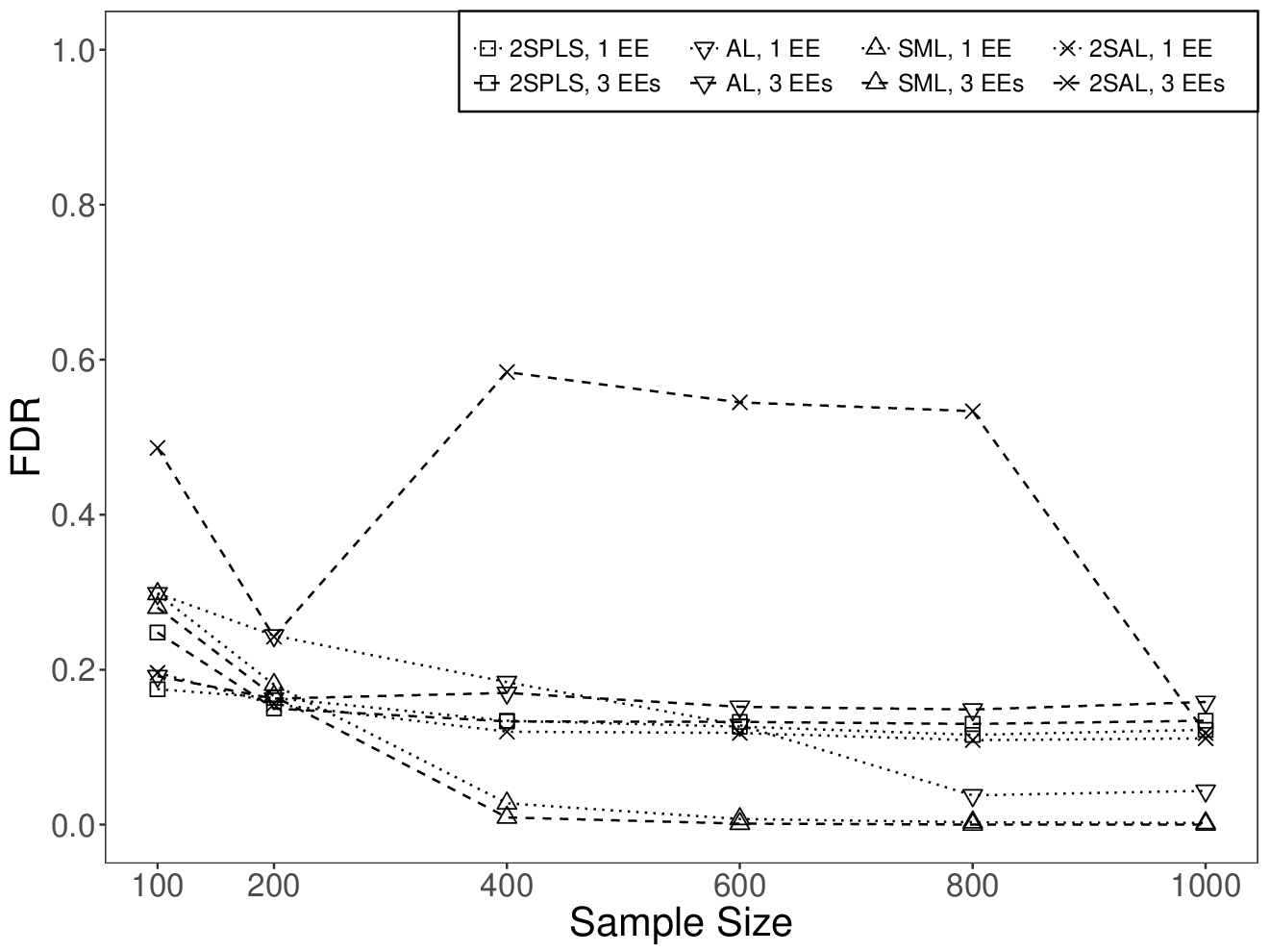}}
\end{minipage}

\vskip6pt
\begin{minipage}[h]{0.5\linewidth}
\centering c. Power of Dense Networks
\end{minipage}
\begin{minipage}[h]{0.5\linewidth}
\centering d. FDR of Dense Networks
\end{minipage}

\begin{minipage}[h]{0.5\linewidth}
\centering
\makebox{\includegraphics[width=3in, height=2.25in]{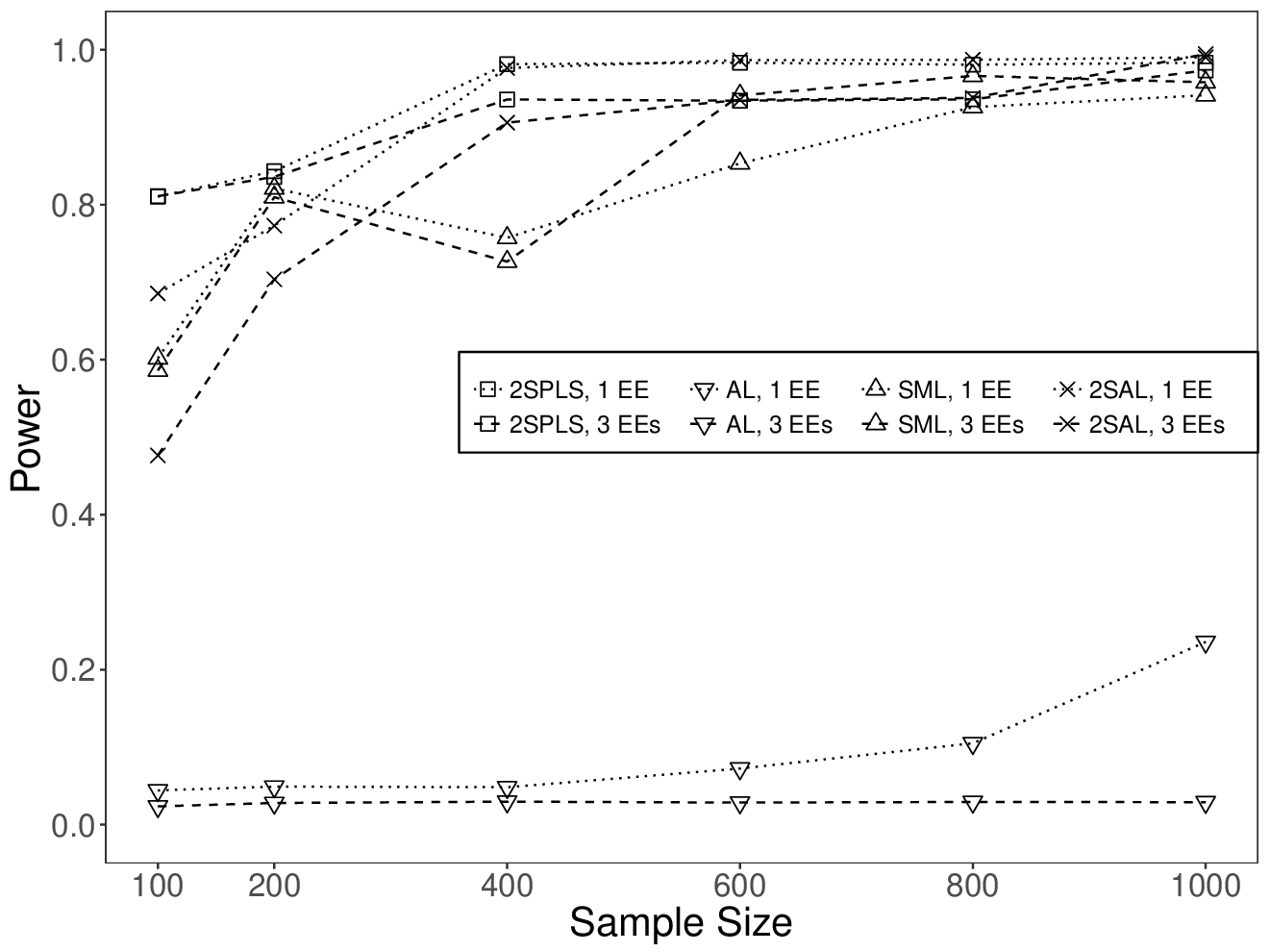}}
\end{minipage}
\begin{minipage}[h]{0.5\linewidth}
\centering
\makebox{\includegraphics[width=3in, height=2.25in]{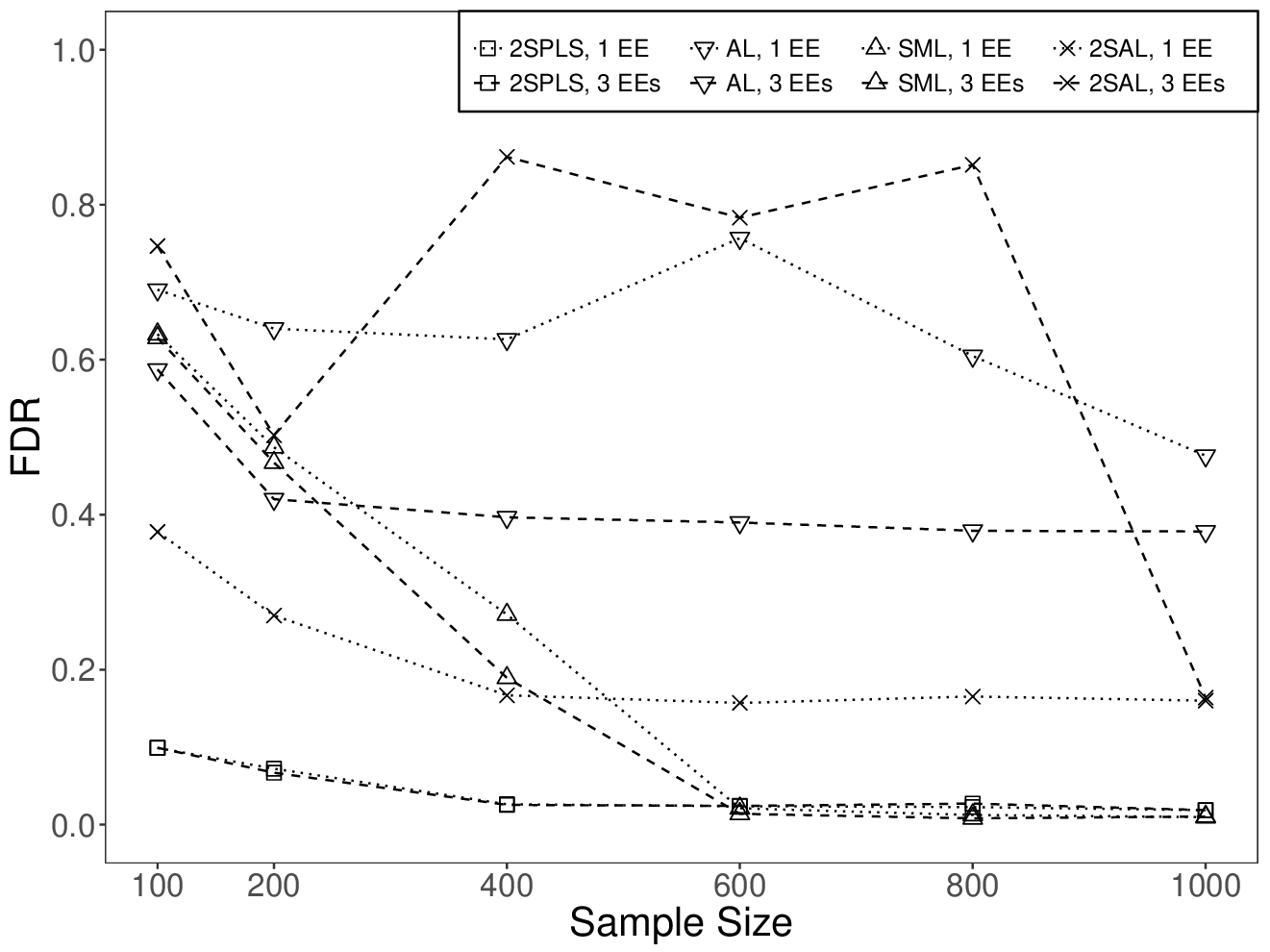}}
\end{minipage}

\caption{\label{Figure-AcyclicLarge} Performance of 2SPLS, AL, SML, and 2SAL when identifying regulatory effects in acyclic networks with one EE or three EEs.}
\end{figure}

For inferring acyclic networks, the power and FDR of the four different algorithms are plotted in Figure~\ref{Figure-AcyclicLarge}. 2SPLS has greater power than the other three algorithms to infer both sparse and dense acyclic networks when the sample size is small or moderate. When the sample size is large, 2SPLS, SML, and 2SAL are comparable for constructing both sparse and dense acyclic networks. In any case, AL has much lower power than other methods. Specifically, AL provides power as low as under $10\%$ when the sample size is small, and its power is still under $50\%$ even when the sample size increases to $1,000$. On the other hand, 2SPLS provides power over $80\%$ for small sample sizes, and over $90\%$ for moderate to large sample sizes.

As shown in Figure~\ref{Figure-AcyclicLarge}, 2SPLS controls the FDR under $20\%$ except for the case which has three available EEs with small sample sizes ($n=100$). Although SML controls the FDR as low as under $5\%$ for sparse acyclic networks when the sample sizes are large, it reports large FDRs when the sample sizes are small. For example, when the sample sizes are under 200, SML reports FDR over $40\%$ for dense acyclic networks. In general, both 2SPLS and SML outperform AL and 2SAL in terms of FDR. Only in the case when inferring sparse acyclic networks with one available EE from data sets of moderate or large sample sizes, AL and 2SAL report FDR lower than 2SPLS.

\begin{figure}[!ht]
\begin{minipage}[h]{0.5\linewidth}
\centering a. Power of Sparse Networks
\end{minipage}
\begin{minipage}[h]{0.5\linewidth}
\centering b. FDR of Sparse Networks
\end{minipage}

\begin{minipage}[h]{0.5\linewidth}
\centering
\makebox{\includegraphics[width=3in, height=2.25in]{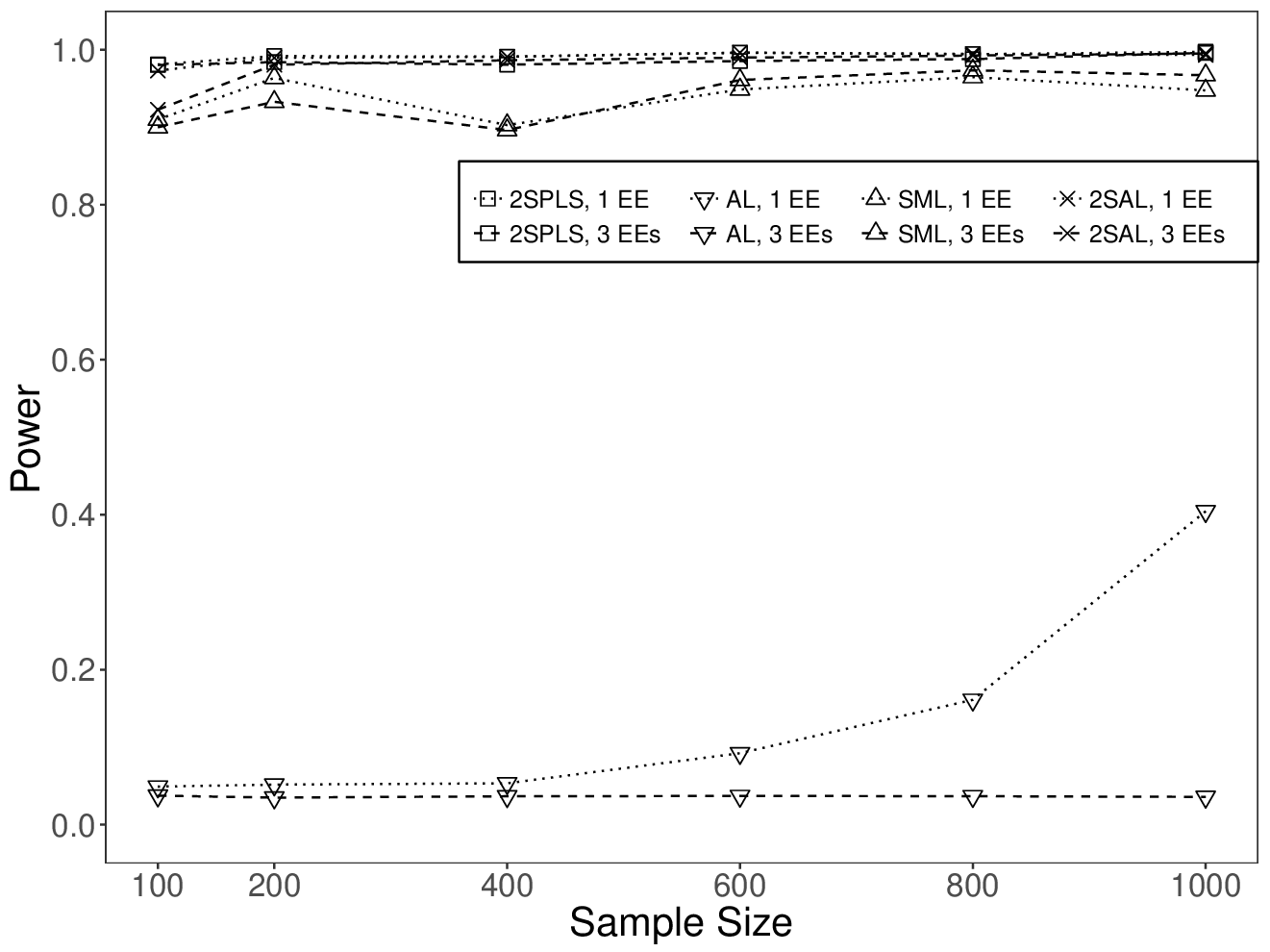}}
\end{minipage}
\begin{minipage}[h]{0.5\linewidth}
\centering
\makebox{\includegraphics[width=3in, height=2.25in]{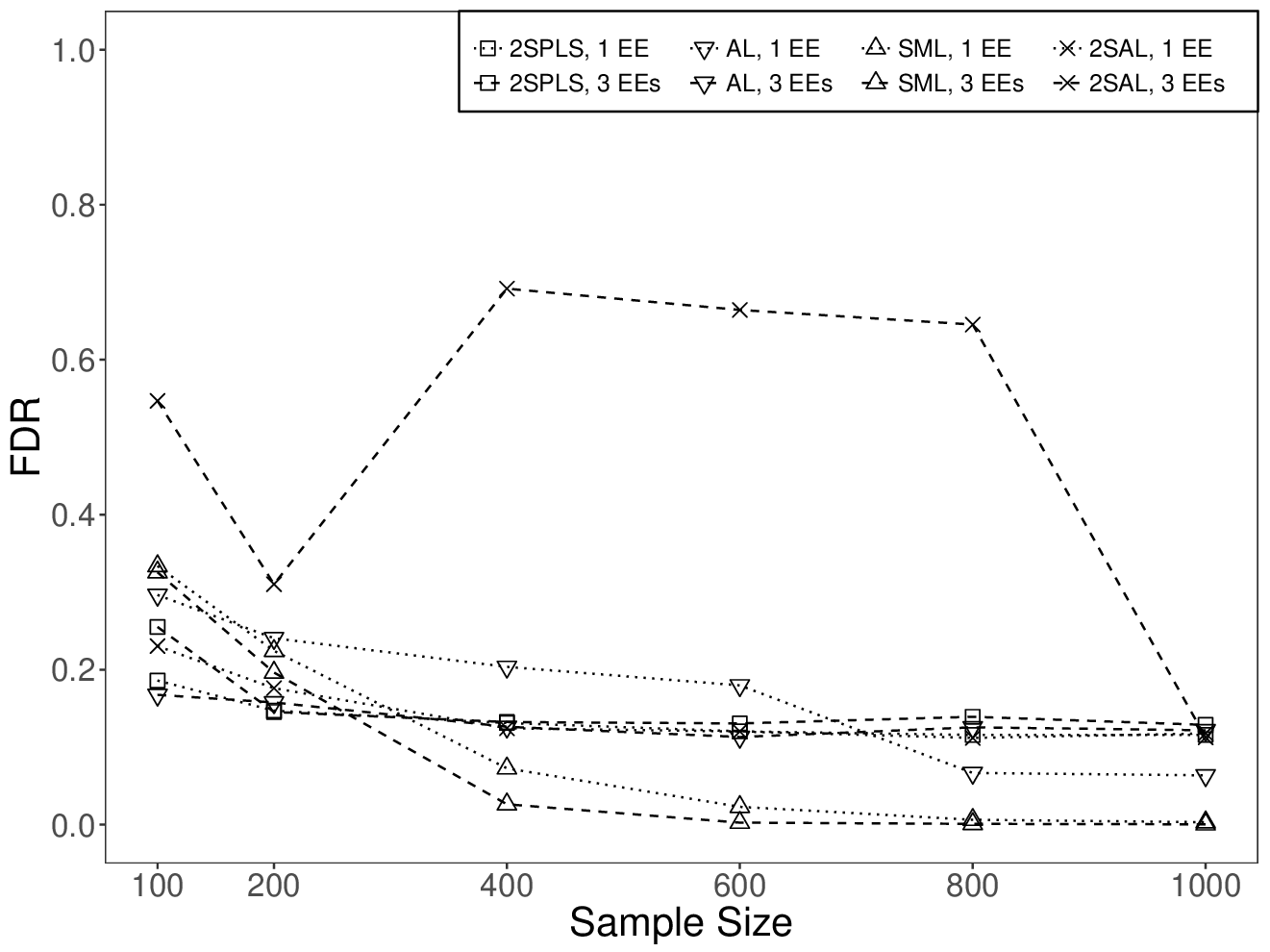}}
\end{minipage}

\vskip6pt
\begin{minipage}[h]{0.5\linewidth}
\centering c. Power of Dense Networks
\end{minipage}
\begin{minipage}[h]{0.5\linewidth}
\centering d. FDR of Dense Networks
\end{minipage}

\begin{minipage}[h]{0.5\linewidth}
\centering
\makebox{\includegraphics[width=3in, height=2.25in]{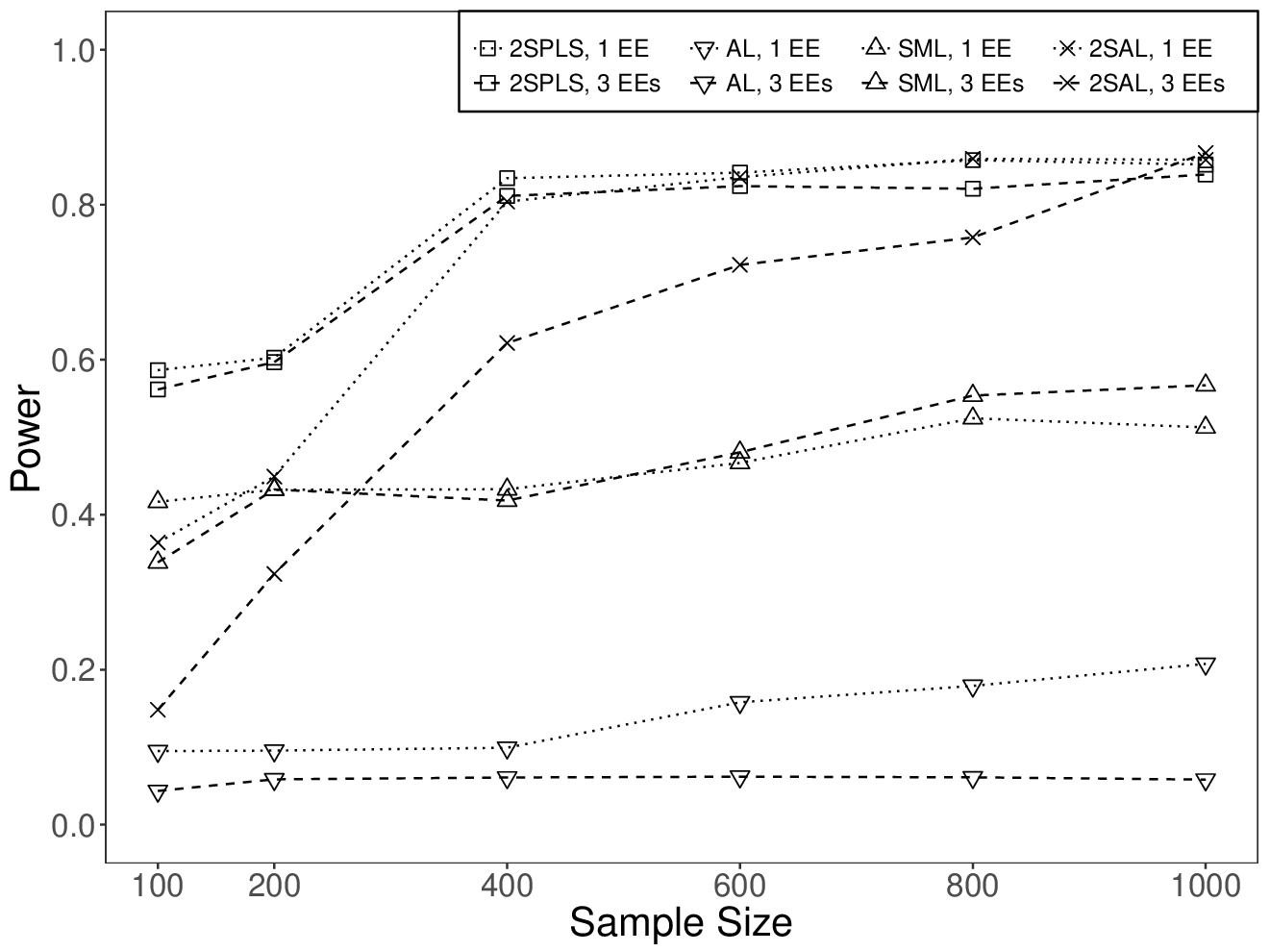}}
\end{minipage}
\begin{minipage}[h]{0.5\linewidth}
\centering
\makebox{\includegraphics[width=3in, height=2.25in]{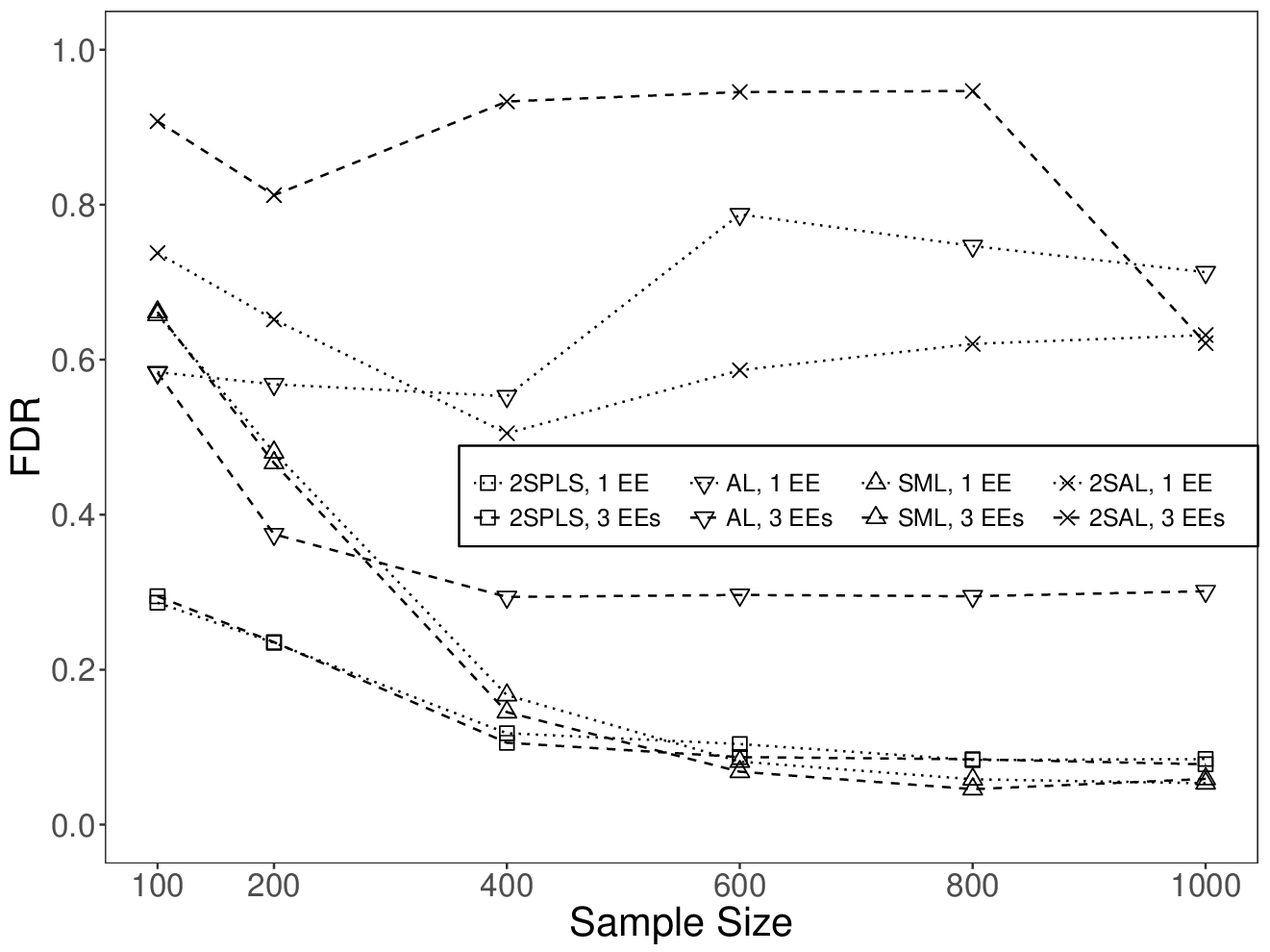}}
\end{minipage}

\caption{\label{Figure-CyclicLarge} Performance of 2SPLS, AL, SML, and 2SAL when identifying regulatory effects in cyclic networks with one EE or three EEs.}
\end{figure}

Plotted in Figure~\ref{Figure-CyclicLarge} are the power and FDR of the four different algorithms when inferring cyclic networks. Similar to the results on acyclic networks, 2SPLS has greater power than SML and AL across all sample sizes and has lower FDR when the sample size is small. 2SPLS has greater power than 2SAL in most scenarios and has much lower FDR than 2SAL except for the case when inferring sparse cyclic networks from data sets of large sample sizes. SML provides power competitive to 2SPLS for sparse cyclic networks, but its power is much lower than that of 2SPLS for dense cyclic networks. Similar to the case of acyclic networks, SML reports much higher FDR for inferring dense networks from data sets with small sample sizes though it reports small FDR when the sample sizes are large. 2SAL reports the highest FDR, especially for networks with three available EEs.

Although not performing as well as 2SPLS, 2SAL reports competitive power to SML when inferring either acyclic or cyclic networks. For the acyclic sparse network with one EE, 2SAL can control FDR at a similar level to 2SPLS because each endogenous variable may be associated to a very small set of exogenous variables in (\ref{Eqn-ReducedForm}). However, we observed high FDR of 2SAL in Figure~\ref{Figure-AcyclicLarge}.b for the acyclic sparse network with three EEs which triples the average number of exogenous variables associated to each endogenous variable. The similar phenomenon of 2SAL appears in Figure~\ref{Figure-CyclicLarge}.b for the cyclic sparse networks. The dense networks also triple the average number of regulatory effects for each endogenous variable, which implies an increased number of exogenous variables associated to each endogenous variable in (\ref{Eqn-ReducedForm}). Therefore, we unsurprisingly observed even higher FDR of 2SAL in Figure~\ref{Figure-AcyclicLarge}.d and Figure~\ref{Figure-CyclicLarge}.d, where the FDR is over 0.8. In summary, variable selection at the first stage seems work well when each endogenous variable is associated to a small set of exogenous variables in (\ref{Eqn-ReducedForm}), but may compromise the identification of regulatory effects at the second stage when the number of exogenous variables associated to an endogenous variable increases.

Both 2SPLS and 2SAL are two-stage methods developed based on the limited-information model (\ref{Eqn-LimitedInfo}), instead of the full-information model used by SML, leading to fast computation and potential implementation of parallel computing. To demonstrate the computational advantage of 2SPLS and 2SAL, we recorded the computing time of all algorithms when inferring the same networks from small data sets ($n=100$). Each algorithm analyzed the same data set using only one CPU in a server with Quad-Core AMD Opteron\texttrademark\ Processor 8380. Reported in Table~\ref{Table-RunTime} are the running times of all four algorithms for inferring different networks. AL is the fastest although it performs with the least power. The running time of 2SPLS usually doubles or triples that of AL, but the computation time of 2SAL generally triples that of 2SPLS because 2SAL employed $K$-fold cross-validation to choose the tuning parameter at the first stage. SML is the slowest algorithm which generally takes more than 40 times longer than 2SPLS to infer different networks. In particular, SML is almost 200 times slower than 2SPLS when inferring acyclic sparse networks.

\begin{table}[!ht]
  \centering
\fbox{ %
\begin{tabular}{ccccccccc}
 & \multicolumn{4}{c}{\underline{\hspace{72pt}Acyclic\hspace{72pt}}} & \multicolumn{4}{c}{\underline{\hspace{75pt}Cyclic\hspace{75pt}}} \\
 & \multicolumn{2}{c}{\underline{\hspace{27pt}Sparse\hspace{27pt}}} & \multicolumn{2}{c}{\underline{\hspace{27pt}Dense\hspace{27pt}}} & \multicolumn{2}{c}{\underline{\hspace{27pt}Sparse\hspace{27pt}}} & \multicolumn{2}{c}{\underline{\hspace{27pt}Dense\hspace{27pt}}} \\
 & 1 EE & 3 EEs & 1 EE & 3 EEs & 1 EE & 3 EEs & 1 EE & 3 EEs \\ \hline
2SPLS & 1303 & 1332 & 1127 & 1112 & 1297 & 1337 & 1125 & 1165 \\
AL & 405 & 652 & 404 & 637 & 443 & 659 & 430 & 781 \\
SML & 258875 & 195739 & 58509 & 43118 & 49393 & 58716 & 67949 & 68081 \\
2SAL & 3239 & 4726 & 3398 & 5357 & 3135 & 4681 & 3686 & 5651 \\
\end{tabular}}
  \caption{\label{Table-RunTime} The running time (in seconds) of inferring networks from a data set with $n=100$.}
\end{table}

\begin{figure}[!ht]
\begin{minipage}[h]{0.5\linewidth}
\centering a. Power of Sparse Networks
\end{minipage}
\begin{minipage}[h]{0.5\linewidth}
\centering b. FDR of Sparse Networks
\end{minipage}

\begin{minipage}[h]{0.5\linewidth}
\centering
\makebox{\includegraphics[width=3in, height=2.25in]{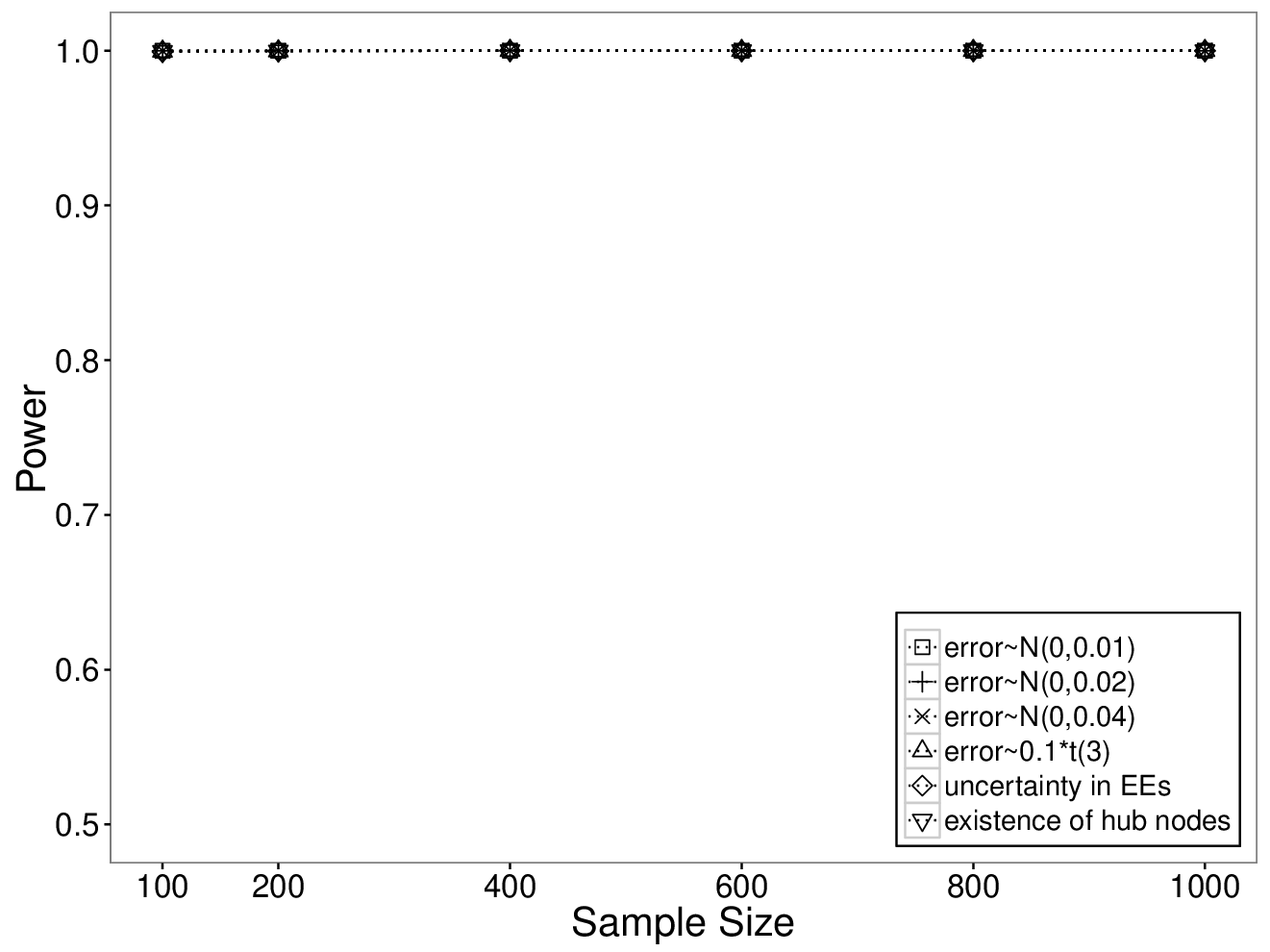}}
\end{minipage}
\begin{minipage}[h]{0.5\linewidth}
\centering
\makebox{\includegraphics[width=3in, height=2.25in]{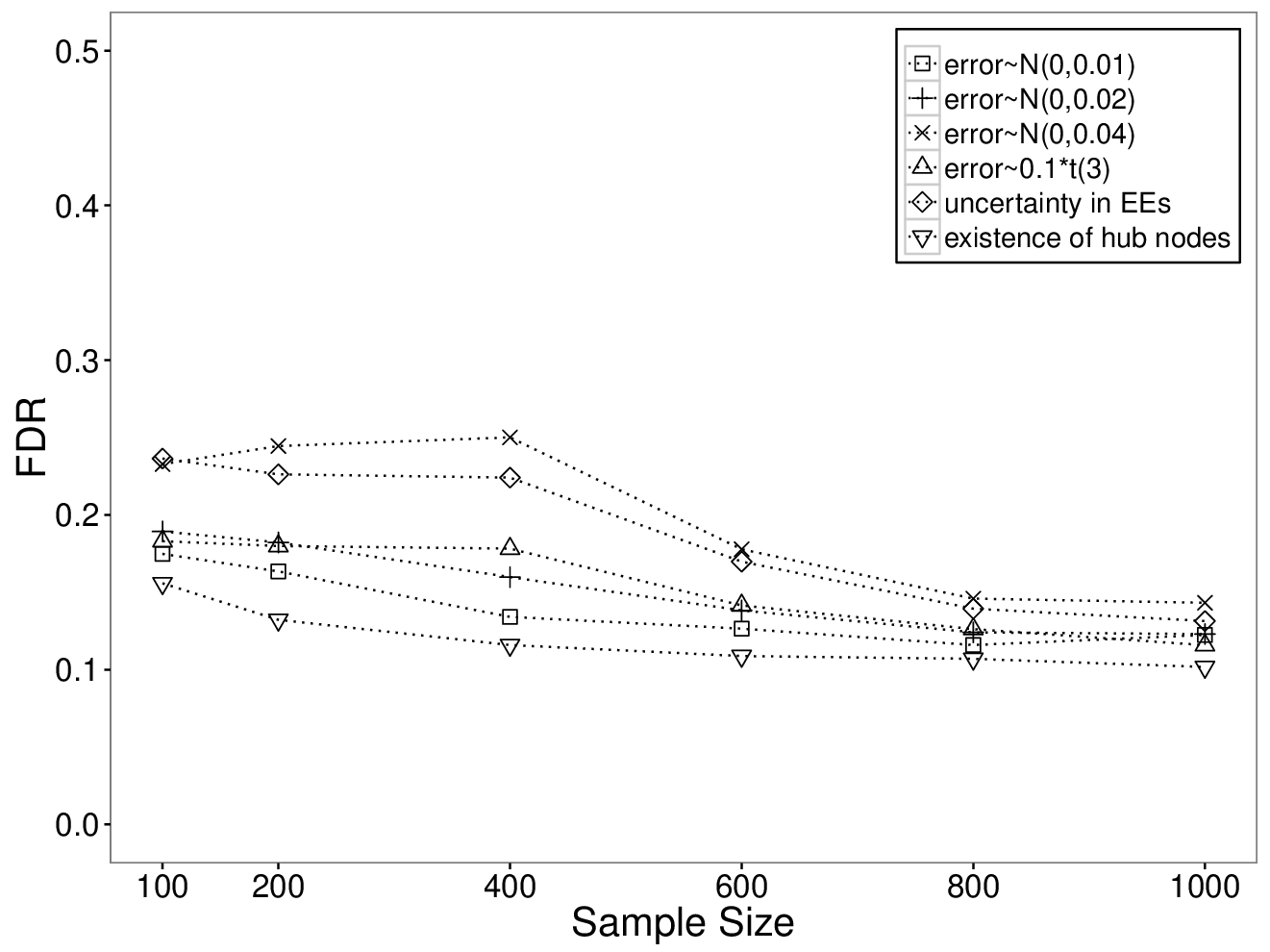}}
\end{minipage}

\vskip6pt
\begin{minipage}[h]{0.5\linewidth}
\centering c. Power of Dense Networks
\end{minipage}
\begin{minipage}[h]{0.5\linewidth}
\centering d. FDR of Dense Networks
\end{minipage}

\begin{minipage}[h]{0.5\linewidth}
\centering
\makebox{\includegraphics[width=3in, height=2.25in]{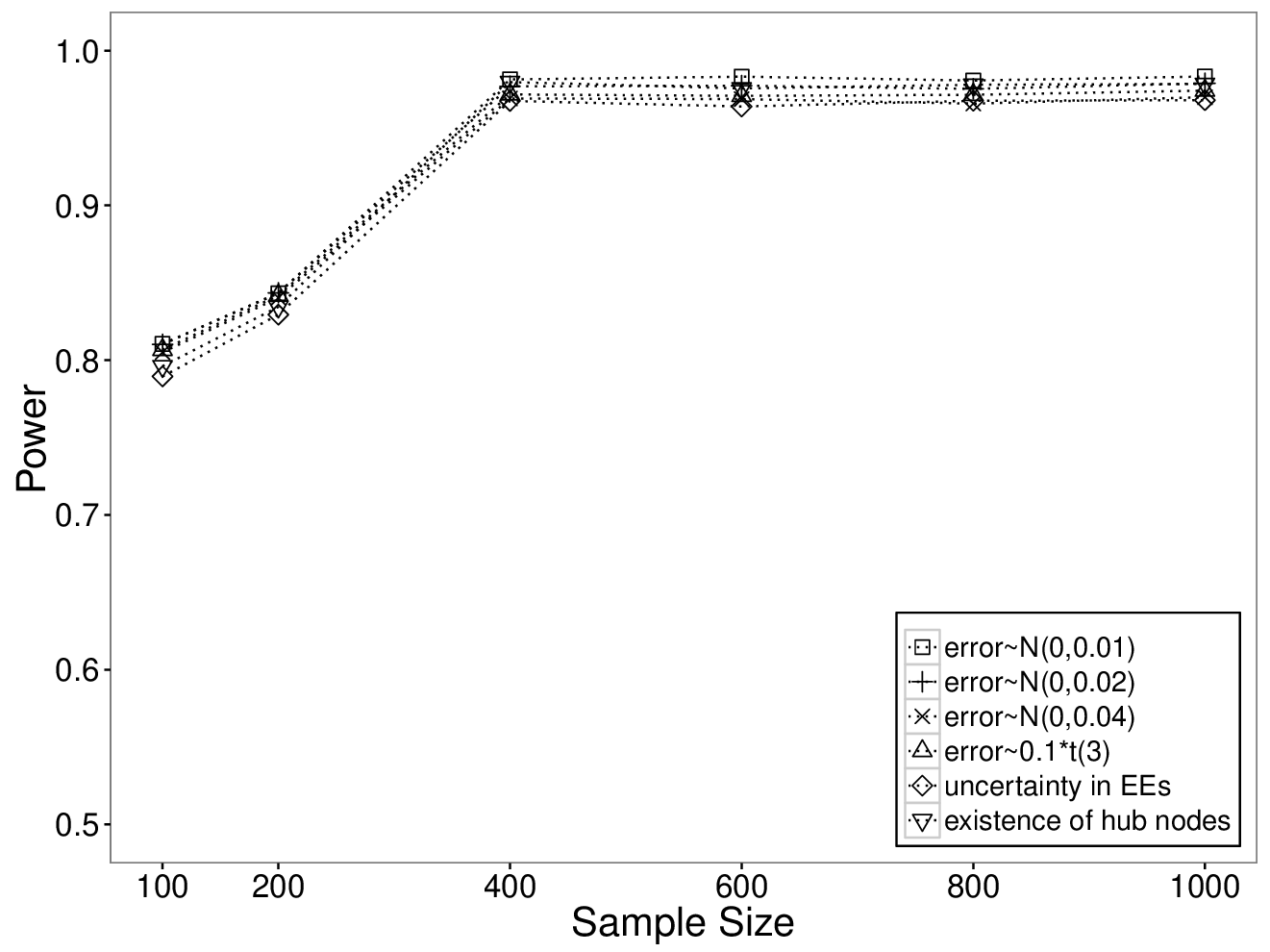}}
\end{minipage}
\begin{minipage}[h]{0.5\linewidth}
\centering
\makebox{\includegraphics[width=3in, height=2.25in]{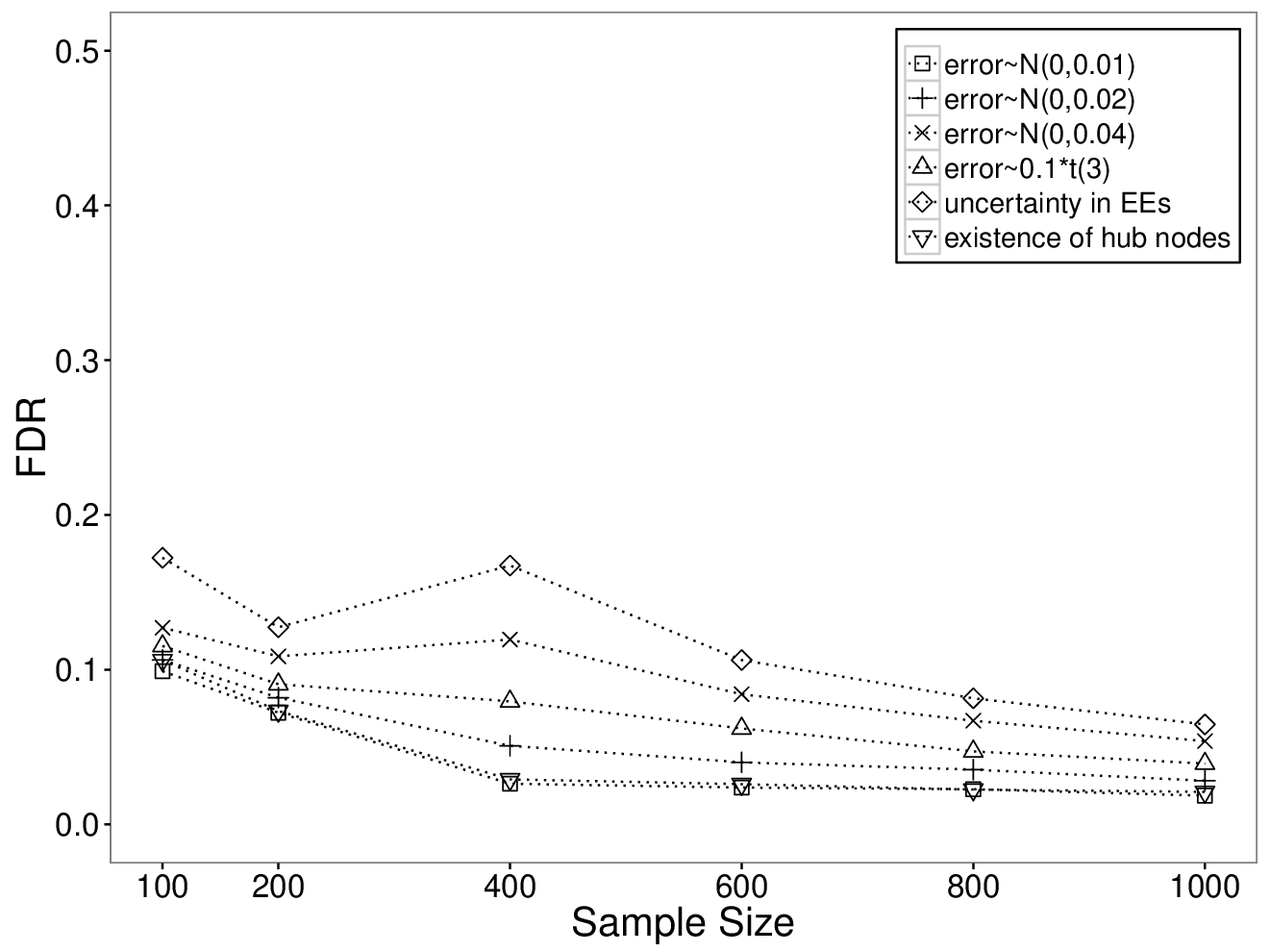}}
\end{minipage}

\caption{\label{Figure-Robustness} Performance of 2SPLS in robustness tests when identifying regulatory effects in acyclic networks with one EE.}
\end{figure}

The robustness of 2SPLS was also evaluated from different aspects: (i) its robustness to different noise levels by doubling or even quadrupling the error variance; (ii) its robustness to non-normality of error terms by simulating errors sampled from a t-distribution, i.e., $t(3)$; (iii) its robustness to uncertainty in the connections between exogenous and endogenous variables by simulating three exogenous effects for each endogenous variable (to emulate the genetical genomics experiment, the three exogenous variables are correlated with correlation coefficients at 0.8, and have effects at 1, 0.5, and -0.3, respectively) but including only one exogenous variable with the strongest estimated effects for each endogenous variable; (iv) its robustness to existence of hub nodes by simulating networks with six hub nodes having five regulatory effects on average while other endogenous variables having on average one regulatory effect for sparse networks, or three regulatory effects for dense networks. All networks include 300 endogenous variables, and the networks with errors following $N(0, 0.01)$ are the same as those shown in Figure~\ref{Figure-AcyclicLarge}. As shown in Figure~\ref{Figure-Robustness}, the 2SPLS method demonstrated robust power while the FDR was slightly affected when the error variance doubled. When the error variance quadrupled, a higher FDR was reported as expected. With errors from $t(3)$, we observed similar power and slightly increased FDR of 2SPLS, which confirms the robustness of 2SPLS to non-normality. The uncertainty in the connections between exogenous and endogenous variables had almost no effect on the power of 2SPLS, and only slightly increased the FDR in constructing sparse networks. The existence of hub nodes rarely affected construction of dense networks, but decreased the FDR in constructing sparse networks. Overall, the performance of 2SPLS is remarkable in demonstrating robustness under a variety of realistic data structures.

\section{Real Data Analysis} \label{Sec-RData}

We analyzed a yeast data set with 112 segregants from a cross between two strains BY4716 and RM11-la \citep{Brem2005}. A total of 5,727 genes were measured for their expression values, and 2,956 markers were genotyped. Each marker within a genetic region (including 1kb upstream and downstream regions) was evaluated for its association with the corresponding gene expression, yielding 722 genes with marginally significant cis-eQTL ($p$-value $<0.05$). The set of cis-eQTL for each gene was filtered to control a pairwise correlation under $0.90$, and then further filtered to keep up to three cis-eQTL which have the strongest association with the corresponding gene expression.

\begin{figure}[!ht]
\begin{minipage}[h]{1\linewidth}
a.
\end{minipage}

\begin{minipage}[h]{1\linewidth}
\centering
\makebox{\includegraphics[width=5.5in,height=2in,clip=true]{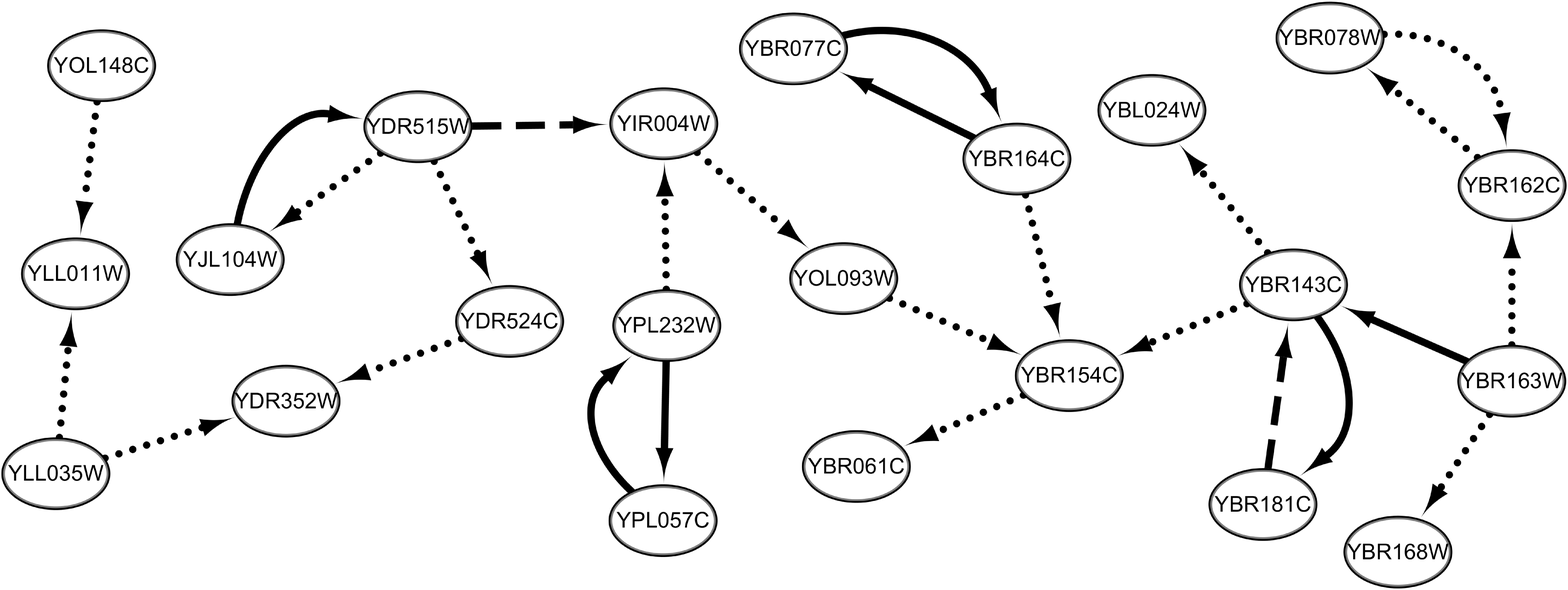}}
\end{minipage}

\vskip6pt
\begin{minipage}[h]{0.5\linewidth}
b.
\end{minipage}
\begin{minipage}[h]{0.5\linewidth}
c.
\end{minipage}

\begin{minipage}[h]{0.5\linewidth}
\centering
\makebox{\includegraphics[width=3in,height=1.75in,clip=true]{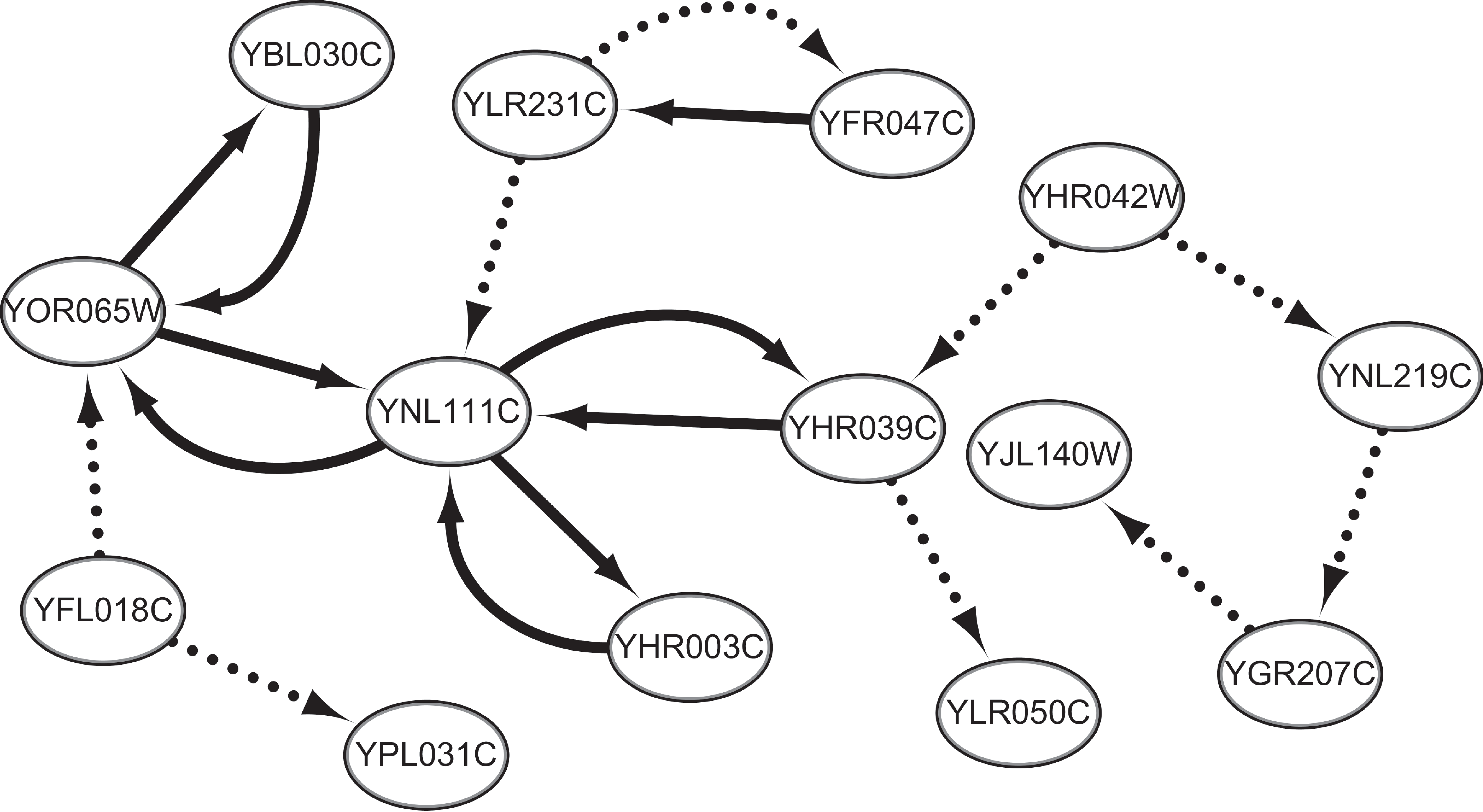}}
\end{minipage}
\begin{minipage}[h]{0.5\linewidth}
\centering
\makebox{\includegraphics[width=3in,height=1.75in,clip=true]{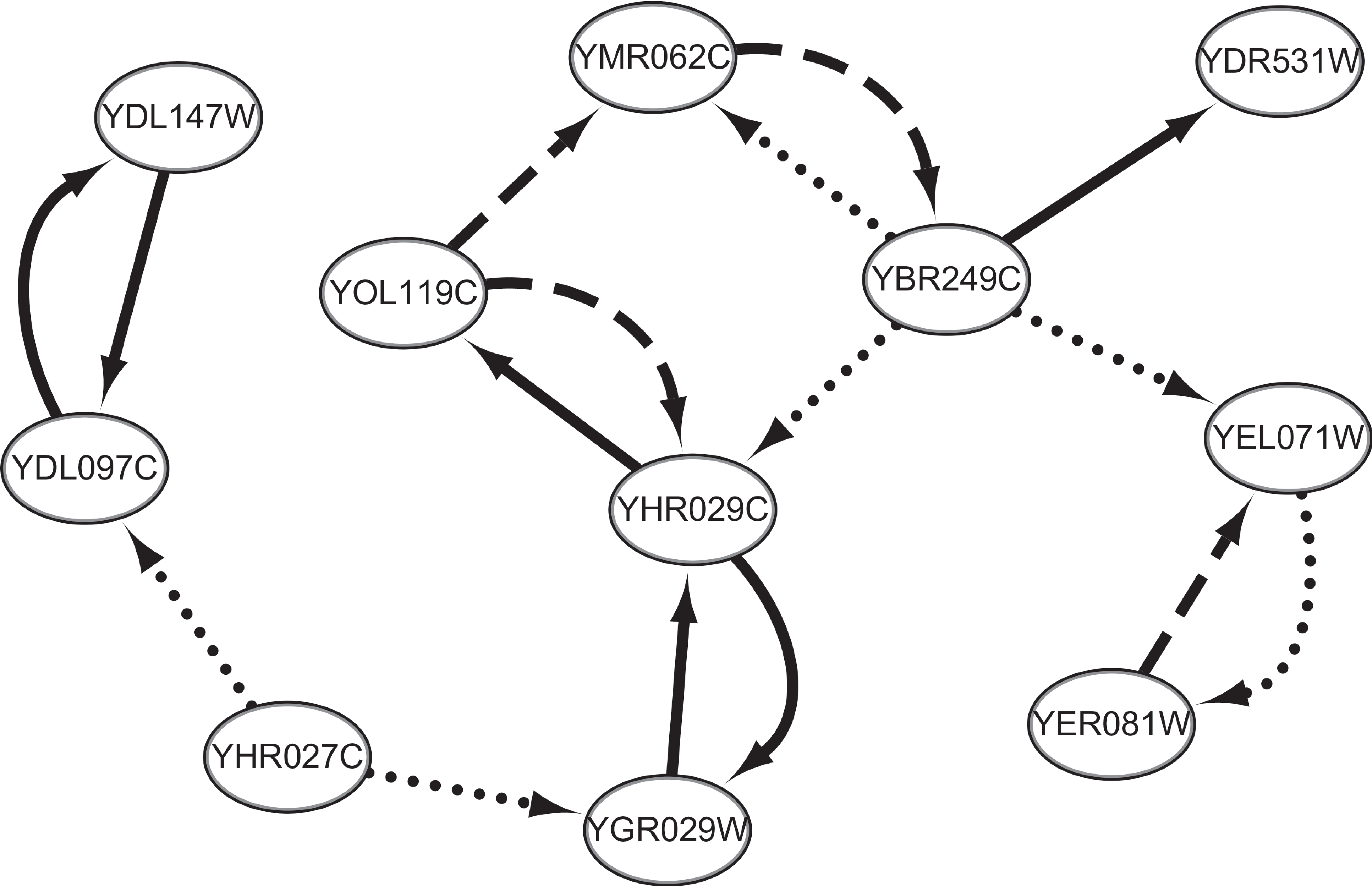}}
\end{minipage}

\caption{\label{Figure-Subnetwork} Three gene regulatory subnetworks in yeast (the dotted, dashed, and solid arrows implied that the corresponding regulations were constructed respectively from over $80\%$, $90\%$, and $95\%$ of the bootstrap data sets).}
\end{figure}

With 112 observations of 722 endogenous variables and 732 exogenous variables, we applied 2SPLS to infer the gene regulatory network in yeast. The constructed network includes 7,300 regulatory effects in total. To evaluate the reliability of constructed gene regulations, we generated 10,000 bootstrap data sets (each with $n=112$) by randomly sampling the original data with replacement, and applied 2SPLS to each data set to infer the gene regulatory network. Among the 7,300 regulatory effects, 323 effects were repeatedly identified in more than 80\% of the 10,000 data sets, and Figure~\ref{Figure-Subnetwork} shows the three largest subnetworks formed by these 323 effects. Specifically, the largest subnetwork consists of 22 endogenous variables and 26 regulatory effects, the second largest one includes 14 endogenous variables and 18 regulatory effects, and the third largest one has 11 endogenous variables and 16 regulatory effects.

A gene-enrichment analysis with DAVID \citep{Huang2009} showed that the three subnetworks are enriched in different gene clusters (controlling $p$-values from Fisher's exact tests under $0.01$). A total of six gene clusters are enriched with genes from the first subnetwork, and four of them are related to either methylation or methyltransferase. Six of 22 genes in the first subnetwork are found in a gene cluster which is related to none-coding RNA processing. The second subnetwork is enriched in nine gene clusters. While three of the clusters are related to electron, one cluster includes half of the genes from the second subnetwork and is related to oxidation reduction. The third subnetwork is also enriched in nine different gene clusters, with seven clusters related to proteasome.

\begin{figure}[!ht]
\centering
\includegraphics[width=4.5in,height=1.75in,clip=true]{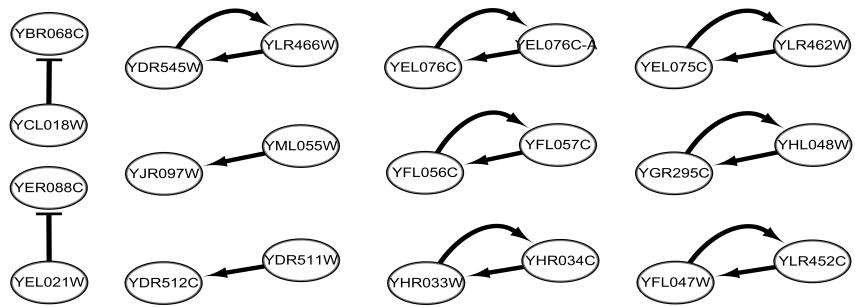}
\caption{\label{Figure-YeastEdge} The yeast gene regulatory subnetworks constructed in each of 10,000 bootstrap data sets (with arrow- and bar-headed lines implying up and down regulations, respectively).}
\end{figure}

A total of 18 regulations were constructed from each of the 10,000 bootstrap data sets, and are shown in Figure~\ref{Figure-YeastEdge}. There are seven pairs of genes which regulate each other. It is interesting to observe that all regulatory genes up-regulate the target genes except two genes, namely, YCL018W and YEL021W.

\section{Discussion} \label{Sec-Disc}

In a classical setting with small numbers of endogenous/exogenous variables, constructing a system of structural equations has been well studied since \citet{Haavelmo1943, Haavelmo1944}. \citet{Anderson1949} first proposed to estimate the parameters of a single structural equation with the limited information maximum likelihood estimator. Later, \citet{Theil1953a, Theil1953b, Theil1961} and \citet{Basmann1957} independently developed the 2SLS estimator, which is the simplest and most common estimation method for fitting a system of structural equations. However, genetical genomics experiments usually collect data in which both the number of endogenous variables and the number of exogenous variables can be very large, invalidating the classical methods for building gene regulatory networks. It is noteworthy that, although each structural equation modeling gene regulation has few exogenous variables, the genome-wide gene regulatory network consists of a large number of structural equations and therefore has a large number of exogenous variables.

The instrumental variables view of 2SLS sheds light on the consistency of 2SLS estimators which is guaranteed by good estimation of the conditional expectations of endogenous variables given exogenous variables. For large systems, we proposed to estimate these conditional expectations via ridge regression coupled with GCV so as to address possible overfitting issues brought by a large number of exogenous variables. We obtained approximately optimal estimation of these conditional expectations at the first stage. At the second stage, we could adopt results from high-dimensional variable selection, e.g., \citet{Fan2001}, \citet{Zou2006}, \citet{Zhang2010}, and \citet{Huang2011}, to consistently identify and further estimate the regulatory effects of the endogenous variables. As a high-dimensional extension of the classical 2SLS method, the 2SPLS method is also computationally fast and easy to implement. As shown in constructing a genome-wide gene regulatory network of yeast, the high computational efficiency of 2SPLS allows us to employ the bootstrap method to calculate the $p$-values of the regulatory effects.

Our simulation studies show a seemingly counterintuitive result that our moment-based method 2SPLS provides higher power than the likelihood-based method SML, because the maximum likelihood method is usually the most efficient method, and dominates moment methods. However, as evidenced in \citet{Bollen1996} and \citet{Kennedy1985} (p.134), 2SLS can perform better than the maximum likelihood method in small samples. Furthermore, SML is not a pure likelihood method but rather a penalized likelihood method, and 2SPLS is not a pure moment method but rather a penalized moment method. Therefore, the theoretical advantage of likelihood methods over moment methods may not carry over to comparing penalized likelihood methods versus penalized moment methods. In fact, SML uses an $L_1$ penalty to penalize nonzero regulatory effects, but 2SPLS employs an $L_2$ penalty on the regression coefficients of the reduced models at the first stage and an $L_1$ penalty on the regulatory effects at the second stage. We conjecture that the different choice of penalty terms may also distinguish the two different methods as shown in the advantage of the elastic net \citep{Zou2005} over lasso \citep{Tibshirani1996}.

Although applicable to diverse fields, our development of 2SPLS is motivated by constructing gene regulatory networks using genetical genomics data. The algorithm is applicable to any population-based studies with either experimental crosses or natural populations. Assumption A means that each gene under investigation has at least one unique polymorphism from its cis-eQTL, which can be detected with classical eQTL mapping methods, e.g., \citet{Kendziorski2006}, \citet{Gelfond2007}, and \citet{Jia2007}. Trans-eQTL (i.e., eQTL outside the regions of their target genes) hold the key to our understanding of gene regulation because their indirect regulations are likely caused by interactions among genes. When the gene regulatory network is modeled with a system of structural equations, classical eQTL mapping methods essentially identify both cis-eQTL and trans-eQTL involved in each reduced-form equation in the reduced model (\ref{Eqn-ReducedForm}). Nonetheless, it is challenging, if not impossible, to recover a large system from the reduced model.

An alternative strategy to construct the whole system is to build undirected graphs first \citep{Spirtes2001, Shipley2002, Fuente2004} and then locally orient the edges in the graphs \citep{Aten2008, Neto2008}. While constructing a small network is much easier and more robust than constructing a large system, we here intend to construct large networks, such as whole-genome gene regulatory networks from genetical genomics data. Furthermore, application of the alternative strategy is contingent on whether the underlying system is composed of unconnected subsystems, because ignoring the regulatory effects from other genes outside a subset of genes may lead to false regulatory interaction \citep{Neto2008, Fuente2004}. Instead, 2SPLS allows to construct a subset of structural equations inside the whole system, ignoring many other structural equations. Therefore, we can apply 2SPLS to investigate the interactive regulation among a subset of genes as well as how these genes are regulated by others.

It is evidenced in different species that effects of trans-eQTL are weaker than those of cis-eQTL and trans-eQTL are more difficult to identify than cis-eQTL \citep{Schadt2003, Dixon2007}. On the other hand, a system of structural equations modeling genome-wide gene regulation may induce a large number of trans-eQTL to each reduced-form equation in (\ref{Eqn-ReducedForm}). While constructing the system is contingent on the accuracy of predicting each endogenous variable on the basis of the corresponding reduced-form equation in (\ref{Eqn-ReducedForm}), the weak effects of a large number of trans-eQTL privilege the use of ridge regression at the first stage of 2SPLS for constructing gene regulatory networks \citep{Frank1993}. By comparing 2SPLS with 2SAL, our simulation studies demonstrated the superiority of using ridge regression over the adaptive lasso at the first stage. In fact, when some genes have a relatively large number of trans-eQTL, selecting variables at the first stage may compromise the identification of regulatory effects at the second stage.

\acks{We thank the action editor and four anonymous reviewers for their helpful comments. This work is partially supported by NSF CAREER award IIS-0844945, NIH R03CA211831, and the Cancer Care Engineering project at the Oncological Science Center of Purdue University.}

\appendix

\section*{Appendix A: Proof of Theorem \ref{Thm-RidgeEst}}

a. Since $\tau_j/\sqrt{n}\to 0$ for any $1\le j\le p$, the different choice of $\tau_j$ for each $j$ does not affect the following asymptotic property involving $\tau_j$,
\begin{eqnarray} \label{Asy-XXtau}
n(\mathbf{X}^T\mathbf{X}+\tau_j\mathbf{I})^{-1}\to\mathbf{C}.
\end{eqnarray}
Without loss of generality, we assume $\tau_1=\tau_2=\dots=\tau_p=\tau$. Then $\hat{\mathbf{Z}}_{-k}=\mathbf{P}_\tau\mathbf{Y}_{-k}$.
\begin{eqnarray*}
n^{-1}\hat{\mathbf{Z}}_{-k}^T\mathbf{H}_k\hat{\mathbf{Z}}_{-k}
&=& n^{-1} (\mathbf{X}\mathbf{\boldsymbol{\pi}}_{-k}+\mathbf{\boldsymbol{\xi}}_{-k})^T \mathbf{P}_{\tau}^T \mathbf{H}_k \mathbf{P}_{\tau} (\mathbf{X}\mathbf{\boldsymbol{\pi}}_{-k}+\mathbf{\boldsymbol{\xi}}_{-k}) \\
&=& n^{-1} \mathbf{\boldsymbol{\pi}}_{-k}^T \mathbf{X}^T\mathbf{P}_{\tau}\mathbf{H}_k\mathbf{P}_{\tau}\mathbf{X}\mathbf{\boldsymbol{\pi}}_{-k} + n^{-1} \mathbf{\boldsymbol{\xi}}_{-k}^T\mathbf{P}_{\tau}\mathbf{H}_k\mathbf{P}_{\tau}\mathbf{X}\mathbf{\boldsymbol{\pi}}_{-k}\\
& & + n^{-1} \mathbf{\boldsymbol{\pi}}_{-k}^T\mathbf{X}^T\mathbf{P}_{\tau}\mathbf{H}_k\mathbf{P}_{\tau}\mathbf{\boldsymbol{\xi}}_{-k} + n^{-1} \mathbf{\boldsymbol{\xi}}_{-k}^T\mathbf{P}_{\tau}\mathbf{H}_k\mathbf{P}_{\tau}\mathbf{\boldsymbol{\xi}}_{-k}
\end{eqnarray*}
We will consider the asymptotic property of each of the above four terms.

First, $n^{-1}\mathbf{X}^T\mathbf{X}\to\mathbf{C}$ implies that
\begin{eqnarray} \label{Asy-XHX}
n^{-1} \mathbf{X}^T\mathbf{H}_k\mathbf{X} =
n^{-1} \mathbf{X}^T \{\mathbf{I}-\mathbf{X}_{\mathcal{S}_k} (\mathbf{X}_{\mathcal{S}_k}^T \mathbf{X}_{\mathcal{S}_k})^{-1}\mathbf{X}_{\mathcal{S}_k}^T\}\mathbf{X} \to \mathbf{C}-\mathbf{C}_{\bullet\mathcal{S}_k}\mathbf{C}_{\mathcal{S}_k,\mathcal{S}_k}^{-1} \mathbf{C}_{\mathcal{S}_k\bullet}.
\end{eqnarray}
The above result and (\ref{Asy-XXtau}) easily lead to the following result,
\begin{eqnarray} \label{Asy-Mk}
\lefteqn{n^{-1} \mathbf{\boldsymbol{\pi}}_{-k}^T\mathbf{X}^T \mathbf{P}_{\tau}\mathbf{H}_k\mathbf{P}_{\tau}\mathbf{X}\mathbf{\boldsymbol{\pi}}_{-k}} \nonumber\\
&=& n^{-1} \mathbf{\boldsymbol{\pi}}_{-k}^T\mathbf{X}^T \mathbf{X} (\mathbf{X}^T\mathbf{X}+\tau\mathbf{I})^{-1} \mathbf{X}^T\mathbf{H}_k\mathbf{X} (\mathbf{X}^T\mathbf{X}+\tau\mathbf{I})^{-1} \mathbf{X}^T\mathbf{X}\mathbf{\boldsymbol{\pi}}_{-k} \nonumber \\
&\to& \mathbf{\boldsymbol{\pi}}_{-k}^T (\mathbf{C}-\mathbf{C}_{\bullet\mathcal{S}_k}\mathbf{C}_{\mathcal{S}_k,\mathcal{S}_k}^{-1} \mathbf{C}_{\mathcal{S}_k\bullet}) \mathbf{\boldsymbol{\pi}}_{-k} = \mathbf{M}_k.
\end{eqnarray}

The other three terms approaching to zero directly follows that $n^{-1} \mathbf{\boldsymbol{\xi}}_{-k}^T\mathbf{X}\to_p\mathbf{0}$. Thus, $\frac{1}{n}\hat{\mathbf{Z}}_{-k}^T\mathbf{H}_k\hat{\mathbf{Z}}_{-k} \to_p \mathbf{M}_k$.

b. Since $\mathbf{H}_k(\mathbf{Y}_{k}-\mathbf{Y}_{-k}\boldsymbol{\gamma}_k)=\mathbf{H}_k\boldsymbol{\epsilon}_k$, we have
\begin{eqnarray*}
\lefteqn{n^{-1/2} (\mathbf{Y}_k-\hat{\mathbf{Z}}_{-k}\boldsymbol{\gamma}_k)^T \mathbf{H}_k \hat{\mathbf{Z}}_{-k}}\\
&=& n^{-1/2} \{(\mathbf{Y}_{k}-\mathbf{Y}_{-k}\boldsymbol{\gamma}_k) + (\mathbf{I}-\mathbf{P}_{\tau})\mathbf{Y}_{-k}\boldsymbol{\gamma}_k\}^T \mathbf{H}_k \hat{\mathbf{Z}}_{-k} \\
&=& n^{-1/2} \boldsymbol{\epsilon}_k^T \mathbf{H}_k\mathbf{P}_{\tau}\mathbf{Y}_{-k} + n^{-1/2} \boldsymbol{\gamma}_k^T \{(\mathbf{I}-\mathbf{P}_{\tau})\mathbf{Y}_{-k}\}^T \mathbf{H}_k\mathbf{P}_{\tau}\mathbf{Y}_{-k}.
\end{eqnarray*}

In the following, we will prove that the second term approaches to zero, and the first term asymptotically approaches to the required distribution, i.e.,
\begin{eqnarray} \label{Asy-EHPX}
n^{-1/2} \boldsymbol{\epsilon}_k^T \mathbf{H}_k\mathbf{P}_{\tau}\mathbf{Y}_{-k} \to_d N(\mathbf{0},\sigma_k^2 \mathbf{M}_k).
\end{eqnarray}

We notice that
\[
n^{-1/2} \boldsymbol{\epsilon}_k^T \mathbf{H}_k\mathbf{P}_{\tau} \mathbf{X}\mathbf{\boldsymbol{\pi}}_{-k} \sim N(\mathbf{0}, n^{-1}\sigma_k^2 \mathbf{\boldsymbol{\pi}}_{-k}^T\mathbf{X}^T\mathbf{P}_{\tau}\mathbf{H}_k\mathbf{P}_{\tau}\mathbf{X}\mathbf{\boldsymbol{\pi}}_{-k}).
\]
Following (\ref{Asy-Mk}), we have
\begin{eqnarray} \label{Asy-EHPXPi}
n^{-1/2} \boldsymbol{\epsilon}_k^T \mathbf{H}_k\mathbf{P}_{\tau} \mathbf{X}\mathbf{\boldsymbol{\pi}}_{-k} \to_d N(\mathbf{0},\sigma_k^2 \mathbf{M}_k).
\end{eqnarray}

Because of (\ref{Asy-XHX}) and
\[
n^{-1/2} \boldsymbol{\epsilon}_k^T\mathbf{H}_k\mathbf{X} \sim N(\mathbf{0}, n^{-1}\sigma_k^2\mathbf{X}^T\mathbf{H}_k\mathbf{X}),
\]
we have
\[
n^{-1/2} \boldsymbol{\epsilon}_k^T\mathbf{H}_k\mathbf{X} \to_d N(\mathbf{0}, \sigma_k^2(\mathbf{C}-\mathbf{C}_{\bullet\mathcal{S}_k} \mathbf{C}_{\mathcal{S}_k,\mathcal{S}_k}^{-1} \mathbf{C}_{\mathcal{S}_k\bullet})).
\]
Since $n^{-1}\boldsymbol{\xi}_{-k}^T\mathbf{X}\to_p \mathbf{0}$, we can apply Slutsky's theorem and obtain that
\begin{eqnarray*}
n^{-1/2} \boldsymbol{\epsilon}_k^T \mathbf{H}_k\mathbf{P}_{\tau}\mathbf{\boldsymbol{\xi}}_{-k} = n^{-1/2} \boldsymbol{\epsilon}_k^T \mathbf{H}_k\mathbf{X}(\mathbf{X}^T\mathbf{X} + \tau\mathbf{I})^{-1}\mathbf{X}^T\mathbf{\boldsymbol{\xi}}_{-k} \to_p \mathbf{0}.
\end{eqnarray*}
Pooling the above result and (\ref{Asy-EHPXPi}) leads to the asymptotic distribution in (\ref{Asy-EHPX}).

To prove that the second term asymptotically approaches to zero, we further partition it as follows,
\begin{eqnarray*}
\lefteqn{n^{-1/2} \boldsymbol{\gamma}_k^T\{(\mathbf{I}-\mathbf{P}_{\tau}) \mathbf{Y}_{-k}\}^T\mathbf{H}_k\mathbf{P}_{\tau}\mathbf{Y}_{-k}}\\
&=& n^{-1/2} \boldsymbol{\gamma}_k^T\mathbf{\boldsymbol{\pi}}_{-k}^T\mathbf{X}^T (\mathbf{I}-\mathbf{P}_{\tau}) \mathbf{H}_k\mathbf{P}_{\tau}\mathbf{X}\mathbf{\boldsymbol{\pi}}_{-k} + n^{-1/2} \boldsymbol{\gamma}_k^T\boldsymbol{\xi}_{-k}^T (\mathbf{I}-\mathbf{P}_{\tau}) \mathbf{H}_k\mathbf{P}_{\tau}\mathbf{X}\mathbf{\boldsymbol{\pi}}_{-k}\\
&& +n^{-1/2} \boldsymbol{\gamma}_k^T\mathbf{\boldsymbol{\pi}}_{-k}^T \mathbf{X}^T (\mathbf{I}-\mathbf{P}_{\tau}) \mathbf{H}_k\mathbf{P}_{\tau}\boldsymbol{\xi}_{-k} + n^{-1/2} \boldsymbol{\gamma}_k^T\boldsymbol{\xi}_{-k}^T (\mathbf{I}-\mathbf{P}_{\tau}) \mathbf{H}_k\mathbf{P}_{\tau}\boldsymbol{\xi}_{-k}.
\end{eqnarray*}
It suffices to prove each of these four parts asymptotically approaches to zero.

First, notice that
\[
\mathbf{X}^T (\mathbf{I}-\mathbf{P}_{\tau}) = \tau(\mathbf{X}^T\mathbf{X}+\tau\mathbf{I})^{-1}\mathbf{X}^T,
\]
we have
\begin{eqnarray} \label{Asy-XIPHPXPi}
\lefteqn{n^{-1/2} \boldsymbol{\gamma}_k^T\mathbf{\boldsymbol{\pi}}_{-k}^T\mathbf{X}^T (\mathbf{I}-\mathbf{P}_{\tau}) \mathbf{H}_k\mathbf{P}_{\tau}\mathbf{X}\mathbf{\boldsymbol{\pi}}_{-k}} \nonumber\\
&=& n^{-1/2}\tau \boldsymbol{\gamma}_k^T\mathbf{\boldsymbol{\pi}}_{-k}^T (\mathbf{X}^T\mathbf{X}+\tau\mathbf{I})^{-1} \mathbf{X}^T\mathbf{H}_k\mathbf{X} (\mathbf{X}^T\mathbf{X}+\tau\mathbf{I})^{-1} \mathbf{X}^T\mathbf{X}\mathbf{\boldsymbol{\pi}}_{-k} \to \mathbf{0},
\end{eqnarray}
which follows (\ref{Asy-XHX}) and that $\tau/\sqrt{n}\to 0$ as $n\to \infty$.

Because $\mathbf{C}_{\mathcal{S}_{k}\bullet}\mathbf{C}^{-1}\mathbf{C}_{\bullet\mathcal{S}_{k}} = \mathbf{C}_{\mathcal{S}_{k}\mathcal{S}_{k}}$, we have
\[
(\mathbf{C}-\mathbf{C}_{\bullet\mathcal{S}_k} \mathbf{C}_{\mathcal{S}_k,\mathcal{S}_k}^{-1} \mathbf{C}_{\mathcal{S}_k\bullet}) \mathbf{C}^{-1} (\mathbf{C}-\mathbf{C}_{\bullet\mathcal{S}_k} \mathbf{C}_{\mathcal{S}_k,\mathcal{S}_k}^{-1} \mathbf{C}_{\mathcal{S}_k\bullet}) = \mathbf{C}-\mathbf{C}_{\bullet\mathcal{S}_k} \mathbf{C}_{\mathcal{S}_k,\mathcal{S}_k}^{-1} \mathbf{C}_{\mathcal{S}_k\bullet},
\]
which implies that
\begin{eqnarray*}
\lefteqn{n^{-1/2} \mathbf{X}^T \mathbf{P}_{\tau}^T \mathbf{H}_k^T (\mathbf{I}-\mathbf{P}_{\tau})^T (\mathbf{I}-\mathbf{P}_{\tau}) \mathbf{H}_k\mathbf{P}_{\tau}\mathbf{X}} \\
&=& n^{-1} \mathbf{X}^T\mathbf{P}_{\tau}\mathbf{H}_k\mathbf{P}_{\tau}\mathbf{X} - 2n^{-1} \mathbf{X}^T\mathbf{P}_{\tau}\mathbf{H}_k\mathbf{P}_{\tau}\mathbf{H}_k\mathbf{P}_{\tau}\mathbf{X} + n^{-1} \mathbf{X}^T\mathbf{P}_{\tau}\mathbf{H}_k\mathbf{P}_{\tau}^2\mathbf{H}_k\mathbf{P}_{\tau}\mathbf{X}\to \mathbf{0}.
\end{eqnarray*}
Since Var$(\boldsymbol{\xi}_{-k}\boldsymbol{\gamma}_k)$ is proportional to an identity matrix, the above result leads to that
\[
\mathrm{Var}\left(n^{-1/2} \boldsymbol{\gamma}_k^T\mathbf{\boldsymbol{\xi}}_{-k}^T (\mathbf{I}-\mathbf{P}_{\tau}) \mathbf{H}_k\mathbf{P}_{\tau}\mathbf{X}\mathbf{\boldsymbol{\pi}}_{-k}\right)\to \mathbf{0},
\]
which implies that
\begin{eqnarray} \label{Asy-GXIPHPXPi}
n^{-1/2} \boldsymbol{\gamma}_k^T\mathbf{\boldsymbol{\xi}}_{-k}^T (\mathbf{I}-\mathbf{P}_{\tau}) \mathbf{H}_k\mathbf{P}_{\tau}\mathbf{X}\mathbf{\boldsymbol{\pi}}_{-k} \to_p \mathbf{0}.
\end{eqnarray}

Similarly, we can prove that, for each $\boldsymbol{\xi}_{j}$,
\[
\mathrm{Var}\left(n^{-1/2} \boldsymbol{\gamma}_k^T \mathbf{\boldsymbol{\pi}}_{-k}^T\mathbf{X}^T (\mathbf{I}-\mathbf{P}_{\tau}) \mathbf{H}_k\mathbf{P}_{\tau}\mathbf{\boldsymbol{\xi}}_j\right)\to \mathbf{0},
\]
which implies that
\begin{eqnarray} \label{Asy-GPiXIPHPX}
n^{-1/2} \boldsymbol{\gamma}_k^T \mathbf{\boldsymbol{\pi}}_{-k}^T\mathbf{X}^T (\mathbf{I}-\mathbf{P}_{\tau}) \mathbf{H}_k\mathbf{P}_{\tau}\mathbf{\boldsymbol{\xi}}_{-k} \to_p \mathbf{0}.
\end{eqnarray}

Note that
\begin{eqnarray*}
n^{-1/2} \boldsymbol{\gamma}_k^T\mathbf{\boldsymbol{\xi}}_{-k}^T(\mathbf{I}-\mathbf{P}_{\tau}) \mathbf{H}_k\mathbf{P}_{\tau}\mathbf{\boldsymbol{\xi}}_{-k} = \left\{n^{-1/2} \boldsymbol{\gamma}_k^T\mathbf{\boldsymbol{\xi}}_{-k}^T(\mathbf{I}-\mathbf{P}_{\tau}) \mathbf{H}_k
\mathbf{X}\right\} \left\{(\mathbf{X}^T\mathbf{X}+\tau \mathbf{I})^{-1} \mathbf{X}^T\mathbf{\boldsymbol{\xi}}_{-k}\right\}.
\end{eqnarray*}
Since
\begin{eqnarray*}
n^{-1}\mathbf{X}^T \mathbf{H}_k (\mathbf{I}-\mathbf{P}_{\tau})
(\mathbf{I}-\mathbf{P}_{\tau}) \mathbf{H}_k\mathbf{X} \to \mathbf{0},
\end{eqnarray*}
we have
\[
\mathrm{Var}\left(n^{-1/2} \boldsymbol{\gamma}_k^T\mathbf{\boldsymbol{\xi}}_{-k}^T(\mathbf{I}-\mathbf{P}_{\tau}) \mathbf{H}_k
\mathbf{X}\right) \to \mathbf{0}.
\]
Therefore,
\[
n^{-1/2} \boldsymbol{\gamma}_k^T\mathbf{\boldsymbol{\xi}}_{-k}^T(\mathbf{I}-\mathbf{P}_{\tau}) \mathbf{H}_k
\mathbf{X} \to _p \mathbf{0},
\]
which, together with $(\mathbf{X}^T\mathbf{X}+\tau \mathbf{I})^{-1}\mathbf{X}^T\boldsymbol{\xi}_{-k}\to_p \mathbf{0}$, leads to that
\begin{eqnarray}\label{Asy-GXIPHPX}
n^{-1/2} \boldsymbol{\gamma}_k^T\mathbf{\boldsymbol{\xi}}_{-k}^T(\mathbf{I}-\mathbf{P}_{\tau}) \mathbf{H}_k\mathbf{P}_{\tau}\mathbf{\boldsymbol{\xi}}_{-k}\to_p \mathbf{0}.
\end{eqnarray}

Pooling (\ref{Asy-XIPHPXPi}), (\ref{Asy-GXIPHPXPi}), (\ref{Asy-GPiXIPHPX}) and (\ref{Asy-GXIPHPX}), we have proved that
$n^{-1/2} \boldsymbol{\gamma}_k^T\{(\mathbf{I}-\mathbf{P}_{\tau}) \mathbf{Y}_{-k}\}^T\mathbf{H}_k\mathbf{P}_{\tau}\mathbf{Y}_{-k}\to_p \mathbf{0}$, which concludes the proof.

\section*{Appendix B: Proof of Theorem~\ref{Thm-Oracle}}

Let $\boldsymbol{\psi}_n{(\boldsymbol{\mu})} = \|\mathbf{H}_k\mathbf{Y}_k-\mathbf{H}_k\hat{\mathbf{Z}}_{-k}(\boldsymbol{\gamma}_k + \boldsymbol{\mu}/\sqrt{n})\|_2^2 + \lambda_k \mathbf{\boldsymbol{\omega}}_k^T |\boldsymbol{\gamma}_k + \boldsymbol{\mu}/\sqrt{n}|$. Let $\hat{\boldsymbol{\mu}} = \arg\min_{\boldsymbol{\mu}}\boldsymbol{\psi}_n(\boldsymbol{\mu})$, then $\hat{\boldsymbol{\gamma}}_k = \boldsymbol{\gamma}_k + \hat{\boldsymbol{\mu}}/\sqrt{n}$ or $\hat{\boldsymbol{\mu}} = \sqrt{n}(\hat{\boldsymbol{\gamma}}_k - \boldsymbol{\gamma}_k)$. Note that $\boldsymbol{\psi}_n(\boldsymbol{\mu})-\boldsymbol{\psi}_n(\mathbf{0})=V_n(\boldsymbol{\mu})$, where
\begin{eqnarray*}
V_n(\boldsymbol{\mu}) & = & \boldsymbol{\mu}^T(n^{-1}\hat{\mathbf{Z}}_{-k}^T\mathbf{H}_k\hat{\mathbf{Z}}_{-k})\boldsymbol{\mu} - 2n^{-1/2} (\mathbf{Y}_{k} - \hat{\mathbf{Z}}_{-k} \boldsymbol{\gamma}_k)^T \mathbf{H}_k \hat{\mathbf{Z}}_{-k} \boldsymbol{\mu} \\
& & + n^{-1/2} \lambda_k \mathbf{\boldsymbol{\omega}}_k^T\times \sqrt{n}(|\boldsymbol{\gamma}_k + n^{-1/2}\boldsymbol{\mu}|-|\boldsymbol{\gamma}_k|).
\end{eqnarray*}

Denote the $j$-th elements of $\boldsymbol{\omega}_k$ and $\boldsymbol{\mu}$ as $\omega_{kj}$ and $\mu_j$, respectively.

If $\gamma_{kj}\neq 0$, then $\omega_{kj}\to_p |\gamma_{kj}|^{-\delta}$ and $\sqrt{n}(|\gamma_{kj}+\mu_j/\sqrt{n}|-|\gamma_{kj}|) \to_p \mu_j \mathrm{sign}({\gamma_{kj}})$. By Slutsky's theorem, we have $(\lambda_k/\sqrt{n})\omega_{kj}\sqrt{n}(|\gamma_{kj}+\mu_j/\sqrt{n}|-|\gamma_{kj}|) \to_p 0$. If $\gamma_{kj}=0$, then $\sqrt{n}(|\gamma_{kj}+\mu_j/\sqrt{n}|-|\gamma_{kj}|)=|\mu_j|$ and $(\lambda_k/\sqrt{n}) \omega_{kj} = (\lambda_k/\sqrt{n}) n^{\delta/2}(|\sqrt{n}\tilde{\gamma}_{kj}|)^{-\delta}$, where $\sqrt{n}\tilde{\gamma}_{kj}=O_p{(1)}$. Thus,
\[
n^{-1/2}\lambda_k \mathbf{\boldsymbol{\omega}}_k^T\times n^{1/2}(|\boldsymbol{\gamma}_k + n^{-1/2} \boldsymbol{\mu}|-|\boldsymbol{\gamma}_k|) \to_p \left\{\begin{array}{ll}
0, &\mathrm{if}\ \|\boldsymbol{\mu}_{\mathcal{A}_k^c}\|_2=0; \\
\infty, &\mathrm{otherwise}.
\end{array}\right.
\]
Hence, following Theorem \ref{Thm-RidgeEst} and Slutsky's theorem, we see that $V_n(\boldsymbol{\mu})\to_d V(\boldsymbol{\mu})$ for every $\boldsymbol{\mu}$, where
\[
V(\boldsymbol{\mu})=
\left\{\begin{array}{ll}
\boldsymbol{\mu}_{\mathcal{A}_k}^T\mathbf{M}_{k,\mathcal{A}_k} \boldsymbol{\mu}_{\mathcal{A}_k}-2 \boldsymbol{\mu}_{\mathcal{A}_k}^T\mathbf{W}_{k,\mathcal{A}_k}, &\mathrm{if}\ \|\boldsymbol{\mu}_{\mathcal{A}_k^c}\|_2=0; \\
\infty, &\mathrm{otherwise}.
\end{array}\right.
\]
$V_n(\boldsymbol{\mu})$ is convex, and the unique minimizer of $V(\boldsymbol{\mu})$ is $(\mathbf{M}_{k,\mathcal{A}_k}^{-1}\mathbf{W}_{k,\mathcal{A}_k},\mathbf{0})^T$.
Following the epi-convergence results of \citet{Geyer1994} and \citet{Fu2000}, we have
\begin{eqnarray*}
\left\{\begin{array}{l}\hat{\boldsymbol{\mu}}_{\mathcal{A}_k} \to_d \mathbf{M}_{k,\mathcal{A}_k}^{-1}\mathbf{W}_{k,\mathcal{A}_k},\\
\hat{\boldsymbol{\mu}}_{\mathcal{A}_k^c} \to_d \mathbf{0}. \end{array}\right.
\end{eqnarray*}
Since $\mathbf{W}_{k,\mathcal{A}_k} \sim N(\mathbf{0},\sigma_k^2\mathbf{M}_{k,\mathcal{A}_k})$, we indeed have proved the asymptotic normality.

Now we show the consistency in variable selection. $\forall j \in \mathcal{A}_k$, the asymptotic normality indicates that $\hat{\gamma}_{kj} \to_p \gamma_{kj}$, thus $P(j\in\hat{\mathcal{A}}_k) \to 1$. Then it suffices to show that $\forall j \notin \mathcal{A}_k$, $P(j\in\hat{\mathcal{A}}_k) \to 0$.

When $j\in\hat{\mathcal{A}}_k$, by the KKT normality conditions, we know that $\hat{\mathbf{Z}}_{j}^T \mathbf{H}_k  (\mathbf{Y}_k-\hat{\mathbf{Z}}_{-k}\hat{\boldsymbol{\gamma}}_k) = \lambda_k \omega_{kj}$. Note that $\lambda_k\omega_{kj}/\sqrt{n} \to_p \infty$, whereas $\hat{\mathbf{Z}}_{j}^T \mathbf{H}_k (\mathbf{Y}_k - \hat{\mathbf{Z}}_{-k} \hat{\boldsymbol{\gamma}}_k)/\sqrt{n} = (\hat{\mathbf{Z}}_{j}^T \mathbf{H}_k \hat{\mathbf{Z}}_{-k}/n) \times \sqrt{n} (\boldsymbol{\gamma}_k-\hat{\boldsymbol{\gamma}}_k) + \hat{\mathbf{Z}}_{j}^T \mathbf{H}_k (\mathbf{Y}_k-\hat{\mathbf{Z}}_{-k}\boldsymbol{\gamma}_k)/\sqrt{n}$. Following Theorem \ref{Thm-RidgeEst} and the asymptotic normality, $\hat{\mathbf{Z}}_{j}^T \mathbf{H}_k (\mathbf{Y}_k - \hat{\mathbf{Z}}_{-k} \hat{\boldsymbol{\gamma}}_k)/\sqrt{n}$ asymptotically follows a normal distribution. Thus, $P(j\in\hat{\mathcal{A}}_k) \leq P(\hat{\mathbf{Z}}_{j}^T \mathbf{H}_k (\mathbf{Y}_k-\hat{\mathbf{Z}}_{-k} \hat{\boldsymbol{\gamma}}_k) = \lambda_k\omega_{kj}) \to 0$. Then we have proved the consistency in variable selection.

\section*{Appendix C: Proof of Theorem \ref{ridgetheoryk}}

Denote $\lambda_{min}(\mathbf{M})$ and $\lambda_{max}(\mathbf{M})$ the minimum and maximum eigenvalues of matrix $\mathbf{M}$, respectively. Follow Assumption B$^{\prime}$ to assume that the singular values of matrix $\textbf{I}-\boldsymbol{\Gamma}$ are positively bounded from below by a constant $c$. Further denote $\tilde{\sigma}_k^{2}=\var(\bxi_{k})$, and $\sigma^2_{p\max} = \underset{1\le k \le p}{\max}(\sigma^2_k)$. Noting that $\boldsymbol{\xi} = \boldsymbol{\epsilon} (\textbf{I}-\boldsymbol{\Gamma})^{-1}$, we have $\tilde{\sigma}^2_k \le \sigma^2_{p\max}/c$.

(a)	From the ridge regression, we have the following closed form solution,
\begin{equation*}
\begin{aligned}
\hat{\bpi}_{k} = (\bX^T\bX +\tau_k I_q)^{-1}\bX^T\bY_k
= (\bX^T\bX +\tau_k I_q)^{-1}\bX^T\bX \bpi_{k}  + (\bX^T\bX +\tau_k I_q)^{-1}\bX^T \bxi_k.
\end{aligned}
\end{equation*}
Note that
\begin{equation*}
\hat{\bpi}_{k} - \bpi_{k} = -\tau_k (\bX^T\bX +\tau_k I_q)^{-1} \bpi_{k} + (\bX^T\bX +\tau_k I_q)^{-1} \bX^T \bxi_k
= \bmu + A_k^T \bxi_k,
\end{equation*}
where $\bmu = -\tau_k (\bX^T\bX +\tau_k I_q)^{-1} $ and $A_{k} = \bX (\bX^T\bX +\tau_k I_q)^{-1}$. Then we have
\begin{equation} \label{eq-expansiona}
\ltwon{\hat{\bpi}_{k}  - \bpi_{k} }^2 = \underbrace{\bmu^T\bmu}_{T_1} + \underbrace{2\bmu^TA_{k}^T\bxi_k}_{T_2}+ \underbrace{\bxi_k^T A_{k} A_{k}^T\bxi_k}_{T_3}.
\end{equation}

Via the singular value decomposition of $\bX$, we can have the decomposition $\bX^T \bX = \mathbf{P}^T \mathbf{U} \mathbf{P}$, where $\mathbf{P}$ is a unitary matrix and matrix $\mathbf{U}$ is a diagonal matrix with non-negative diagonal elements $u_i$. Therefore,
\begin{equation*}
(\bX^T\bX +\tau_k I_q)^{-2} =  \mathbf{P}^T(\mathbf{U}+\tau_k I_q)^{-2}\mathbf{P}.
\end{equation*}

Following Assumption B$^\prime$, we have $\lambda_{min}(\bX^T \bX) > c_{2}^2n$  and  $\lambda_{max}(\bX^T \bX) < c_{1}^2n$, which implies that $u_i \asymp n $ for all $i$. Therefore,
\begin{equation} \label{term1bound}
T_1 =  \tau_k^2 \bpi_{k}^T \mathbf{P}^T (\mathbf{U}+\tau_k I_q)^{-2} \mathbf{P}\bpi_{k}  = \sum_{i=1}^{q} \frac{\tau_k^2 a_{ik}^2}{(u_i + \tau_k)^2} = \bigO{\tau_k^2\ltwon{\bpi_{k} }^2/n^2}=\bigO{r_{nk}/n},
\end{equation}
where $a_{ik}$ is the $i$-th element of $\mathbf{a}_k = \mathbf{P}\bpi_{k} $ with $\ltwon{\mathbf{a}_k}=\ltwon{\bpi_{k} }$.

For the term $T_2$, we have that
\begin{equation*}
E[T_2]=0, \ \ \ \ \ \ \text{Var}(T_2) = 4\tilde{\sigma}_k^2 \bmu^TA_{k}^TA_{k}\bmu.
\end{equation*}
By the classical Gaussian tail probability, we have
\begin{equation*}
\mathbb{P}\left(T_2\le t\right) \ge 1-\exp\left\{-t^2\big/\left(8\tilde{\sigma}_k^2\bmu^TA_{k}^TA_{k}\bmu\right)\right\}.
\end{equation*}
Note that,
\begin{equation*}
\bmu^TA_{k}^TA_{k}\bmu = \tau_k^2 \bpi_{k}^T \mathbf{P}^T (\mathbf{U}+\tau_k I_q)^{-2} \mathbf{U} (\mathbf{U}+\tau_k I_q)^{-2} \mathbf{P}\bpi_{k}
= \sum_{i=1}^{q} \frac{\tau_k^2 u_i a_{ik}^2}{(u_i+\tau_k)^4} = \bigO{\tau_k^2 \ltwon{\bpi_{k}}^2\big/n^3}.
\end{equation*}
Letting $t = \sqrt{8\tilde{\sigma}_k^2\bmu^TA_{k}^TA_{k}\bmu (f_n+\log 2)}$, we have, with probability at least $1-e^{-f_n}/2$,
\begin{equation} \label{t2probbound}
T_2 = \bigO{\sqrt{r_{nk} f_n}/n}.
\end{equation}

For the term $T_3$, we can invoke the Hanson-Wright inequality \citep{rudelson2013hanson} to have, for some constant $t_1>0$,
\begin{equation*}
P(T_3 \le E[T_3] + t) \ge 1- \text{exp}\left\{-t_1 \min\left(\frac{t^2}{\tilde{\sigma}_k^4\frobn{A_{k}A_{k}^T}^2}, \frac{t}{\tilde{\sigma}_k^2\opn{A_{k}A_{k}^T}}\right)\right\},
\end{equation*}
where $\opn{\cdot} = \underset{x\ne 0}{\max} \ltwon{\cdot x}/\ltwon{x}$ is the operator norm and $\frobn{\cdot}$ is the Frobenius norm.

Since
\begin{equation*}
A_{k}A_{k}^T = \bX(\bX^T\bX +\tau_k I_q)^{-2}\bX^T = \bX \mathbf{P}^T(\mathbf{U}+\tau_k I_q)^{-2}\mathbf{P}\bX^T,
\end{equation*}
we have
\begin{equation*}
\begin{aligned}
E[T_3] &= \tilde{\sigma}_k^2 \text{trace}(A_{k}A_{k}^T) =\tilde{\sigma}_k^2 \text{trace}(\bX^T\bX(\bX^T\bX +\tau_k I_q)^{-2})\\
&= \tilde{\sigma}_k^2 \text{trace}(\mathbf{U}(\mathbf{U}+\tau_k I_q)^{-2}) = \sum_{i=1}^{q} \frac{\tilde{\sigma}_k^2  u_i}{(u_i+\tau_k)^2} = \bigO{\tilde{\sigma}_k^2 q/n}, \\
\frobn{A_{k}A_{k}^T}^2 &= \text{trace}(A_{k}A_{k}^TA_{k}A_{k}^T)= \text{trace}(A_{k}^TA_{k}A_{k}^TA_{k})\\
&= \text{trace}(P^T\mathbf{U} (\mathbf{U}+\tau_k I_q)^{-2}U(U+\tau_k I_q)^{-2} ) = \sum_{i=1}^{q} \frac{u_i^2}{(u_i+\tau_k)^4} = \bigO{q\big/n^2},\\
\opn{A_{k}A_{k}^T} & = \bigO{\lambda_{max}\left({\bX\bX^T}\right)\big/n^2} =\bigO{n^{-1}}.
\end{aligned}
\end{equation*}
Letting $t = \max\left(\sqrt{\tilde{\sigma}_k^4\frobn{A_{k}A_{k}^T}^2(f_n+\log 2)/t_1}, \tilde{\sigma}_k^2\opn{A_{k}A_{k}^T}(f_n+\log 2)/t_1\right)$, we obtain that, with probablity at least $1-e^{-f_n}/2$,
\begin{equation} \label{t3probbound2}
T_3 = \bigO{q/n} + \bigO{\sqrt{f_n q}/n} +\bigO{f_n/n}.
\end{equation}

Collecting the bounds in (\ref{term1bound}), (\ref{t2probbound}), and (\ref{t3probbound2}), we conclude that there exist a positive constant $C_1$ such that, with probability at least $1-e^{-f_n}$,
\begin{equation*}
\ltwon{\hat{\bpi}_{k} - \bpi_{k} }^2 \le C_1 (r_{nk} \lor q \lor f_n)/n.
\end{equation*}

(b)	Similar to (\ref{eq-expansiona}), we have
\begin{equation*}
\ltwon{\bX(\hat{\bpi}_{k} - \bpi_{k} )}^2 = \underbrace{\bmu^T\bX^T\bX \bmu}_{T_4} + \underbrace{2\bmu^T\bX^T\bX A_{k}^T\bxi_k}_{T_5}+ \underbrace{\bxi_k^T A_{k}\bX^T\bX A_{k}^T\bxi_k}_{T_6}.
\end{equation*}

For the term $T_4$, we have
\begin{equation} \label{t4bound}
\begin{aligned}
T_4 &= \tau_k^2 \mathbf{a}_k^T \mathbf{U}(\mathbf{U}+\tau_k I_q)^{-1} \mathbf{U}(\mathbf{U}+ \tau_k I_q)^{-1}\mathbf{a}_k\\
&= \tau_k^2 \sum_{i=1}^{q} \frac{u_i a_{ik}^2}{(u_i+\tau_k)^2} = \bigO{\tau_k^2\ltwon{\bpi_{k} }^2\big/n} = \bigO{r_{nk}}.
\end{aligned}
\end{equation}

For the term $T_5$, by the classical Gaussian tail inequality, we have
\begin{equation*}
P\left(T_5\le t\right)\ge 1-\exp\left\{-t^2\big/(2\text{Var}(T_5))\right\},
\end{equation*}
where
\begin{equation*}
\begin{aligned}
\text{Var}(T_5) &= 4\tilde{\sigma}_k^2 \bmu^T\bX^T\bX A_{k}^TA_{k}\bX^T\bX\bmu\\
&=4\tilde{\sigma}_k^2 \tau_k^2 \mathbf{a}_k^T (\mathbf{U}+\tau_k I_q)^{-1} \mathbf{U}(\mathbf{U}+\tau_k I_q)^{-1} \mathbf{U}(\mathbf{U}+\tau_k I_q)^{-1} \mathbf{U}(\mathbf{U}+\tau_k I_q)^{-1}\mathbf{a}_k\\
&= 4\tilde{\sigma}_k^2\tau_k^2 \sum_{i=1}^{q}\frac{u_i^3 a_{ik}^2}{(u_i+\tau_k)^4} =\bigO{\tilde{\sigma}_k^2 \tau_k^2\ltwon{\bpi_{k}}^2/n}.
\end{aligned}
\end{equation*}
Taking $t = \sqrt{2\text{Var}(T_5) (f_n+\log 2)}$, we can obtain that, with probability at least $1- e^{-f_n}/2$,
\begin{equation} \label{t5probbound}
T_5 = \bigO{\sqrt{r_{nk} f_n}}.
\end{equation}	

For the term $T_6$, by the Hanson-Wright inequality, we have, for some constant $t_2>0$,
\begin{equation*}
P(T_6 \le \Ex{T_6} + t) \ge 1- \text{exp}\left\{-t_2 \min\left(\frac{t^2}{\tilde{\sigma}_k^4\frobn{A_{k}\bX^T\bX A_{k}^T}^2}, \frac{t}{\tilde{\sigma}_k^2\opn{A_{k}\bX^T\bX A_{k}^T}}\right)\right\}.
\end{equation*}
Similar to managing the term $T_3$ in (a), we have
\begin{equation*}
\begin{aligned}
E[T_6] &= \tilde{\sigma}_k^2\text{trace}(A_{k}\bX^T\bX A_{k}^T) = \tilde{\sigma}_k^2\text{trace}(U(U+\tau_k I_q)^{-1}U(U+\tau_k I_q)^{-1})\\
	& = \tilde{\sigma}_k^2\sum_{i=1}^{q}\frac{u_i^2}{(u_i+\tau_k)^{2}} = \bigO{\tilde{\sigma}_k^2 q}, \\
\frobn{A_{k}\bX^T\bX A_{k}^T}^2 &= \text{trace}( A_{k}\bX^T\bX A_{k}^T A_{k}\bX^T\bX A_{k}^T) = \text{trace}( \bX^T\bX A_{k}^T A_{k}\bX^T\bX A_{k}^TA_{k})\\
	&= \text{trace}(\mathbf{U}(\mathbf{U} +\tau_k I_q)^{-1} \mathbf{U}(\mathbf{U}+\tau_k I_q)^{-1} \mathbf{U}(\mathbf{U}+\tau_k I_q)^{-1} \mathbf{U}(\mathbf{U}+\tau_k I_q)^{-1} ) \\
	&= \sum_{i=1}^{q}\frac{u_i^4}{(u_i+\tau_k)^{4}} = \bigO{q}, \\
\opn{A_{k}\bX^T\bX A_{k}^T} &= \opn{\bX(\bX^T\bX+\tau_k I_q)^{-1} \bX^T \bX(\bX^T\bX+\tau_k I_q)^{-1}\bX^T} \\
	& = \bigO{\lambda_{\max}\left(\bX \bX^T \bX \bX^T\right)\big/n^2} = \bigO{1}.
\end{aligned}
\end{equation*}
Letting $t = \max\left(\sqrt{\tilde{\sigma}_k^4\frobn{A_{k}\bX^T\bX A_{k}^T}^2(f_n+\log 2)/t_2}, \tilde{\sigma}_k^2\opn{A_{k}\bX^T\bX A_{k}^T}(f_n+\log 2)/t_2\right)$, we have that, with probability at least $1-e^{-f_n}/2$,
\begin{equation} \label{t6probbound}
T_6 = \bigO{q} + \bigO{\sqrt{q \,f_n}} + \bigO{f_n}.
\end{equation}

Collecting the bounds in (\ref{t4bound}), (\ref{t5probbound}), and (\ref{t6probbound}), we conclude that there exists a positive constant $C_2$ such that, with probability at least $1-e^{-f_n}$,
\begin{equation*}
n^{-1}\ltwon{\bX(\hat{\bpi_{k}} - \bpi_{k})}^2 \le C_2 (r_{nk}\lor q \lor f_n)/n.
\end{equation*}

\section*{Appendix D: Proof of Theorem~\ref{theoremAdaConsisit}}

Let
\[
g_n = C_2 \left(r_{\max}\lor q \lor f_n\right)/n + 2c_1 C_2 \lonen{\bpi} \sqrt{\left(r_{\max}\lor q\lor f_n\right)/n}.
\]
We will first prove some lemmas, and then proceed to prove Theorem~\ref{theoremAdaConsisit}.

\begin{lemma} \label{prophatc5} Suppose that there exists a positive constant $\phi_0$  such that $\phivarmin{k}{\bH_k\bX\bpi_{-k}} \ge \phi_0$ for all $k$. If
	\begin{equation} \label{ineq:RERegularity1}
		\sqrt{(r_{\text{max}}\lor q \lor f_n)\big/n} +c_1\lonen{\bpi} \le \sqrt{c_1^2\lonen{\bpi}^2+\phi_0^2\big/(64C_2|\mathcal{A}_{k}|)}¡£
	\end{equation}
then, with probability at least $1- e^{-(f_n -\log p)}$, we have $\phivarmin{k}{\bH_k\bX\hat{\bpi}_{-k}} \ge \phi_0/2$.
\end{lemma}

\begin{proof} \label{proofhatc5}
	Note that the inequality~(\ref{ineq:RERegularity1}) implies that $g_n \le \frac{\phi_0^2}{64|\mathcal{A}_{k}|}$.  Then, for any index $i$ and $j$, we first investigate the bound of
\begin{eqnarray*}
\lefteqn{(\bH_k\bX\hat{\bpi}_{i})^T (\bH_k\bX\hat{\bpi}_{j}) - (\bH_k\bX \bpi_{i})^T (\bH_k\bX\bpi_{j})} \\
&=& \underbrace{(\hat{\bpi}_i-\bpi_{i})^T\bX^T\bH_k\bX(\hat{\bpi}_j-\bpi_{j})}_{T_7} + \underbrace{(\hat{\bpi}_i-\bpi_{i})^T\bX^T\bH_k\bX\bpi_{j}}_{T_8} + \underbrace{(\bX\bpi_{i})^T\bH_k\bX(\hat{\bpi}_j-\bpi_{j})}_{T_9}.
\end{eqnarray*}

Note that $\lambda_{max}(\bH_k) = 1$. By Theorem~\ref{ridgetheoryk}, we have, with probability at least $1-e^{-f_n}$,
\begin{equation} \label{eq-T7}
\begin{aligned}
|T_7| &\le \ltwon{\bH_k\bX (\hat{\bpi}_i - \bpi_{i} )}\times \ltwon{\bH_k\bX (\hat{\bpi}_j -  \bpi_{j} )}\\
& \le \lambda_{max}(\bH_k)\times \ltwon{\bX (\hat{\bpi}_i - \bpi_{i} )} \times \ltwon{\bX (\hat{\bpi}_j - \bpi_{j} )}
\le C_2 \left(r_{\max}\lor q\lor f_n\right).
\end{aligned}
\end{equation}

Following that $\ltwon{\bX\bpi_{j} } \le c_1\sqrt{n} \ltwon{\bpi_{j}}$, we have,
\begin{equation} \label{eq-T8}
\begin{aligned}
|T_8| &\le \ltwon{\bX\bpi_{j} }\times \ltwon{\bH_k\bX (\hat{\bpi}_i - \bpi_{i})} \le c_1\sqrt{n}\ltwon{\bpi_{j}}\times \ltwon{\bX (\hat{\bpi}_i - \bpi_{i} )}\\
& \le c_1C_2 \lonen{\bpi} \sqrt{n \left(r_{\max} \lor q \lor f_n\right)}.\\
\end{aligned}
\end{equation}
Similarly, we have,
\begin{equation} \label{eq-T9}
|T_9|  \le c_1\sqrt{n}\ltwon{\bpi_{i}} \ltwon{\bX (\hat{\bpi}_j - \bpi_{j} )} \le c_1C_2 \lonen{\bpi} \sqrt{n\left(r_{\max} \lor q\lor f_n\right)}.
\end{equation}

Putting together the bounds in (\ref{eq-T7}), (\ref{eq-T8}), and (\ref{eq-T9}), we have, with probability at least $1-e^{-f_n}$,
\begin{equation}
|(\bH_k\bX\hat{\bpi}_{i})^T (\bH_k\bX\hat{\bpi}_{j}) - (\bH_k\bX \bpi_{i})^T (\bH_k\bX\bpi_{j})|\le n g_n.
\end{equation}

By definition, for any set $\mathcal{A}_{k}$£¬ and any $\beta$, we have
\[
\lonen{\beta}^2 \le ( \lonen{\beta_{\mathcal{A}_{k}^c}}  + \lonen{\beta_{\mathcal{A}_{k}}})^2 \le ( 3\sqrt{|\mathcal{A}_k| }\ltwon{\beta_{\mathcal{A}_{k}}} + \sqrt{|\mathcal{A}_k| }\ltwon{\beta_{\mathcal{A}_{k}}}   )^2 = 16 |\mathcal{A}_k| \ltwon{\beta_{\mathcal{A}_{k}}}^2.
\]
We then have, with probability at least $1-p e^{-f_n}$,
\begin{equation*}
\begin{aligned}
\lefteqn{|\beta^T((\bH_k\bX\hat{\bpi}_{-k})^T (\bH_k\bX\hat{\bpi}_{-k}) - (\bH_k\bX\bpi_{-k})^T (\bH_k\bX\bpi_{-k})) \beta|\big/(n\ltwon{\beta_{\mathcal{A}_{k}} }^2)}\\
&\le \lonen{\beta}^2 \ltwon{\beta_{\mathcal{A}_{k} }}^{-2} \underset{i,j}{\text{max}}| (\bH_k\bX\hat{\bpi}_{i})^T  (\bH_k\bX\hat{\bpi}_{j}) -  (\bH_k\bX \bpi_{i} )^T (\bH_k\bX \bpi_{j} )  |/n\\
& \le 16 |\mathcal{A}_k| \times g_n \le 16 |\mathcal{A}_k| \times \phi_0^2\big/\left(64|\mathcal{A}_k|\right) = \phi_0^2/4,
\end{aligned}
\end{equation*}
which implies that $\phivarmin{k}{\bH_k\bX\hat{\bpi}_{-k}} \ge \phi_0/2$.
\end{proof}

\begin{lemma}(Basic Inequality) \label{lemmaBI}
Let random vector $\boldsymbol{J}_{k} = 2n^{-1} \hat{\bZ}_{-k}^T  \bH_k\bepsilon_{k} - 2n^{-1} \hat{\bZ}_{-k}^T \bH_k (\hat{\bZ}_{-k} - \bY_{-k}) \bgamma_{k} $ and $W_{k}^{-1} = diag(\boldsymbol{w}_{k}^{-1})$, then, for the event
\[
\mathcal{J}_k(\lambda_{k}) = \left\{ \infn{ W_{k}^{-1} \boldsymbol{J}_{k} } \le \lambda_{k}/n \right\},
\]
there exists a constant $C_3>0$ such that
\begin{equation*}
P(\mathcal{J}_k(\lambda_{k})) \ge 1- e^{-C_3 h_n+\log(4pq)}-e^{-f_n+\log(p)}.
\end{equation*}
Furthermore, concurring with the random vector $\boldsymbol{J}_{k}$, we have the following basic inequality,
\begin{equation} \label{basicinequality}
n^{-1} \ltwon{\bH_k\hat{\bZ}_{-k} (\hat{\bgamma}_k - \bgamma_{k})}^2 + 2n^{-1} \lambda_k \bweight_k^T |\hat{\bgamma}_k|
\le 2n^{-1} \lambda_k \bweight_k^T |\bgamma_{k}| +\boldsymbol{J}_{k}^T |\hat{\bgamma}_k - \bgamma_{k}|.
\end{equation}
\end{lemma}

\begin{proof} \label{Prof-lemmaBI}
With $\bY_{-k} = \bX \bpi_{-k} +\bxi_{-k}$ and $\hat{\bZ}_{-k} = \bX \hat{\bpi}_{-k}$, we have
\begin{eqnarray*}
\boldsymbol{J}_{k} &=& 2n^{-1} \hat{\bZ}_{-k}^T \bH_k\bepsilon_{k} - 2n^{-1} \hat{\bZ}_{-k}^T \bH_k (\hat{\bZ}_{-k} - \bY_{-k}) \bgamma_{k} \\
&=& 2n^{-1} \hat{\bpi}_{-k}^T \bX^T \bH_k\bepsilon_{k} - \frac{2}{n} \hat{\bpi}_{-k}^T \bX^T \bH_k (\bX\hat{\bpi}_{-k} - \bX \bpi_{-k} -\bxi_{-k}) \bgamma_{k } \\
&=& \underbrace{2n^{-1} (\hat{\bpi}_{-k} - \bpi_{-k})^T\bX^T\bH_k\bepsilon_{k}}_{T_{10}}
+ \underbrace{2n^{-1} \bpi_{-k}^T\bX^T\bH_k\bepsilon_{k}}_{T_{11}} + \underbrace{2n^{-1} (\hat{\bpi}_{-k} - \bpi_{-k})^T\bX^T \bH_k \bxi_{-k} \bgamma_{k}}_{T_{12}}\\
&& + \underbrace{2n^{-1} \bpi_{-k}^T \bX^T \bH_k \bxi_{-k} \bgamma_{k}}_{T_{13}} \underbrace{- 2n^{-1}(\hat{\bpi}_{-k} - \bpi_{-k})^T\bX^T\bH_k\bX(\hat{\bpi}_{-k}-\bpi_{-k}) \bgamma_{k}}_{T_{14}}\\
&& \underbrace{-2n^{-1} \bpi_{-k}^T\bX^T\bH_k\bX(\hat{\bpi}_{-k}-\bpi_{-k}) \bgamma_{k}}_{T_{15}}.
\end{eqnarray*}

Denote $\bX = (X_{\cdot 1}, X_{\cdot 2},\ldots,X_{\cdot q})$, then $X_{\cdot j}^T X_{\cdot j} = n$ due to standardization. With $\sigma^2_{pmax} = \underset{1\le k \le  p}{\text{max}} \sigma_k^2$, we have $\text{Var}(X_{\cdot j}^T \bH_k \bepsilon_{k}) = X_{\cdot j}^T \bH_k X_{\cdot j} \sigma_k^2 \le n \sigma_k^2 \le n\sigma_{pmax}^2$. Further let, for some constant $t_{\lambda}>0$,
\[
\lambda_{k} = t_{\lambda} \|\omega_{k}\|_{-\infty}^{-1} \lonen{\bGamma} \lonen{\bpi}\sqrt{n (r_{\text{max}}\lor q\lor f_n )\log p}.
\]
By the Gaussian tail inequality, we have
\begin{eqnarray*}
\lefteqn{P\left(\infn{W_{k}^{-1} T_{10}}\ge \lambda_{k}/(6n) \right) \le
	P\left(\infn{T_{10}}\ge \lambda_{k} \|\omega_{k}\|_{-\infty}/(6n) \right)} \\
&=& P\left(\infn{2n^{-1} (\hat{\bpi}_{-k} - \bpi_{-k})^T \bX^T\bH_k\bepsilon_{k}} \ge \lambda_{k} \|\omega_{k}\|_{-\infty}/(6n) \right) \\
&\le& P\left(\infn{(\hat{\bpi}_{-k} - \bpi_{-k})^T}\times \infn{2n^{-1} \bX^T\bH_k\bepsilon_{k}} \ge \lambda_{k} \|\omega_{k}\|_{-\infty}/(6n)\right)\\
&\le& P\left(\infn{2n^{-1} \bX^T\bH_k\bepsilon_{k}} \ge \lambda_{k} \|\omega_{k}\|_{-\infty}\big/(6n \delta_{\pi}) \right) \\
&\le&  q \exp\left\{-\lambda_{k}^2 \|\omega_{k}\|_{-\infty}^2\big/(288 n\sigma^2_{pmax} \delta_{\pi}^2) \right\}
= q\cdot p^{-\frac{n}{q} t_3 \lonen{\Gamma}^2\lonen{\pi}^2},
\end{eqnarray*}
where $t_3 = t_{\lambda}^2\big/\left(288 C_1 \sigma_{pmax}^2\right)$ and
\begin{equation*}
\delta_{\pi} = \underset{k}{\max}\lonen{\hat{\bpi}_{k} - \bpi_k} \le \underset{k}{\max}\sqrt{q}\ltwon{\hat{\bpi}_{k} - \bpi_k} = \sqrt{C_1 q(r_{\text{max}}\lor q \lor f_n)/n}.
\end{equation*}

Similarly, letting $t_4 = t_{\lambda}\big/\left(288\sigma_{pmax}^2\right)$, we have
\begin{eqnarray*}
\lefteqn{P\left(\infn{W_{k}^{-1}T_{11}}\ge \lambda_{k}\big/(6n)\right) \le P\left(\infn{T_{11}}\ge \lambda_{k} \|\omega_{k}\|_{-\infty}\big/(6n)\right)}\\
&=& P\left(\infn{2n^{-1} \bpi_{-k}^T \bX^T\bH_k\bepsilon_{k}} \ge \lambda_{k} \|\omega_{k}\|_{-\infty}\big/(6n)\right) \\ &\le& \mathbb{P}\left(\infn{ \bpi_{-k}^T} \infn{2n^{-1}\bX^T\bH_k\bepsilon_{k}} \ge \lambda_{k} \|\omega_{k}\|_{-\infty}\big/(6n) \right) \\
&\le& P\left(\infn{2n^{-1} \bX^T\bH_k\bepsilon_{k}} \ge \lambda_{k} \|\omega_{k}\|_{-\infty} \infn{\bpi_{-k}^T}^{-1}\big/(6n) \right) \\
&\le& q \exp \left\{-\lambda_{k}^2 \|\omega_{k}\|_{-\infty}^2 \infn{\bpi_{-k}^T }^{-2} \big/(288n \sigma^2_{pmax}) \right\} \\
&=& q \cdot p^{-t_4 \lonen{\Gamma}^2(r_{\max}\lor q \lor f_n)}.
\end{eqnarray*}

Let $\tilde{\sigma}_{pmax}^2 = \underset{k}{\text{max}}\ \text{Var}(\bxi_k)$ and $t_5 = t_{\lambda}\big/\left(288 C_1\tilde{\sigma}_{pmax}^2\right)$. For the term $T_{12}$,  we have
\begin{eqnarray*}
\lefteqn{P\left(\infn{W_{k}^{-1}T_{12}}\ge \lambda_{k}\big/(6n) \right) \le P\left(\infn{T_{12}}\ge \lambda_{k} \|\omega_{k}\|_{-\infty}\big/(6n) \right)} \\
&\le& P\left(\infn{(\hat{\bpi}_{-k} -\bpi_{-k})^T} \lonen{2n^{-1} \bX^T\bH_k \bxi_{-k}\bgamma_{k}} \ge \lambda_{k} \|\omega_{k}\|_{-\infty}\big/(6n) \right) \\
&\le& P\left(\delta_{\pi} \underset{i,j}{\max} \left|2n^{-1} \bx_i^T\bH_k\bxi_j\right| \lonen{\gamma_k} \ge \lambda_{k} \|\omega_{k}\|_{-\infty}\big/(6n)\right) \\
&\le& P\left(\underset{i,j}{\max} \left|2n^{-1} \bx_i^T\bH_k\bxi_j\right| \ge \lambda_{k} \|\omega_{k}\|_{-\infty}\lonen{\gamma_k}^{-1}\big/(6n \delta_{\pi}) \right) \\
&\le& q p \exp\left\{-\lambda_{k}^2 \|\omega_{k}\|_{-\infty}^2 \tilde{\sigma}_{pmax}^{-2} \delta_{\pi}^{-2}\lonen{\gamma_k}^{-2}\big/(288 n) \right\} = q p^{1- t_5\lonen{\pi}^2 n/q}.
\end{eqnarray*}

Letting $t_6 = t_{\lambda}\big/\left(288\tilde{\sigma}_{pmax}^2\right)$, we similarly have
\begin{eqnarray*}
\lefteqn{P\left(\infn{W_{k}^{-1}T_{13}}\ge \lambda_{k}\big/(6n) \right)}\\
&\le& q p \exp \left\{-\lambda_{k}^2 \tilde{\sigma}_{pmax}^{-2} \infn{\bpi_{-k}^T}^{-2}\lonen{\gamma_k}^{-2} \big/(288 n) \right\} = q p ^{1-t_6 (r_{max}\lor q \lor f_n)}.
\end{eqnarray*}

When $t_{\lambda}$ is sufficiently large, say $t_{\lambda}\ge 6 C_2\lonen{\bpi}^{-1} \sqrt{(r_{max}\lor q \lor f_n)/(n \log p)}$, we have
\begin{eqnarray*}
\lefteqn{\infn{W_{k}^{-1}T_{14}} \le n^{-1}\|\omega_{k}\|_{-\infty}^{-1} \lonen{\gamma_k} \underset{i,j}{\max} | (\hat{\bpi}_{i} - \bpi_{i})^T\bX^T\bH_k\bX(\hat{\bpi}_{j} -\bpi_{j})|} \\
&\le& n^{-1}\|\omega_{k}\|_{-\infty}^{-1} \lonen{\gamma_k} \underset{i,j}{\max} \ltwon{\bH_k\bX(\hat{\bpi}_{i} -\bpi_{i})}\ltwon{\bH_k\bX(\hat{\bpi}_{j} -\bpi_{j})}\\
&\le&  n^{-1}\|\omega_{k}\|_{-\infty}^{-1} \lonen{\gamma_k} \underset{i,j}{\max} \lambda_{\text{max}}(\bH_k) \ltwon{\bX(\hat{\bpi}_{i} -\bpi_{i})}\ltwon{\bX(\hat{\bpi}_{j} -\bpi_{j})}\\
&\le& n^{-1}\|\omega_{k}\|_{-\infty}^{-1}\lonen{\gamma_k} \underset{i,j}{\max} \ltwon{\bX(\hat{\bpi}_{i} -\bpi_{i})}\ltwon{\bX(\hat{\bpi}_{j} -\bpi_{j})} \\
&\le& C_2 \|\omega_{k}\|_{-\infty}^{-1} \lonen{\gamma_k} n^{-1}(r_{\max}\lor q \lor f_n) \\
&\le& \left\{\lambda_{k}\big/(6n)\right\} \times \left\{6 C_2 t_{\lambda}^{-1}\lonen{\bpi}^{-1} \sqrt{n^{-1}(\log p)^{-1}(r_{max}\lor q \lor f_n)} \right\} \le \lambda_{k}\big/(6n).
\end{eqnarray*}

Similarly, when $t_{\lambda}\ge 12\sqrt{C_2/\log p}$,
\begin{eqnarray*}
\lefteqn{\infn{W_{k}^{-1}T_{15}} \le 2 n^{-1} \lonen{\gamma_k} \infn{\bpi_{-k}^T} \|\omega_{k}\|_{-\infty}^{-1} \underset{i,j}{\max} |X_{\cdot i}^T\bH_k\bX(\hat{\bpi}_{j} -\bpi_{j})|}\\
&\le& 2n^{-1/2} \lonen{\gamma_k}\infn{\bpi_{-k}^T} \|\omega_{k}\|_{-\infty}^{-1}
\underset{j}{\max} \ltwon{\bH_k\bX(\hat{\bpi}_{j} -\bpi_{j})}\\
&\le& 2n^{-1/2} \lonen{\gamma_k}\infn{\bpi_{-k}^T} \|\omega_{k}\|_{-\infty}^{-1} \underset{j}{\max} \ltwon{\bX(\hat{\bpi}_{j} -\bpi_{j})} \\
&\le& \left\{\lambda_{k}\big/(6n)\right\} \times \left\{12t_{\lambda}^{-1}\sqrt{C_2\big/\log p} \right\} \le \lambda_{k}\big/(6n).
\end{eqnarray*}

Putting together all the above results, we have, for some constant $C_3>0$,
\begin{equation*}
P(\mathcal{J}_{k}(\lambda_{k})) \ge 1- e^{-C_3 h_n+\log(4pq)}-e^{-f_n+\log(p)}.
\end{equation*}

Concurring with the random vector $\boldsymbol{J}_{k}$, we have the following inequality based on the optimality of $\hat{\bgamma}_k$,
\begin{equation} \label{optimalityinequlity}
\ltwon{\bH_k\bY_k -\bH_k \hat{\bZ}_{-k} \hat{\bgamma}_k} + 2\lambda_{k} \bweight_k^T |\hat{\bgamma}_k| \le  \ltwon{\bH_k\bY_k -\bH_k \hat{\bZ}_{-k} \bgamma_{k}} + 2\lambda_{k} \bweight_k^T |\bgamma_{k}|.
\end{equation}
With $\bH_k\bY_k = \bH_k \bY_{-k} \bgamma_{k} + \bH_k\bepsilon_{k}$, we also have
\begin{eqnarray}
\lefteqn{\ltwon{ \bH_k\bY_k -\bH_k \hat{\bZ}_{-k} \hat{\bgamma}_k }^2} \nonumber\\
&=& \ltwon{\bH_k \bY_{-k}\bgamma_{k } + \bH_k\bepsilon_{k} -\bH_k \hat{\bZ}_{-k} \hat{\bgamma}_k }^2 \nonumber\\
&=& \ltwon{\bH_k\bepsilon_{k}}^2 - 2\bepsilon_{k}^T\bH_k(\hat{\bZ}_{-k} \hat{\bgamma}_k -\bY_{-k} \bgamma_{k }) + \ltwon{\bH_k \hat{\bZ}_{-k} \hat{\bgamma}_k -\bH_k \bY_{-k} \bgamma_{k}}^2 \nonumber\\
&=& \ltwon{\bH_k\bepsilon_{k}}^2 - 2\bepsilon_{k}^T\bH_k(\hat{\bZ}_{-k} \hat{\bgamma}_k -\bY_{-k} \bgamma_{k }) + \ltwon{\bH_k\hat{\bZ}_{-k} (\hat{\bgamma}_k - \bgamma_{k})}^2 \nonumber\\
& & + \ltwon{\bH_k(\hat{\bZ}_{-k} -\bY_{-k}) \bgamma_{k}}^2 + 2\bgamma_{k}^T (\hat{\bZ}_{-k} -\bY_{-k})^T\bH_k\hat{\bZ}_{-k}(\hat{\bgamma}_k - \bgamma_{k}), \label{ltwoleft} \\
\lefteqn{\ltwon{\bH_k\bY_k - \bH_k\hat{\bZ}_{-k} \bgamma_{k}}^2}\nonumber\\
&=& \ltwon{\bH_k\bY_{-k} \bgamma_{k} + \bH_k\bepsilon_{k} -\bH_k\hat{\bZ}_{-k} \bgamma_{k}}^2 \nonumber \\
&=& \ltwon{\bH_k\bepsilon_{k}}^2 + \ltwon{\bH_k(\hat{\bZ}_{-k}-\bY_{-k}) \bgamma_{k}}^2 - 2\bepsilon_{k}^T\bH_k(\hat{\bZ}_{-k} -\bY_{-k}) \bgamma_{k}. \label{ltworight}
\end{eqnarray}
Combining the equations (\ref{optimalityinequlity}), (\ref{ltwoleft}), and (\ref{ltworight}), we obtain that
\begin{eqnarray*}
\lefteqn{n^{-1}\ltwon{\bH_k\hat{\bZ}_{-k} (\hat{\bgamma}_k - \bgamma_{k})}^2 + 2n^{-1} \lambda_{k} \bweight_k^T |\hat{\bgamma}_k|} \\
&\le&  2n^{-1} \lambda_{k} \bweight_k^T |\bgamma_ {k}| + \left(\frac{2}{n} \hat{\bZ}_{-k}^T \bH_k\bepsilon_{k} - 2n^{-1} \hat{\bZ}_{-k}^T \bH_k (\hat{\bZ}_{-k} - \bY_{-k}) \bgamma_{k}\right)^T (\hat{\bgamma}_k - \bgamma_{k}) \\
&=& 2n^{-1} \lambda_{k} \bweight_k^T |\bgamma_ {k}|+ \boldsymbol{J}_{k}^T (\hat{\bgamma}_k - \bgamma_{k}),
\end{eqnarray*}
which concludes the proof.
\end{proof}

By the basic inequality we just proved above and condition on the event $\mathcal{J}_{k}(\lambda_{k})$, we have that
\begin{eqnarray*}
\lefteqn{n^{-1} \ltwon{\bH_k\hat{\bZ}_{-k} (\hat{\bgamma}_k - \bgamma_{k})}^2 \le 2n^{-1} \lambda_{k} \bweight_k^T |\bgamma_ {k}| - 2n^{-1} \lambda_{k} \bweight_k^T |\hat{\bgamma}_k| + \boldsymbol{J}_{k}^T (\hat{\bgamma}_k - \bgamma_{k})}\\
&\le& 2n^{-1} \lambda_{k} \boldsymbol{\omega}^T_{k,\mathcal{A}_{k}} |\gamma_{k,\mathcal{A}_{k}}| -
2n^{-1} \lambda_{k} \boldsymbol{\omega}^T_{k,\mathcal{A}_{k}} |\hat{\gamma}_{k,\mathcal{A}_{k}}| - 2n^{-1} \lambda_{k} \boldsymbol{\omega}^T_{k,\mathcal{A}_{k}^{c}} |\hat{\gamma}_{k,\mathcal{A}_{k}^{c}}| \\
&& + \boldsymbol{J}_{k,\mathcal{A}_{k}^{c}}^T(\hat{\gamma}_{k,\mathcal{A}_{k}^{c}}) +  \boldsymbol{J}_{k,\mathcal{A}_{k}}^T(\hat{\gamma}_{k,\mathcal{A}_{k}} -\gamma_{k,\mathcal{A}_{k}}) \\
&\le& 2n^{-1} \lambda_{k} \boldsymbol{\omega}^T_{k,\mathcal{A}_{k}} |\hat{\gamma}_{k,\mathcal{A}_{k}} -
\gamma_{k,\mathcal{A}_{k}}| - 2n^{-1} \lambda_{k} \boldsymbol{\omega}^T_{k,\mathcal{A}_{k}^{c}} |\hat{\gamma}_{k,\mathcal{A}_{k}^{c}}| \\
&& + n^{-1}\lambda_{k} \boldsymbol{\omega}^T_{k,\mathcal{A}_{k}^{c}}|\hat{\gamma}_{k,\mathcal{A}_{k}^{c}}| + n^{-1}\lambda_{k} \boldsymbol{\omega}^T_{k,\mathcal{A}_{k}} |\hat{\gamma}_{k,\mathcal{A}_{k}} -\gamma_{k,\mathcal{A}_{k}}|
\\
&\le& 3n^{-1}\lambda_{k} \boldsymbol{\omega}^T_{k,\mathcal{A}_{k}} |\hat{\gamma}_{k,\mathcal{A}_{k}} -
\gamma_{k,\mathcal{A}_{k}}|- n^{-1}\lambda_{k} \boldsymbol{\omega}^T_{k,\mathcal{A}_{k}^{c}} |\hat{\gamma}_{k,\mathcal{A}_{k}^{c}}|\\
&\le& 3n^{-1} \lambda_{k} \|\omega_{k,\mathcal{A}_{k}}\|_{\infty} \lonen{\hat{\gamma}_{k,\mathcal{A}_{k}} -
\gamma_{k,\mathcal{A}_{k}}} - n^{-1}\lambda_{k} \|\omega_{k,\mathcal{A}_{k}^{c}}\|_{-\infty} \lonen{\hat{\gamma}_{k,\mathcal{A}_{k}^{c}}},
\end{eqnarray*}
which implies that
\begin{equation} \label{eqn-lambdaineq}
n^{-1}\lambda_{k} \|\omega_{k,\mathcal{A}_{k}^{c}}\|_{-\infty}
\lonen{\hat{\gamma}_{k,\mathcal{A}_{k}^{c}}} \le 3n^{-1}\lambda_{k} \|\omega_{k,\mathcal{A}_{k}}\|_{\infty} \lonen{\hat{\gamma}_{k,\mathcal{A}_{k}} - \gamma_{k,\mathcal{A}_{k}}}.
\end{equation}
Note that $\|\omega_{k,\mathcal{A}_{k}}\|_{\infty} \|\omega_{k,\mathcal{A}_{k}^{c}}\|_{-\infty}^{-1} \le1$, we have that
\begin{eqnarray}
\lefteqn{\lonen{\hat{\gamma}_{k,\mathcal{A}_{k}^{c}} -\gamma_{k,\mathcal{A}_{k}^{c}}} } \nonumber\\
 &\le&  3\|\omega_{k,\mathcal{A}_{k}}\|_{\infty}\|\omega_{k,\mathcal{A}_{k}^{c}}\|_{-\infty}^{-1} \lonen{\hat{\gamma}_{k,\mathcal{A}_{k}} - \gamma_{k,\mathcal{A}_{k}}}
 \le 3\lonen{\hat{\gamma}_{k,\mathcal{A}_{k}} - \gamma_{k,\mathcal{A}_{k}}}.
\end{eqnarray}

On the other hand, following Lemma \ref{prophatc5}, we have, with $C_4 = 6t_{\lambda}$,
\begin{eqnarray*}
\lefteqn{n^{-1}\ltwon{\bH_k\hat{\bZ}_{-k} (\hat{\bgamma}_k - \bgamma_{k})}^2
\le 3n^{-1}\lambda_{k} \|\omega_{k,\mathcal{A}_{k}}\|_{\infty} \sqrt{|\mathcal{A}_{k}|} \ltwon{\hat{\gamma}_{k,\mathcal{A}_{k}} - \gamma_{k,\mathcal{A}_{k}}} }\\
&\le& 3n^{-1}\lambda_{k} \|\omega_{k,\mathcal{A}_{k}}\|_{\infty} \sqrt{|\mathcal{A}_{k}|}
\times 2 n^{-1/2}\phi_0^{-1} \ltwon{\bH_k\hat{\bZ}_{-k} (\hat{\bgamma}_k - \bgamma_{k})}\\
&\le& 36 n^{-2}\phi_0^{-2} \|\omega_{k,\mathcal{A}_{k}}\|_{\infty}^2 |\mathcal{A}_{k}| \lambda_{k}^2\\
&=& C_4^2 \phi_0^{-2} \|\omega_{k}\|_{-\infty}^{-2}  \|\omega_{k,\mathcal{A}_{k}}\|_{\infty}^2 \lonen{\bpi}^2\lonen{\bGamma}^2 |\mathcal{A}_{k}| (r_{\text{max}}\lor q\lor f_n )\log p\big/n.
\end{eqnarray*}

Employing the inequality (\ref{eqn-lambdaineq}), along with $\|\omega_{k,\mathcal{A}_{k}}\|_{\infty} \|\omega_{k,\mathcal{A}_{k}^{c}}\|_{-\infty}^{-1} \le1$, we have
\begin{eqnarray*}
\lefteqn{\lonen{\hat{\gamma}_{k} -\gamma_{k }}\le \left(3\|\omega_{k,\mathcal{A}_{k}}\|_{\infty} \|\omega_{k,\mathcal{A}_{k}^{c}}\|_{-\infty}^{-1} +1\right) \lonen{\hat{\gamma}_{k,\mathcal{A}_{k}} -
	\gamma_{k,\mathcal{A}_{k}}}} \\
&\le& \left(3\|\omega_{k,\mathcal{A}_{k}}\|_{\infty} \|\omega_{k,\mathcal{A}_{k}^{c}}\|_{-\infty}^{-1} +1\right)\sqrt{|\mathcal{A}_{k}|} \ltwon{\hat{\gamma}_{k,\mathcal{A}_{k}} -
	\gamma_{k,\mathcal{A}_{k}}} \\
&\le& \left(6\|\omega_{k,\mathcal{A}_{k}}\|_{\infty} \|\omega_{k,\mathcal{A}_{k}^{c}}\|_{-\infty}^{-1} +2\right) \sqrt{|\mathcal{A}_{k}|}\times n^{-1/2}\ltwon{\bH_k\hat{\bZ}_{-k} (\hat{\bgamma}_k - \bgamma_{k})} \phi_0^{-1}\\
&\le& 8C_4 \times \|\omega_{k,\mathcal{A}_{k}}\|_{\infty} \lonen{\bpi} \lonen{\bGamma} \phi_0^{-2} \|\omega_{k}\|_{-\infty}^{-1} \\
&& \times |\mathcal{A}_{k}| \sqrt{(r_{\text{max}}\lor q\lor f_n)\log p \big/n}.
\end{eqnarray*}
Since we condition on event $\mathcal{J}_{k}(\lambda_{k})$, the above prediction and estimation bounds hold with probability at least $1-e^{-C_3 h_n+\log(4pq)}-e^{-f_n+\log(p)}$.

\section*{Appendix E: Proof of Theorem~\ref{selectionconsistency}}

Denote $\hat{V}_{k} = (\hat{v}_{ij})_{(p-1)\times(p-1)} \triangleq n^{-1} \hat{\bpi}_{-k}^T\bX^T\bH_k\bX\hat{\bpi}_{-k} $, $\hsto = (\hat{v}_{ij})_{i\in \mathcal{A}_{k}^c,j\in \mathcal{A}_{k}}$, and $\hsoo = (\hat{v}_{ij})_{i\in \mathcal{A}_{k},j\in \mathcal{A}_{k}}$. The proof of Theorem~\ref{selectionconsistency} will be presented after the following lemma.

\begin{lemma} \label{predirrepresentable}
Assume that, for each node $i$, the following inequality holds.
\begin{eqnarray} \label{ineq:RERegularity2}
	\lefteqn{\sqrt{(r_{\text{max}}\lor q \lor f_n)/n} +c_1\lonen{\bpi}} \nonumber \\
	&\le& \sqrt{c_1^2\lonen{\bpi}^2+\min(\phi_0^2/64,\zeta(4-\zeta)^{-1}\|\bomega_{k}\|_{-\infty}/\theta_{i})/(C_2|\mathcal{A}_{k}|)}.
\end{eqnarray}
Under the assumptions and conditions of Theorem~\ref{selectionconsistency}, we have that, with probability at least $1- p e^{-f_n}$,
\[
\infn{W^{-1}_{k,\mathcal{A}^c_k}\left(\hsto \hsoo[-1]\right)W_{k,\mathcal{A}_{k}}}\le 1-\zeta/2.
\]
\end{lemma}

\begin{proof}
Following Theorem~\ref{ridgetheoryk}, we have, with probability at least $1-p e^{-f_n}$,
\begin{equation*}
n^{-1} \underset{i,j}{\max}\left|(\bH_k\bX\hat{\bpi}_{i})^T (\bH_k\bX\hat{\bpi}_{j}) - (\bH_k\bX \bpi_{i})^T (\bH_k\bX \bpi_{j})\right| \le g_n.
\end{equation*}

The inequality~(\ref{ineq:RERegularity2}) implies that $\theta_{k} \|\omega_{k,\mathcal{A}_{k}}\|_{-\infty}^{-1} |\mathcal{A}_k| g_n\le  \zeta/(4-\zeta)$, we have
\begin{equation*}
\infn{W_{k,\mathcal{A}_{k}}^{-1}(\hsoo - \soo)} \le \|\omega_{k,\mathcal{A}_{k}}\|_{-\infty}^{-1}  |\mathcal{A}_k| g_n\le  \zeta\big/\{(4-\zeta)\theta_{k}\}.
\end{equation*}

Similarly we have that
\begin{equation*}
\infn{W_{k,\mathcal{A}_{k}^c}^{-1}(\hsto[] - \sto)} \le \zeta\big/\{(4-\zeta)\theta_{k}\}.
\end{equation*}

Applying the matrix inversion error bound in \cite{horn2012matrix}, we obtain
\begin{eqnarray*}
\lefteqn{\infn{\hsoo[-1] W_{k,\mathcal{A}_{k}} } \le \infn{\soo[-1] W_{k,\mathcal{A}_{k}} } + \infn{\hsoo[-1] W_{k,\mathcal{A}_{k}} -\soo[-1]W_{k,\mathcal{A}_{k}}}}\\
&\le& \theta_{k} + \theta_{k}\infn{W_{k,\mathcal{A}_{k}}^{-1}(\hsoo[] - \soo)}\left(1-\theta_{k}\infn{W_{k,\mathcal{A}_{k}}^{-1}(\hsoo[] - \soo)}\right)^{-1} \theta_{k} \\
&\le& \theta_{k}(4-\zeta)\big/(4-2\zeta).
\end{eqnarray*}

Therefore,
\begin{eqnarray*}
\lefteqn{\infn{W^{-1}_{k,\mathcal{A}_{k}^c}\left(\hsto \hsoo[-1]  - \sto \soo[-1]  \right)W_{k,\mathcal{A}_{k}}}}\\
&\le& \infn{W^{-1}_{k,\mathcal{A}_{k}^c}\left(\hsto-\sto\right) (\hsoo[-1])W_{k,\mathcal{A}_{k}}} \\
&& + \infn{W^{-1}_{k,\mathcal{A}_{k}^c}\sto\soo[-1]W_{k,\mathcal{A}_{k}} W_{k,\mathcal{A}_{k}}^{-1}\left(\hsoo-\soo\right) (\hsoo[-1])W_{k,\mathcal{A}_{k}}} \\
&\le& \infn{W^{-1}_{k,\mathcal{A}_{k}^c}\left(\hsto-\sto\right)} \infn{(\hsoo[-1])W_{k,\mathcal{A}_{k}}} \\
&& + \infn{W^{-1}_{k,\mathcal{A}_{k}^c}\sto\soo[-1]W_{k,\mathcal{A}_{k}}} \infn{W_{k,\mathcal{A}_{k}}^{-1}\left(  \hsoo   - \soo  \right)}\infn{(\hsoo[-1])W_{k,\mathcal{A}_{k}}} \\
&\le& \zeta/2,
\end{eqnarray*}
which implies that $\infn{W^{-1}_{k,\mathcal{A}_{k}^c}(\hsto \hsoo[-1])W_{k,\mathcal{A}_{k}}}\le 1-\zeta/2$.
\end{proof}

By the optimality of $\hat{\gamma}_k$, it must satisfy the KKT condition as follows,
\begin{equation}
\label{equ-kktraw}
-2n^{-1} (\bH_{k}\hat{\bZ}_{-k})^T(\bH_{k}\bY_k - \bH_{k}\hat{\bZ}_{-k}\hat{\gamma}_{k}) + 2n^{-1}\lambda_{k} W_{k} \alpha_{k} =0,
\end{equation}
where $\infn{\alpha_{k}} \le 1$ and $\alpha_{kj}I[\hat{\gamma}_{kj}\ne 0] = sign(\hat{\gamma}_{kj})$. Plug in the equation $\bH_{k}\bY_{k} = \bH_{k}\bY_{-k}\bgamma_k + \bH_k\epsilon_k$, we can have that
\begin{eqnarray} \label{equ-errorexpand}
\lefteqn{\bH_{k}\bY_{k} - \bH_{k} \hat{\bZ}_{-k} \hat{\gamma}_{k}=\bH\bY_{-k}\gamma_{k} +\bH_{k}\epsilon_k - \bH_{k} \hat{\bZ}_{-k}\hat{\gamma}_{k}} \nonumber\\
&=& \bH_{k}\epsilon_k + \bH_{k}\bY_{-k}\gamma_{k} - -\bH_{k}\hat{\bZ}_{-k}\gamma_{k} + \bH_{k}\hat{\bZ}_{-k}\gamma_{k} - \bH_{k} \hat{\bZ}_{-k}\hat{\gamma}_{k} \nonumber\\
&=& \bH_{k} \epsilon_{k} - \bH_{k}(\hat{\bZ}_{-k} - \bY_{-k})\gamma_{k} -\bH_{k}\hat{\bZ}_{-k}(\hat{\gamma}_{k}-\gamma_{k}).
\end{eqnarray}
Combining (\ref{equ-kktraw}) and (\ref{equ-errorexpand}), we can get that
\begin{equation} \label{equ-KKT}
2\hat{V}_{k}(\hat{\gamma}_{k} - \gamma_{k}) -\boldsymbol{J}_k = -2 \lambda_{k} W_{k} \alpha_{k}\big/n,
\end{equation}
where $\boldsymbol{J}_k =  2n^{-1} \hat{\bZ}_{-k}^T  \bH_k\bepsilon_{k} - 2n^{-1} \hat{\bZ}_{-k}^T \bH_k (\hat{\bZ}_{-k} - \bY_{-k}) \bgamma_{k}$. For an estimator satisfying $\hat{\gamma}_{k,\mathcal{A}_{k}^{c}} = \gamma_{k,\mathcal{A}_{k}^{c}} = 0$, the above equation implies that
\begin{eqnarray} \label{equ-Aksplit}
\left\{\begin{array}{l}
2\hat{V}_{k,11}(\hat{\gamma}_{k,\mathcal{A}_{k}} - \gamma_{k,\mathcal{A}_{k}}) -\boldsymbol{J}_{k,\mathcal{A}_{k}} = -\lambda_{k}W_{k,\mathcal{A}_{k}} \alpha_{k,\mathcal{A}_{k}}\big/n, \\
2\hat{V}_{k,21}(\hat{\gamma}_{k,\mathcal{A}_{k}} - \gamma_{k,\mathcal{A}_{k}}) -\boldsymbol{J}_{k,\mathcal{A}_{k}^{c}} = -\lambda_{k}W_{k,\mathcal{A}_{k}^{c} } \alpha_{k,\mathcal{A}_{k}^{c}}\big/n. \end{array}\right.
\end{eqnarray}
Manipulating the above equations, we have that
\begin{eqnarray} \label{equ-gammadiff}
\hat{\gamma}_{k,\mathcal{A}_{k}} - \gamma_{k,\mathcal{A}_{k}} &=& 2^{-1} \hat{V}_{k,11}^{-1} (\boldsymbol{J}_{k,\mathcal{A}_{k}} -\lambda_{k} W_{k,\mathcal{A}_{k}}^T \alpha_{k,\mathcal{A}_{k}}) \nonumber \\
&=& 2^{-1} \hat{V}_{k,11}^{-1} W_{k, \mathcal{A}_{k}} (W_{k,\mathcal{A}_{k}}^{-1} \boldsymbol{J}_{k,\mathcal{A}_{k}}   -\lambda_{k} \alpha_{k,\mathcal{A}_{k}}).
\end{eqnarray}
Following the similar strategy in proving Lemma \ref{lemmaBI}, we can prove that there exists a constant $C_5>0$ such that $\infn{W_k^{-1}\boldsymbol{J}_k} \le 2\lambda_{k}\zeta\big/\{n(4-\zeta)\}$ with probability at least $1- e^{-C_5 h_n+\log(4pq)}-e^{-f_n+\log(p)}$. Therefore, with $\infn{\alpha_{k,\mathcal{A}_{k}}} \le 1$, we have that
\begin{eqnarray*}
\lefteqn{\infn{\hat{\gamma}_{k,\mathcal{A}_{k}} - \gamma_{k,\mathcal{A}_{k}} } \le 2^{-1} \infn{\hat{V}_{k,11}^{-1} W_{k,\mathcal{A}_{k}}} (\infn{W_{k,\mathcal{A}_{k}}^{-1} \boldsymbol{J}_{k,\mathcal{A}_{k}} } + 2n^{-1}\lambda_{k})}\\
&\le& \{\theta_{k} (4-\zeta)\big/(2-\zeta)\} \times \{4/(4-\zeta)\} \times \{2\lambda_{k}\big/n\} = 2\lambda_{k}\theta_k\big/\{n(2-\zeta)\} \le \underset{j\in \mathcal{A}_{k}}{\text{min}} |\gamma_{kj}|.
\end{eqnarray*}
The above inequality implies that $sign(\hat{\gamma}_{k,\mathcal{A}_{k}}) = sign(\gamma_{k,\mathcal{A}_{k}})$.

Combining (\ref{equ-Aksplit}) and (\ref{equ-gammadiff}), we can also verify that
\begin{eqnarray*}
\lefteqn{\infn{W_{k,\mathcal{A}_{k}^{c}}^{-1} \hat{V}_{k,21} (\hat{V}_{k,11})^{-1}(\boldsymbol{J}_{k,\mathcal{A}_{k}} - 2\lambda_{k}W_{k,\mathcal{A}_{k}} \alpha_{k,\mathcal{A}_{k}}/n) - W_{k,\mathcal{A}_{k}^{c} }^{-1} J_{k,\mathcal{A}_{k}^{c}}} }\\
&\le& \infn{W_{k,\mathcal{A}_{k}^{c}}^{-1} \hat{V}_{k,21} (\hat{V}_{k,11})^{-1} W_{k,\mathcal{A}_{k}}}
 (\infn{ W^{-1}_{k,\mathcal{A}_{k}} \boldsymbol{J}_{k}} +2\lambda_{k}/n) + \infn{W_{k,\mathcal{A}_{k}^{c}}^{-1} J_{k,\mathcal{A}_{k}^{c}}} \\
&\le& (1-\zeta/2)(4/(4-\zeta))2\lambda_{k}/n + \zeta/(4-\zeta) 2\lambda_{k}/n = 2\lambda_{k}/n.
\end{eqnarray*}
Therefore, there exists an estimator $\hat{\gamma}_k$ satisfying the KKT condition (\ref{equ-KKT}) as well as $sign(\hat{\gamma}_k) = sign(\gamma_k)$ which implies $\hat{\mathcal{A}}_k = \mathcal{A}_k$.

\vskip 0.2in

\begin{thebibliography}{}

\bibitem[Akaike(1974)]{Akaike1974}
Hirotugu Akaike.
\newblock A new look at the statistical model identification.
\newblock {\it IEEE Transactions on Automatic Control}, 19(6): 716-723, 1974.

\bibitem[Anderson and Rubin(1949)]{Anderson1949}
Theodore W. Anderson and Herman Rubin.
\newblock Estimation of the parameters of a single equation in a complete system of stochastic equations.
\newblock {\it The Annals of Mathematical Statistics}, 20(1): 46-63, 1949.

\bibitem[Aten {\it et al.}(2008)]{Aten2008}
Jason E. Aten, Tova F. Fuller, Aldons J. Lusis, and Steve Horvath.
\newblock Using genetic markers to orient the edges in quantitative trait networks: The NEO software.
\newblock {\it BMC Systems Biology}, 2: 34, 2008.

\bibitem[Basmann(1957)]{Basmann1957}
Robert L. Basmann.
\newblock A generalized classical method of linear estimation of coefficients in a structural equation.
\newblock {\it Econometrica}, 25(1): 77-83, 1957.

\bibitem[Belloni {\it et al.}(2012)]{Belloni2012}
Alexandre Belloni, Daniel Chen, Victor Chernozhukov, and Christian Hansen.
\newblock Sparse models and methods for optimal instruments with an application to eminent domain.
\newblock {\it Econometrics}, 80(6): 2369-2429, 2012.

\bibitem[Bollen(1996)]{Bollen1996}
Kenneth A. Bollen.
\newblock An alternative two stage least squares (2SLS) estimator for latent variable equations.
\newblock {\it Psychometrika}, 61(1): 109-121, 1996.

\bibitem[Brem and Kruglyak(2005)]{Brem2005}
Rachel B. Brem and Leonid Kruglyak.
\newblock The landscape of genetic complexity across 5,700 gene expression traits in yeast.
\newblock {\it Proceedings of the National Academy of Sciences of the United States of America}, 102(5): 1572-1577, 2005.

\bibitem[Broman and Speed(2002)]{Broman2002}
Karl W. Broman and Terence P. Speed.
\newblock A model selection approach for the identification of quantitative trait loci in experimental crosses.
\newblock{\it Journal of the Royal Statistical Society, Series B}, 64(4): 641-656, 2002.

\bibitem[Cai {\it et al.}(2013)]{Cai2013}
Xiaodong Cai, Juan Andr\'{e}s Bazerque, and Georgios B. Giannakis.
\newblock  Inference of gene regulatory networks with sparse structural equation models exploiting genetic perturbations.
\newblock {\it PLoS Computational Biology}, 9(5): e1003068, 2013.

\bibitem[de la Fuente {\it et al.}(2004)]{Fuente2004}
Alberto de la Fuente, Nan Bing, Ina Hoeschele, and Pedro Mendes.
\newblock Discovery of meaningful associations in genomic data using partial correlation coefficients.
\newblock {\it Bioinformatics}, 20(18): 3565-3574, 2004.

\bibitem[Dixon {\it et al.}(2007)]{Dixon2007}
Anna L. Dixon, Liming Liang, Miriam F. Moffatt, Wei Chen, Simon Heath, Kenny C. C. Wong, Jenny Taylor, Edward Burnett, Ivo Gut, Martin Farrall, G Mark Lathrop, Gon\c{c}alo R. Abecasis, and William O. C. Cookson.
\newblock A genome-wide association study of global gene expression.
\newblock {\it Nature Genetics}, 39(10): 1202-1207, 2007.

\bibitem[Fan and Li(2001)]{Fan2001}
Jianqing Fan and Runze Li.
\newblock Variable selection via nonconcave penalized likelihood and its oracle properties.
\newblock {\it Journal of the American Statistical Association}, 96(456): 1348-1360, 2001.

\bibitem[Frank and Friedman(1993)]{Frank1993}
Ildiko E. Frank and Jerome H. Friedman.
\newblock A statistical view of some chemometrics regression tools.
\newblock {\it Technometrics}, 35(2): 109-135, 1993.

\bibitem[Fu and Knight(2000)]{Fu2000}
Wenjiang Fu and Keith Knight.
\newblock Asymptotics for  lasso-type estimators.
\newblock {\it The Annals of Statistics}, 28(5): 1356-1378, 2000.

\bibitem[Gelfond {\it et al.}(2007)]{Gelfond2007}
Jonathan A. L. Gelfond, Joseph G. Ibrahim, and Fei Zou.
\newblock Proximity model for expression quantitative trait loci (eQTL) detection.
\newblock {\it Biometrics}, 63(4): 1108-1116, 2007.

\bibitem[Geyer(1994)]{Geyer1994}
Charles J. Geyer.
\newblock On the asymptotics of constrained M-estimation.
\newblock {\it The Annals of Statistics}, 22(4): 1993-2010, 1994.

\bibitem[Golub {\it et al.}(1979)]{Golub1979}
Gene H. Golub, Michael Heath, and Grace Wahba.
\newblock Generalized cross-validation as a method for choosing a good ridge parameter.
\newblock {\it Technometrics}, 21(2): 215-223, 1979.

\bibitem[Haavelmo(1943)]{Haavelmo1943}
Trygve Haavelmo.
\newblock The statistical implications of a system of simultaneous equations.
\newblock {\it Econometrica}, 11(1): 1-12, 1943.

\bibitem[Haavelmo(1944)]{Haavelmo1944}
Trygve Haavelmo.
\newblock The probability approach in econometrics.
\newblock {\it Econometrica}, 12: S1-S115, 1944.

\bibitem[Horn and Johnson(2012)]{horn2012matrix}
Roger A. Horn and Charles R. Johnson.
\newblock {\it Matrix Analysis}.
\newblock Cambridge University Press, 2012.

\bibitem[Huang {\it et al.}(2009)]{Huang2009}
Da Wei Huang, Brad T. Sherman, and Richard A. Lempicki.
\newblock Bioinformatics enrichment tools: paths toward the comprehensive functional analysis of large gene lists.
\newblock {\it Nucleic Acids Research}, 37(1): 1-13, 2009.

\bibitem[Huang {\it et al.}(2011)]{Huang2011}
Jian Huang, Shuangge Ma, Hongzhe Li, and Cun-Hui Zhang.
\newblock The sparse Laplacian shrinkage estimator for high-dimensional regression.
\newblock {\it The Annals of Statistics}, 39(4): 2021-2046, 2011.

\bibitem[Jansen and Nap(2001)]{Jansen2001}
Ritsert C. Jansen and Jan-Peter Nap.
\newblock Genetical genomics: the added value from segregation.
\newblock{\it Trends in Genetics}, 17(7): 388-391, 2001.

\bibitem[Jia and Xu(2007)]{Jia2007}
Zhenyu Jia and Shizhong Xu.
\newblock Mapping quantitative trait loci for expression abundance.
\newblock{\it Genetics}, 176(1): 611-623, 2007.

\bibitem[Kendziorski {\it et al.}(2006)]{Kendziorski2006}
Christina Kendziorski, Meng Chen, Ming Yuan, Hong Lan, and Alan D. Attie.
\newblock Statistical methods for expression quantitative trait loci (eQTL) mapping.
\newblock{\it Biometrics}, 62(1): 19-27, 2006.

\bibitem[Kennedy(1985)]{Kennedy1985}
Peter E. Kennedy.
\newblock {\it A Guide to Econometrics}. Cambridge, MA: MIT Press, 1985.

\bibitem[Lin {\it et al.}(2015)]{Lin2015}
Wei Lin, Rui Feng, and Hongzhe Li.
\newblock Regularization methods for high-dimensional instrumental variables regression with an application to genetical genomics.
\newblock {\it Journal of the American Statistical Association}, 110(509): 270-288, 2015.

\bibitem[Liu {\it et al.}(2008)]{Liu2008}
Bing Liu, Alberto de la Fuente, and Ina Hoeschele.
\newblock Gene network inference via structural equation modeling in genetical genomics experiments.
\newblock {\it Genetics}, 178(3): 1763-1776, 2008.

\bibitem[Logsdon and Mezey(2010)]{Logsdon2010}
Benjamin A. Logsdon and Jason Mezey.
\newblock Gene expression network reconstruction by convex feature selection when incorporating genetic perturbations.
\newblock {\it PLoS Computational Biology}, 6(12): e1001014, 2010.

\bibitem[Neto {\it et al.}(2008)]{Neto2008}
Elias Chaibub Neto, Christine T. Ferrara, Alan D. Attie, and Brian S. Yandell.
\newblock Inferring causal phenotype networks from segregating populations.
\newblock {\it Genetics}, 179(2): 1089-1100, 2008.

\bibitem[Reiers{\o}l(1941)]{Reiersol1941}
Olav Reiers{\o}l.
\newblock Confluence analysis by means of lag moments and other methods of confluence analysis.
\newblock {\it Econometrica}, 9(1): 1-24, 1941.

\bibitem[Reiers{\o}l(1945)]{Reiersol1945}
Olav Reiers{\o}l.
\newblock Confluence analysis by means of instrumental sets of variables.
\newblock {\it Arkiv for Mathematik, Astronomi och Fysik}, 32A(4), 1945.

\bibitem[Rudelson and Vershynin(2013)]{rudelson2013hanson}
Mark Rudelson and Roman Vershyninand.
\newblock Hanson-Wright inequality and sub-gaussian concentration.
\newblock {\it Electronic Communications in Probability}, 18: 1-9, 2013.

\bibitem[Schadt {\it et al.}(2003)]{Schadt2003}
Eric E. Schadt, Stephanie A. Monks, Thomas A. Drake, Aldons J. Lusis, Nam Che, Veronica Colinayo, Thomas G. Ruff, Stephen B. Milligan, John R. Lamb, Guy Cavet, Peter S. Linsley, Mao Mao, Roland B. Stoughton, and Stephen H. Friend.
\newblock Genetics of gene expression surveyed in maize, mouse and man.
\newblock {\it Nature}, 422: 297-302, 2003.

\bibitem[Schmidt(1976)]{Schmidt1976}
Peter Schmidt.
\newblock {\it Econometrics}. New York: Marcel Dekker, 1976.

\bibitem[Schwartz(1978)]{Schwartz1978}
Gideon Schwarz.
\newblock Estimating the dimension of a model.
\newblock {\it The Annals of Statistics}, 6(2): 461-464, 1978.

\bibitem[Shipley(2002)]{Shipley2002}
Bill Shipley.
\newblock {\it Cause and Correlation in Biology: A User's Guide to Path Analysis, Structural Equations and Causal Inference}. New York: Cambridge University Press, 2002.

\bibitem[Spirtes {\it et al.}(2001)]{Spirtes2001}
Peter Spirtes, Clark Glymour, and Richard Scheines.
\newblock {\it Causation, Prediction, and Search}. Boston: The MIT Press, 2001.

\bibitem[Theil(1953a)]{Theil1953a}
Henri Theil.
\newblock Repeated least-squares applied to complete equation systems.
\newblock {\it Mimeo. The Hague: Central Planning Bureau.}, 1953a.

\bibitem[Theil(1953b)]{Theil1953b}
Henri Theil.
\newblock Estimating and simultaneous correlation in complete equation systems.
\newblock {\it Mimeo. The Hague: Central Planning Bureau.}, 1953b.

\bibitem[Theil(1961)]{Theil1961}
Henri Theil.
\newblock {\it Economic Forecasts and Policy.} Amsterdam: North Holland, 1961.

\bibitem[Tibshirani(1996)]{Tibshirani1996}
Robert Tibshirani.
\newblock Regression shrinkage and selection via the lasso.
\newblock {\it Journal of the Royal Statistical Society, Series B}, 58(1): 267-288, 1996.

\bibitem[Xiong {\it et al.}(2004)]{Xiong2004}
Momiao Xiong, Jun Li, and Xiangzhong Fang.
\newblock Identification of genetic networks.
\newblock {\it Genetics}, 166(2): 1037-1052, 2004.

\bibitem[Zhang(2010)]{Zhang2010}
Cun-Hui Zhang.
\newblock Nearly unbiased variable selection under minimax concave penalty.
\newblock {\it The Annals of Statistics}, 38(2): 894-942, 2010.

\bibitem[Zhu(2015)]{Zhu2015}
Ying Zhu.
\newblock Sparse linear models and l$_1$-regularized 2SLS with high-dimensional endogenous regressors and instruments.
\newblock {\it MPRA} Paper No. 65703, 2015.

\bibitem[Zou(2006)]{Zou2006}
Hui Zou.
\newblock The adaptive lasso and its oracle properties.
\newblock {\it Journal of the American Statistical Association}, 101(476): 1418-1429, 2006.

\bibitem[Zou and Hastie(2005)]{Zou2005}
Hui Zou and  Trevor Hastie.
\newblock Regularization and variable selection via the elastic net.
\newblock {\it Journal of the Royal Statistical Society, Series B}, 67(2): 301-320, 2005.

\end{thebibliography}

\end{document}